\documentclass[graybox]{svmult}

\usepackage{mathptmx}        
\usepackage{helvet}          
\usepackage{courier}         

\usepackage{makeidx}         
\usepackage{graphicx}        
\usepackage{multicol}        
\usepackage[bottom]{footmisc}
\usepackage{amsmath}
\usepackage{amssymb}
\usepackage{aas_macros}

\makeindex             

\newcommand{\propdelay}[1]{\mathbf{D}_{#1}}
\newcommand{\tdidelay}[1]{D_{#1}}

\begin{document}

\tableofcontents


\title{Space-based Gravitational Wave Observatories}
\author{Jonathan Gair \thanks{corresponding author}, Martin Hewitson, Antoine Petiteau, Guido Mueller}
\institute{Jonathan R. Gair \at Max Planck Institute for Gravitational Physics (Albert Einstein Institute), Am M\"{u}hlenberg 1, Potsdam-Golm 14476, Germany, \email{jgair@aei.mpg.de}
\and Martin Hewitson \at Institut f\"ur Gravitationsphysik der Leibniz Universit\"at Hannover, Callinstr. 38, 30167 Hannover, Germany, \email{hewitson@aei.mpg.de}
\and Guido Mueller \at Department of Physics, 2001 Museum Rd, University of Florida, Gainesville, Florida 32611, USA, \email{gum1@ufl.edu}
\and Antoine Petiteau \at AstroParticule et Cosmologie (APC), Universit\'{e} de Paris/CNRS, 75013 Paris, France, \email{antoine.petiteau@cea.fr}}

%
%

\maketitle

\abstract*{In this article, which will appear as a chapter in the \textit{Handbook of Gravitational Wave Astronomy}, we will describe the detection of gravitational waves with space-based interferometric gravitational wave observatories. We will provide an overview of the key technologies underlying their operation, illustrated using the specific example of the Laser Interferometer Space Antenna (LISA). We will then give an overview of data analysis strategies for space-based detectors, including a description of time-delay interferometry, which is required to suppress laser frequency noise to the necessary level. We will describe the main sources of gravitational waves in the millihertz frequency range targeted by space-based detectors and then discuss some of the key science investigations that these observations will facilitate. Once again, quantitative statements given here will make reference to the capabilities of LISA, as that is the best studied mission concept. Finally, we will describe some of the proposals for even more sensitive space-based detectors that could be launched further in the future.}


\section{Introduction}
\label{sec:space_gw_intro}

As described in the chapter \textit{Terrestrial Laser Interferometers} of this volume, ground-based detectors will
soon be sensitive to gravitational waves (GWs) above a few Hertz generated
for example by systems involving a few thousand solar mass black holes
just before merger. To listen to systems which involve even larger
black holes such as the black holes in the centers of most galaxies,
or to study systems well before merger, like the hundreds of thousands
of compact binary systems in our own Milky way, we need observatories
which are sensitive between roughly $10\,\mu\text{Hz}$ and $100\,\text{mHz}$. This is the frequency range which is targeted by space-based observatories such as the Laser Interferometer Space Antenna or LISA.

LISA will be the first space-based gravitational-wave observatory with a launch scheduled for the first half of the 2030s. It is a European Space Agency (ESA) led space mission with major contributions expected from NASA and several other national space agencies. While LISA is also based on measuring minute changes in the separation of free falling test masses, its low frequency sensitivity, the necessary long arm lengths and the space environment require approaches which differ significantly from the techniques used in ground-based observatories. The LISA concept is essentially defined by five key features:
\begin{enumerate}
\item The observatory consists of a triangular constellation of three spacecraft in passive, heliocentric orbits.
\item Free-falling test masses inside each spacecraft serve as inertial reference points defining the ends of each measurement arm.
\item Each measurement arm is millions of kilometers long.
\item The primary measurement is continuous, one-way, interferometric, laser-ranging between the test masses or spacecraft.
\item Laser frequency noise is cancelled by combining time-delayed laser phases from different arms in post-processing (Time Delay Interferometry).
\end{enumerate}
In the first half of this document we will shed some light on each of these features and discuss their impact on the mission design and operation. The discussion on the first four points will have to be rather superficial; the devil is in the details and those are covered in the references given throughout this chapter. The last point, Time Delay Interferometry (TDI), is discussed in more detail starting with section~\ref{sec:TDI}. The TDI algorithm produces the data streams which are then used in the data analysis; the basics of which is described in section~\ref{sec:DA}. Section~\ref{sec:space_gw_sources} gives an overview of the gravitational wave sources of LISA followed by a discussion in section~\ref{sec:space_gw_science} of the science questions LISA will try to answer. Section~\ref{sec:space_gw_future} discusses prospects for proposed future missions beyond LISA which are loosely based on LISA technologies and concepts. Note that during the last two decades concepts based on atom interferometry have been developed and are discussed in a different chapter in this book. However, we will start with a brief historic review.\\

\noindent Following initial discussions within a small NASA working group~\cite{NASA_1975},
the first space-based gravitational wave observatory, LAGOS, was proposed
to NASA in the early 1980's by Peter Bender and colleagues~\cite{FirstOrbits_1980,Orbits_1984}.
In the early 1990's, two similar missions called LISA~\cite{LISA_Proposal_1993}
and Sagittarius~\cite{Sagittarius93} (which later evolved into Omega~\cite{Omega_98}) were proposed
to ESA. The LISA mission was selected as an M3 mission and later upgraded
to a cornerstone mission in ESA's Horizon 2000 program. In 1997, 
NASA and ESA joined forces and the by now almost classical LISA design
emerged~\cite{LISA_Y-Book98}. It is based on three spacecraft forming
a nearly equilateral triangle  initially five million kilometer arm lengths.
This constellation will be placed into a heliocentric orbit
trailing or leading Earth by about 20~deg. Each spacecraft
will host two free falling test masses in the form of 4~cm gold platinum
cubes. These test masses will form the end points of three interferometer
arms. Gravitational waves will change their distances and laser interferometers
will measure these changes. The shot noise limit defines LISA's sensitivity 
above a few mHz while acceleration noise will limit LISA below a few mHz. 

In the late '90s, the required acceleration noise was several orders of magnitude
better than the performance demonstrated by any prior space mission, 
calling for a technology demonstration before moving forward
with LISA proper. This LISA Pathfinder (LPF) mission was finally launched
in 2015 and its results exceeded all expectations~\cite{LPF_Results16}.
Also in 2015, the two LIGO observatories discovered the first of many
gravitational waves pushing the field to the forefront of science
\cite{PhysRevLett.116.061102}. In the mean time, programmatic constraints led to
a restructuring of the original ESA and NASA partnership and a resizing
of LISA to 2.5 million kilometer arm length \cite{LISA_Proposal2017}. This
new LISA mission will be the first space-based gravitational wave
mission with a launch scheduled for the first half of the 2030's. 

Starting around 2000, first proposals for beyond LISA space-based
gravitational wave observatories such as ALIA \cite{ALIA_2005,ALIA_2005_cqg}, BBO
\cite{BBO}, and DECIGO \cite{DECIGO} were published. These missions
were typically optimized to bridge the gap between LISA and ground-based
observatories and often targeted the expected primordial gravitational
wave background radiation. Later, several missions were proposed as
an alternative to LISA which at that time struggled to take the necessary
programmatic hurdles. These missions often promised lower cost using
for example geocentric orbits \cite{Gadfli2011,GEOGRAWI_11,Lagrange_11_Conklin}
or tried to provide back up options in case LPF would fail and the
acceleration noise requirements of LISA could not be met. While most
of these missions used the same approach of laser interferometry between
free falling macroscopic test masses some were based on atom interferometry.
Follow up studies first at NASA \cite{SGO_Report} and then at ESA
\cite{GOAT_Report} suggested that the original LISA approach was
still the most promising one and the latest discussion ended with
the success of LPF and the detections made by advanced LIGO. The following
sections will focus on LISA as it will be the first of its kind and will set the bar for future missions. However, we note that China is formulating plans for a space-based gravitational wave detector that could be launched on a similar timescale to LISA. The leading proposals are TianQin~\cite{TianQin:2015yph} and Taiji~\cite{Hu:2017mde}. These would both operate on the same principles to LISA, and Taiji would also be in heliocentric orbit, with comparable sensitivity to LISA. TianQin would be in Earth-orbit and have somewhat shorter armlengths, leading to slightly lower sensitivity, and an  optimal frequency range somewhat higher than LISA or Taiji. We will not discuss these further in this article, but the principles on which they would operate and the astrophysical sources they would observe are the same as those we will describe with reference to LISA.

\section{LISA}
\label{sec:space_gw_design}

Ground-based observatories reach their astonishing sensitivity by
maintaining very tight control of the positions of or distances between
their mirrors at frequencies below their measurement band. They use
(near-) equal arm Michelson interferometers to be insensitive to laser
frequency noise. Applying this design principle to space would require
station-keeping or continuously actuating on the spacecraft and the
test masses. The applied forces would have to be calibrated beyond
our current capabilities and have to have virtually zero amplitude
in the frequency range of interest to not limit the sensitivity of
the observatory. Such a mission would also consume significant amounts of fuel 
driving up the cost. Instead, LISA uses a technology that
is known as drag-free inertial sensing. 

A drag-free spacecraft encloses a free-falling test mass
that is shielded from external forces, like drag. The position and often also the orientation of the test mass within the 
spacecraft is sensed, and the spacecraft propulsion system is commanded to keep the spacecraft centered on 
the test mass in one or more degrees of freedom thereby forcing the spacecraft to follow a more perfect inertial orbit. 
In all other degrees of freedom, forces and torques are applied to the test mass to follow the spacecraft. The perfection of the inertial orbit or of the free falling test mass is determined by the residual forces on the test mass and the limitations of the control or actuation system. After minimizing those, the test mass can then serve as the inertial reference point needed in a gravitational wave observatory, benefiting from its fixed position within the spacecraft and shielded from most external forces. 

\begin{figure}
\begin{centering}
\includegraphics[width=8cm]{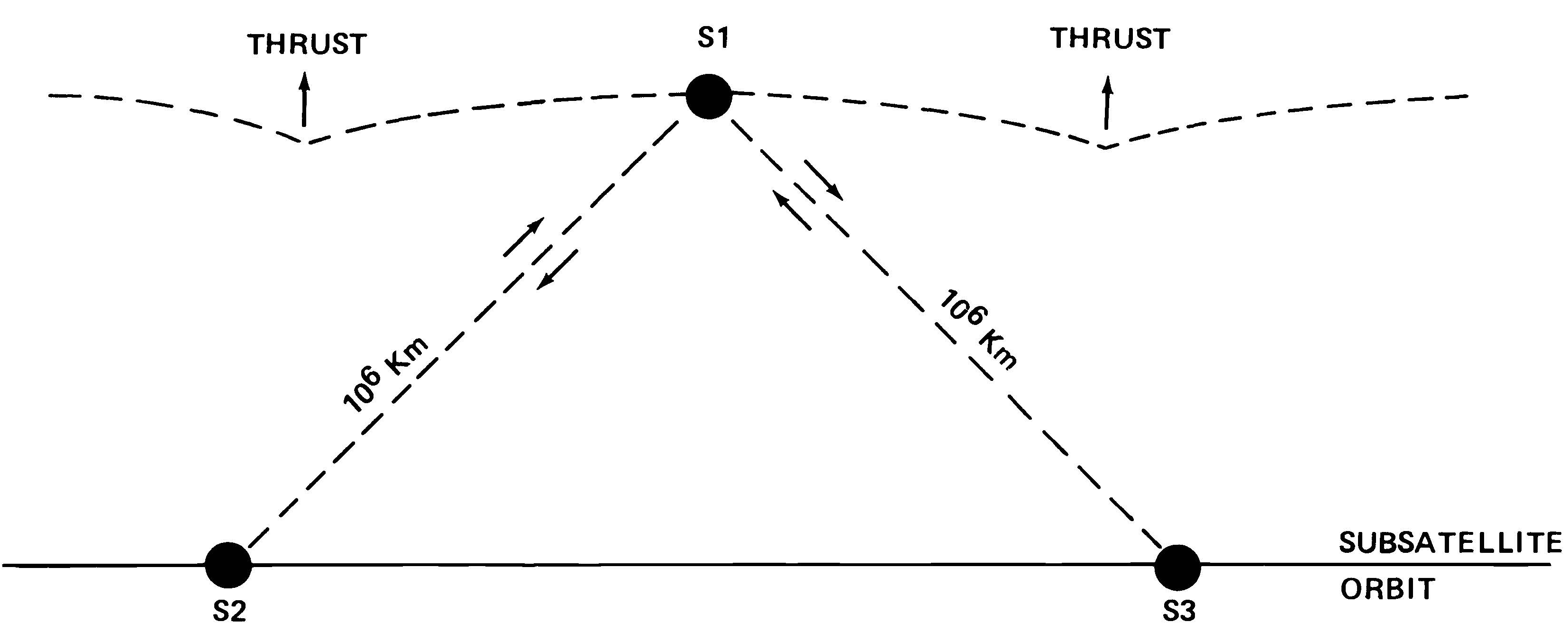}
\par\end{centering}
\vspace{0.5cm}
\begin{centering}
\includegraphics[width=8cm]{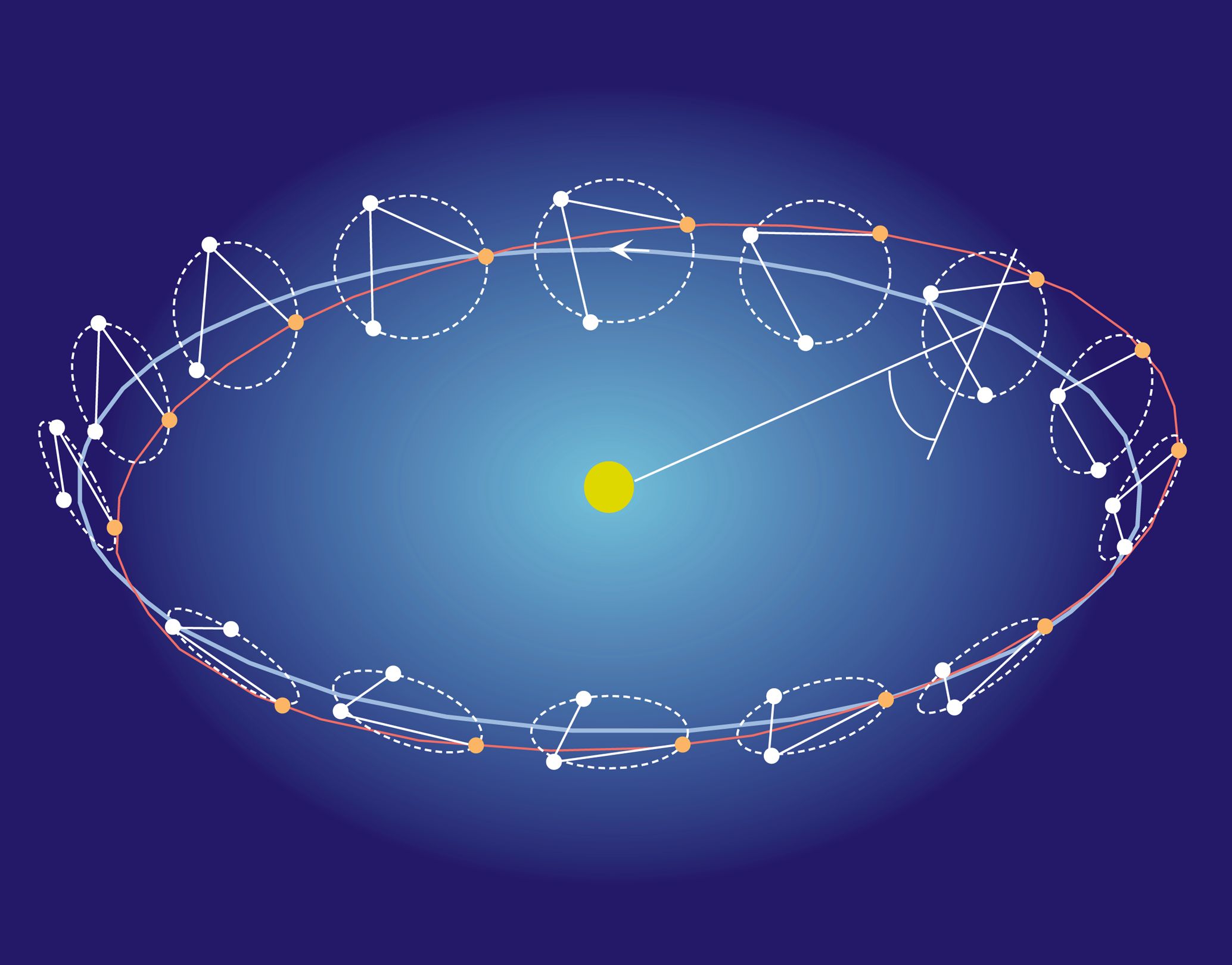}
\par\end{centering}
\caption{\label{Orbits} The top graph shows a concept of the initially proposed
orbits as published in \cite{FirstOrbits_1980}; two satellites S2 and S3 were placed
in one orbit, S3 trailing S2 by $\sqrt{2}L$ while S1 is placed in
an orbit $L/\sqrt{2}$ above the ecliptic which would require regular
station keeping maneuvers. The LISA orbits circumvent this issue by
using three identical orbital planes with an ellipticity of $\epsilon=2L/\sqrt{3}a.u.$
offset by $120^{\circ}$ from each other. Each plane is inclined by
an angle $\iota=\sqrt{3}\epsilon$ to minimize the relative motion
between the spacecraft. Credit for the lower part of the figure: JPL/NASA.}
\end{figure}

LISA will use three spacecraft in a (near-) equilateral triangular configuration. Maintaining this configuration over the decade long lifetime of the mission; ideally with little to no station-keeping maneuvers, requires very specific orbits. These orbits should minimize the relative velocities to minimize Doppler shifts of the frequencies of the exchanged laser beams. They should also minimize changes in the opening angles to reduce the requirements on each spacecraft internal pointing system which tracks each of the far spacecraft. Furthermore, the orbits need to be within range of the deep space network to communicate with the constellation from ground. Other constraints include the need that all orbital frequencies should ideally be well below the LISA measurement band and that the thermal environment should be extremely stable as temperature changes will be a driving noise source in low frequency gravitational-wave observatories; flying through Earth's shadow on a regular basis should be avoided if possible.

In the late 1970's, the orbit geometry shown at the top of figure \ref{Orbits} was being considered. It was based on a main spacecraft and two sub-spacecraft flying in nearly circular orbits around the sun with roughly one million km separations \cite{FirstOrbits_1980}. In order to keep the two arm lengths nearly constant, frequent active corrections to the orbits would be needed. However, in 1981, a new set of orbits was presented at a NIST meeting which avoided the station-keeping maneuver 
but allowed for larger differential arm lengths and opening angle changes.
In this design the main spacecraft was placed in an Earth trailing orbit and
the other two were placed in orbits that let them rotate around the
main spacecraft \cite{Orbits_1984}. This constellation then evolved
into the baseline LISA orbits \cite{Vincent_87,Orbits_LISA} shown
in the bottom part of figure \ref{Orbits}. 
In this configuration, the three spacecraft form an equilateral triangle 
of length $L$. The center of the constellation stays in the plane of the ecliptic trailing or leading
Earth. The orbital planes of each spacecraft are fine tuned to have
an eccentricity of $\epsilon=2L/\sqrt{3}a.u.$ and are inclined by
an angle $\iota=\sqrt{3}\epsilon$ with respect to the ecliptic. Finally,
rotating these three planes by $120^{\circ}$ with respect to each
other around the normal on the ecliptic centered on the sun forms
a near equilateral triangle with residual length changes on the order
of $2\epsilon^{2}\cdot a.u.$ \cite{Orbits_LISA}. 

These baseline orbits are the starting point for the final optimization
which takes into account non-centrosymmetric gravitational forces
from the planets that over time will pull the constellation apart
\cite{Orbit_Optimization_2008}. During LISA's 10+ years in orbit,
each arm will change by about $\pm35000\,\text{km}$ twice a year
resulting in changes of the opening angles by less than $1\,\text{deg}$.
Still, the changes in the opening angles are much larger
than the divergence angle of the laser beams and require constant
realignment of the constellation. The initial Sagittarius and Omega
proposals solved that issue by using two spacecraft in each corner
of the triangle. Each of these two spacecraft would be aligned to
direct the main laser beam to one of the far spacecraft while a secondary
laser link with actuated mirrors connected the two spacecraft in
each corner and allowed to  compare the laser phases in each interferometer arm.
This comparison allows to form what is sometimes called an artificial beam splitter.
This beam splitter does not physically combine the interfering laser beams from the two long interferometer arms, but, as described in section \ref{sec:TDI}, it combines the measured phase evolutions of different laser beat signals 
using the TDI algorithm to form the interferometer signal. 

\begin{figure}

\begin{centering}
\includegraphics[width=8cm]{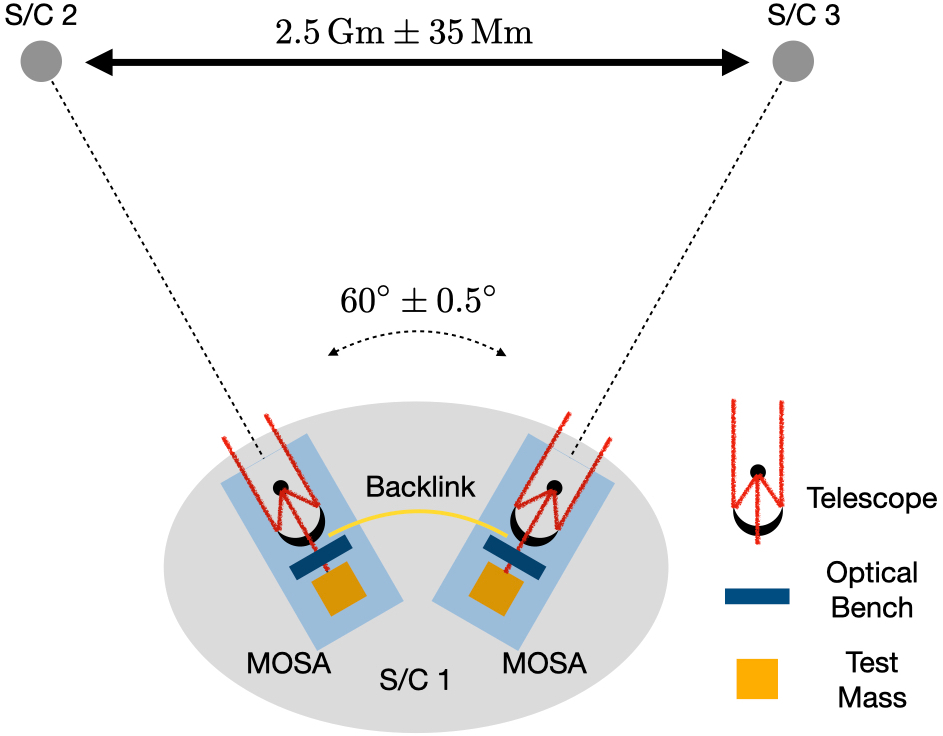}
\par\end{centering}
\caption{\label{fig:MOSA}In LISA, each spacecraft hosts two movable optical
sub-assemblies (MOSA) each containing a test mass, an optical bench
and a telescope. The opening angle between the two MOSAs will change
over the course of the year to track the far spacecraft.}
\end{figure}

In contrast to this, LISA is using two movable optical sub-assemblies (MOSAs)
comprised of a test mass, an optical bench, and a telescope in each
spacecraft. As shown in figure \ref{fig:MOSA}, the opening angle
between the two MOSAs can be changed with a mechanism to independently
track the far spacecraft. In LISA, the phase comparison needed to form the artificial beam splitter is
realized using an optical fiber link between the two MOSAs \cite{Backlink_18}.
The  range of the MOSA actuation system of order one degree will allow to operate LISA for more than a decade.
However, without any major station keeping maneuvers, latest after
about 20 years, the distances, relative velocities and opening angles
will likely be too large to continue operation and LISA will cease to exist
as an instrument. 

In any case, the need for alignment of the constellation but also
for alignment of the test masses within the spacecraft means that
LISA can not use truly free falling test masses but will have to
actuate on the test masses even during science operation in five of
the six degrees of freedom while the test mass is only in free-fall
along the interferometer arm axis. This approach also allows to use two
test masses, one in each MOSA, and to steer the spacecraft simultaneously
around both of them. It has the additional advantage that it adds redundancy
to LISA as the failure of a single test mass would still allow the operation of
one full interferometer (2 of the 3 arms).

\subsection{Interferometry}

The interferometric measurement from test mass to test mass is typically
broken up into three semi-independent interferometric measurements.
The local test mass interferometer in each MOSA measures changes in the position
and orientation of each test mass (TM) with respect to its adjacent
optical bench (OB). The third interferometer measures changes in the
distance between the two widely separated OBs. This interferometer
is typically called the science or inter-spacecraft interferometer. 
A reference signal on each OB on each spacecraft measures the beat signal between both local lasers. Linear combinations of the data streams from all these signals are
needed to measure differential changes in the TM to TM distances. This is described in detail in \ref{sec:TDI}.

The relative velocities of several meters per second between the spacecraft
shifts the frequencies of the exchanged laser beams by several MHz,
making it impossible to maintain a dark fringe and to use read-out
technologies developed for ground-based observatories. Instead, LISA
measures the phase evolution of laser beat signals to extract the
gravitational wave signal. This method is called heterodyne interferometry.
One laser field acts as a local oscillator which provides the reference
phase against which the phase evolution of the signal field will be
measured:
\begin{equation}
   S_{1}=\left|E_{\rm LO}e^{i\left(\omega_{1}t+\phi_{1}\right)}+E_{\rm S}e^{i\left(\omega_{2}t+\phi_{2}\right)}\right|^{2}=E_{\rm LO}^{2}+E_{\rm S}^{2}+2E_{LO}E_{\rm S}\cos\left(\Omega_{12}t+\phi_{1}-\phi_{2}\right).
\end{equation}
The first two terms are proportional to the power in both laser fields
while the third term oscillates at the difference frequency $\Omega_{12}$.
A phasemeter measures the phase of this signal, which depends on the
phase variations $\phi_{i}\,(i=1,2)$ of each laser field. These phase
variations depend on the laser frequency noise $\left(\delta\nu_{i}(t)=\delta\omega_{i}(t)/2\pi\right)$
and all optical pathlength changes $\delta L_{i}$ between some reference
planes, typically realized by a few beam splitters, and the photodetector:
\begin{equation}
\phi_{i}^{0}+\delta\phi_{i}(t)=\frac{\omega_{i}}{c}L_{i}+\frac{1}{c}\left(\delta\omega_{i}(t)L_{i}-\omega_{i}\delta L_{i}(t)\right)\label{eq:heterodyne phase}.
\end{equation}
The gravitational wave signal is contained in one of the length changes,
here $\delta L_{2}$, while all other fluctuations either have to
be minimized or measured elsewhere and then subtracted from this signal.

The fundamental limit of interferometric phase measurements is the
intrinsic phase uncertainty or noise of a coherent field which is, in this context,
also known as shot noise: 
\begin{equation}
\delta\tilde{\phi}_{SN} = \frac{1}{\sqrt{n}}\quad\Rightarrow\quad\delta \tilde{l}_{SN} =  \frac{\lambda}{2\pi}\sqrt{\frac{h\nu}{P_{\rm coh}}},
\label{eq:shotnoise}
\end{equation}
where $n$ is the photon rate of the coherent field. The $\sim$ in $\delta \tilde{l}$ indicates that this is the Fourier transform of the time series normalized to a one second measurement time. This is also known as the linear spectral density measured in units $\rm{m}/\sqrt{\rm{Hz}}$ which is the square root of the power spectral density $S_{\delta} = (\delta \tilde{l})^2$ which will show up in later sections again.

In any case, the larger the amplitude the lower the phase uncertainty of the coherent field. The resulting fundamental shot noise limit $\delta \tilde{l}_{SN}$ of any interferometer depends on the wavelength of the laser field $\lambda$ or its frequency $\nu$ and the power $P_{\rm coh}$ of the coherent field. The inter-spacecraft interferometer measures the weak received field against a much stronger local oscillator field such that the shot noise limit in LISA is determined by the amplitude of the received field only. Similarly, the amplitudes in the test mass interferometer and in the reference signal are significantly higher such that this shot noise limit is only relevant for the inter-spacecraft interferometer.

\begin{figure}

\begin{centering}
\includegraphics[width=8cm]{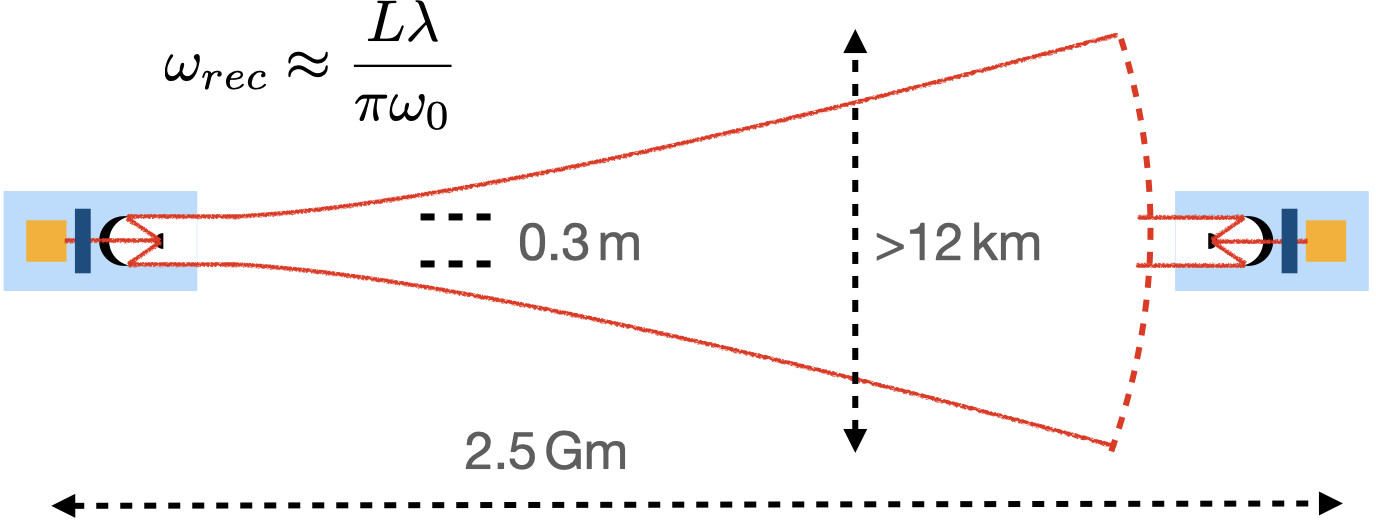}
\par\end{centering}
\caption{\label{fig:beam_prop}Due to diffraction, the laser beam will have grown in radius $\omega_{rec}$ by several orders of magnitude before it reaches the far spacecraft. The final size depends on the initial Gaussian beam radius $\omega_0 = 0.446\,D$ where $D$ is the diameter of the telescope, $\lambda$ is the wavelength and $L$ is the distance. Shown here are the current LISA design parameters ($\lambda = 1064\,\rm{nm}$).}

\end{figure}

The received power on the far spacecraft is maximized for a Gaussian beam when the radius of the injected fundamental Gaussian mode has a width of $\omega=0.446\,D$ overfilling
the exit aperture of the transmitting telescope. In that case, the maximum received power on the far spacecraft depends on the transmitted power $P_{T}$, the distance $L$ between the spacecraft, and the diameter $D$ of the transmitting and receiving diffraction limited telescopes:
\begin{equation}
P_{rec}=\frac{1}{2}\frac{D^{4}}{\lambda^{2}L^{2}}P_{T}\,.
\end{equation}
The scaling with $L^{-2}$ has the interesting effect that the shot
noise limit of the strain sensitivity $h=\delta l/L$ for $\lambda_{\rm GW}>L/2$
is independent of the spacecraft distance. In this regime longer arms
increase the displacement $\delta l=hL$, but as the shot noise limit
scales with $P^{-1/2}\propto L$, the strain sensitivity does not change.
When $\lambda_{\rm GW}$ becomes comparable to, or shorter than, the arm
length, the effect of the gravitational wave averages away following
a sinc-function with zeros at $\lambda_{\rm GW}=L/2n\,\left(n\in\text{\ensuremath{\mathbb{{N}}}}\right)$
for optimally aligned observatories. Averaging over all angles will
wash out these zeros and give rise to the `wiggles' in LISA's sensitivity
curve at higher frequencies.

As shown in figure \ref{fig:beam_prop}, the current LISA design uses a telescope diameter of $D=0.3\,\text{m}$ to send a $P_{T} = 2\,\text{W}$ laser beam of wavelength $\lambda=1064\,\text{nm}$ to an identical telescope on the receiving spacecraft. 
If we assume an efficiency of $\eta\approx0.5$ which takes
unavoidable optical losses, quantum efficiencies, contrast defects and also power lost to additional frequency components modulated onto the field in the final beat signal into account, the effective received power in the relevant frequency component
is then:
\begin{equation}
P_{rec}\approx570\,\text{pW}\left[\frac{\eta}{0.5}\right]\left[\frac{D}{0.3\,\text{m}}\right]^{4}\left[\frac{2.5\,\text{Gm}}{L}\right]^{2}\left[\frac{1064\,\text{nm}}{\lambda}\right]^{2}\left[\frac{P_{T}}{2\,\text{W}}\right]\label{eq:Prec}
\end{equation}
leading to a shot noise limit of $\delta\tilde{l}_{SN}(f)\approx3\,\text{pm}/\sqrt{\text{Hz}}$.

Shot noise is neither the only limitation of the interferometric measurement
system nor would the system even remotely be able to detect gravitational
waves without subtracting several other contributions to the beat signals.
Laser frequency noise was already added in equation \ref{eq:heterodyne phase}.
The historical Michelson interferometer used equal arm lengths $\left(L_{1}=L_{2}\right)$
and a single thermal light source $\left(\delta\omega_{1}=\delta\omega_{2}\right)$
to become insensitive to frequency variations. Ground-based observatories
use near equal arm lengths and a single laser whose frequency is stabilized
to the average or common arm of the interferometer. As discussed before, in LISA, the arm lengths are constantly changing and will reach
macroscopic length differences of up to 35\,000\,km which would place
unrealistic demands on the levels of laser frequency noise needed. 

Still, LISA will use state of the art laser frequency stabilization systems and is expected to
reach a laser frequency noise floor of below $30\,\text{Hz}/\sqrt{\text{Hz}}$
above 2\,mHz, increasing with $f^{-2}$ below 2\,mHz. This is about 8 orders of magnitude to high for a direct strain measurement. Instead, LISA will eliminate laser frequency noise in post-processing using Time Delay Interferometry (TDI) to form a quasi-equal arm Michelson interferometer signal based on the knowledge of the light travel time between the spacecraft~\cite{TDI_Tinto, TDI_Vinet, TDI_McKenzie,TDI_Mitryk, TDI_LRR}. Laser ranging will be used to measure the distances between the spacecraft with sub-meter accuracy which then leads to an apparent length noise caused by laser frequency noise of 
\begin{equation}
\delta \tilde{l}<\frac{\delta\tilde{\nu}\cdot\Delta L_{\rm eff}}{\nu}=0.1\frac{\text{pm}}{\sqrt{\text{Hz}}}\left[\frac{\delta\nu}{30\,\text{Hz}/\sqrt{\text{Hz}}}\right]\left[\frac{\Delta L_{\rm eff}}{1\,\text{m}}\right]\,.
\label{eq:TDI_Ranging}
\end{equation}
still using the simple Michelson interferometer geometry as a baseline. 

The phase measurement devices or phasemeter have to measure the phase
evolution of the laser beat signals with respect to an onboard ultra-stable
oscillator (USO) \cite{Phasemeter_06,Phasemeter_15}. Fundamentally,
this can be described by a multiplication of two signals with identical
frequencies and then averaging the result to eliminate the second
harmonic:
\begin{equation}
    \left<\cos\left(\Omega_{12}t+\phi_{1}-\phi_{2}\right)\cos\left(\Omega_{12}t+\phi_{\rm PM}\right)\right>_{t}=\frac{1}{2}\cos\left(\phi_{1}-\phi_{2}-\phi_{\rm PM}\right).
\end{equation}
The second factor is generated from the USO, and the phasemeter itself
is designed to track the frequency $\Omega_{12}$ of the incoming
signal. This multiplication measures the phase evolution of the beat
signal in fractions of clock cycles of the USO. Any differential noise
between the USOs on the three spacecraft would result in additional
noise in the interferometer read out:
\begin{equation}
\delta\phi_{\rm PM}=\frac{f_{12}}{f_{\rm USO}}\delta\phi_{\rm USO}
\end{equation}
for a given clock phase noise $\delta\phi_{\rm USO}$ at the clock frequency $f_{\rm USO}$.
Lower beat frequencies $f_{12}$ will reduce the phase noise due to clock noise. However, as mentioned before, the Doppler shifts caused by the orbital motion place
a natural lower limit of a few MHz on these beat frequencies. LISA's
current frequency plan uses frequencies between 5 and 20\,MHz which
are about two to three orders of magnitude too high
for state of the art USOs to ignore this issue. The solution proposed
for LISA is to exchange the clock noise between the spacecraft by
modulating the phase of each laser field with a tone that is directly
derived from the local USO. This will create additional frequency
components in the laser beat signals which allow the extraction of differential
clock noise, enabling a clock-noise correction in TDI. 
\begin{figure}

\begin{centering}
\includegraphics[width=8cm]{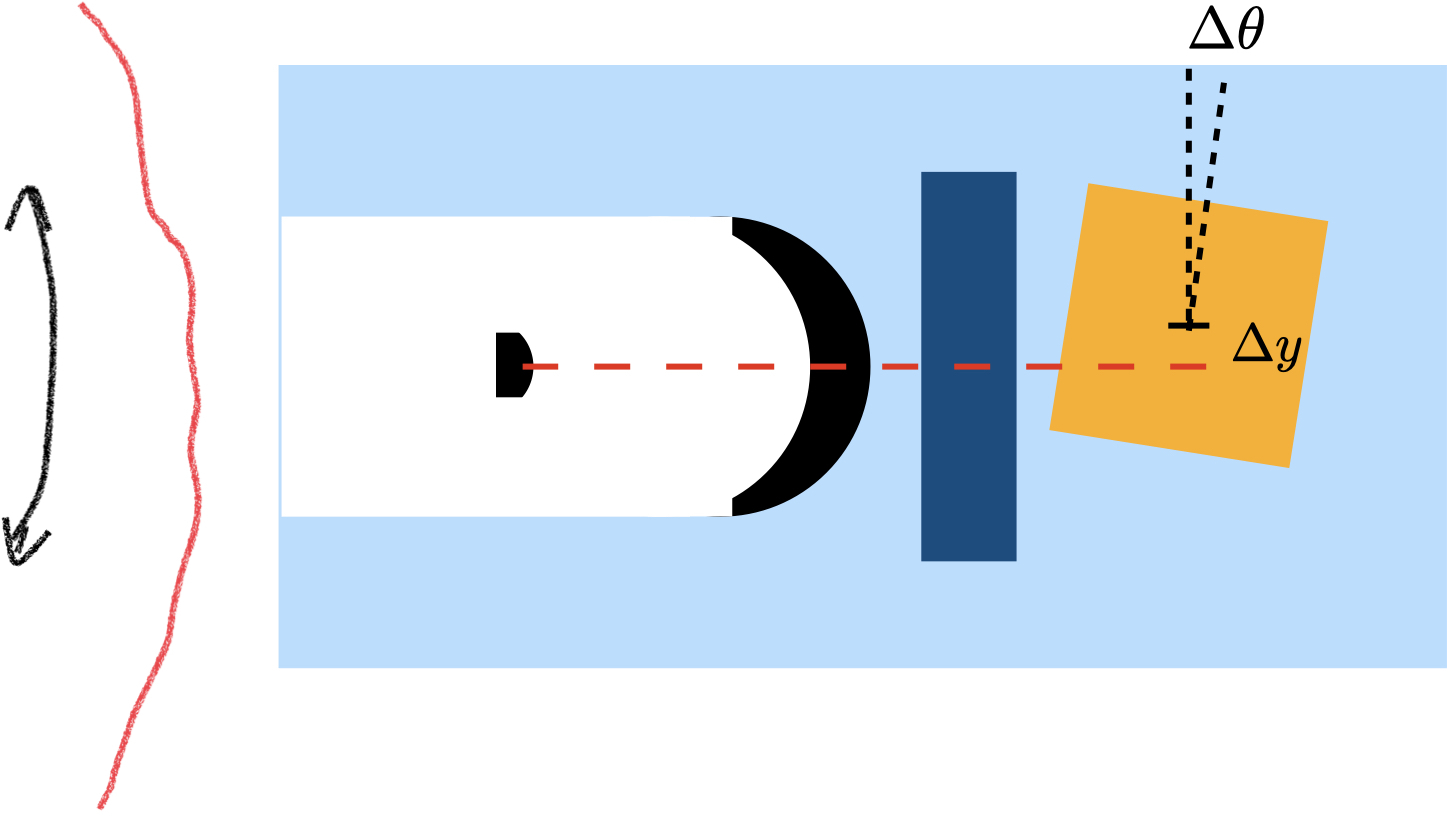}
\par\end{centering}
\caption{\label{fig:TTL}Tilt to length (TTL) coupling is a generic term that is used to describe all noise sources which are associated with dynamic changes in the alignments between different parts of the interferometer. The figure sketches two examples: The wavefront from the far spacecraft will not be perfectly spherical but will have some distortions or phase gradients caused by imperfections in for example the telescope. As the sending spacecraft jitters, this phase gradient will be scanned over the receiving spacecraft. The second effect is associated with a lateral shift $\Delta y$ between the center of mass of the test mass and a tilt $\Delta \theta$ of the test mass. A change in any of these two parameters will look like a center of mass motion of the test mass in the sensitive direction. These are only two examples of a large range of potential noise sources summed up under TTL.}

\end{figure}

Another notorious noise source is known as Tilt-To-Length coupling
or TTL  \cite{TTL_20}. Two of the main mechanisms are depicted in figure \ref{fig:TTL}.
In a drag-free system, the spacecraft is steered around the
free-falling test masses using $\mu\text{N}$-thrusters. The response
time of the thrusters and the large inertia of the spacecraft results
in residual lateral and angular motions of each spacecraft of up to
a few $\text{nm}/\sqrt{\text{Hz}}$ and $\text{nrad}/\sqrt{\text{Hz}}$.
This will lead to beam pointing at the far spacecraft such that the
12~km-wide beam is moving by several $\text{m}/\sqrt{\text{Hz}}$
across the receiving telescope. Any gradient in the phase front caused
by wavefront errors in the transmitted field will look like apparent
length changes. Similarly, any misalignment between the optical axis
of the telescope with the test mass will turn the motion of the spacecraft
in other degrees of freedom into apparent length changes in the sensitive
direction. 

These effects are all second order in the sense that they are the product of two misalignments:
\begin{equation}
    \Delta x \sim M \Delta y \Delta \theta 
\end{equation}
which both have typically {\it fairly large} rms or quasi-static components and minute dynamic or in-band components. In LISA the quasi-static components are typically in the few $\mu\rm{m}$ or $\mu\rm{rad}$ range while the minute dynamic or in-band components are in the few $\rm{nm}/\sqrt{\rm{Hz}}$ or $\rm{nrad}/\sqrt{\rm{Hz}}$ range. The scaling factor $M$ is in the depicted case simply one but can be equal to the magnification of the telescope, in LISA's case 134, when misalignments between the optical bench and the telescope are involved.  

In each interferometer the relative angular motion of the two interfering
beams is measured using quadrant photo-detectors which measure the
phase difference between the fields in each quadrant and extract from
that data the angle and the angular motion between the wavefronts \cite{WFS}. The wavefront sensing
system allows the measurement of the relative motion of the spacecraft with
respect to the free-falling TM and of the spacecraft with respect
to each other. Properly calibrated and delayed in time, this data
can then be used to subtract TTL from the data streams as well.

Other noise sources associated with the interferometry include temperature
dependent dispersion in the photo receivers and cables \cite{Cable_09}, timing jitter
in the ADCs \cite{Phasemeter_15}, scattered light \cite{Scatter_18,Spector_12}, and optical pathlength changes on the
optical benches and within the telescopes \cite{SiC_Telescope_12}. Most of these noise sources
will likely be driven by temperature variations; LISA will likely require temperature
fluctuations within each MOSA to be below $10\,\mu\text{K}/\sqrt{\text{Hz}}$
and $\text{mK}/\sqrt{\text{Hz}}$ for most of the other systems.

\subsection{Gravitational reference sensor}

As discussed earlier, LISA will use a drag free system in which two
test masses inside each spacecraft are ideally in perfect free fall
along the optical axis of their respective interferometer arms. 
LISA will build on the gravitational reference sensors (GRS) that
have been flown on the LISA Pathfinder (LPF). The 1.9\,kg test masses are
46\,mm gold-coated cubes. They are formed from a gold platinum alloy
with a composition that has been fine tuned to minimize the magnetic susceptibility and magnetic moments.

\begin{figure}
\begin{centering}
\includegraphics[width=8cm]{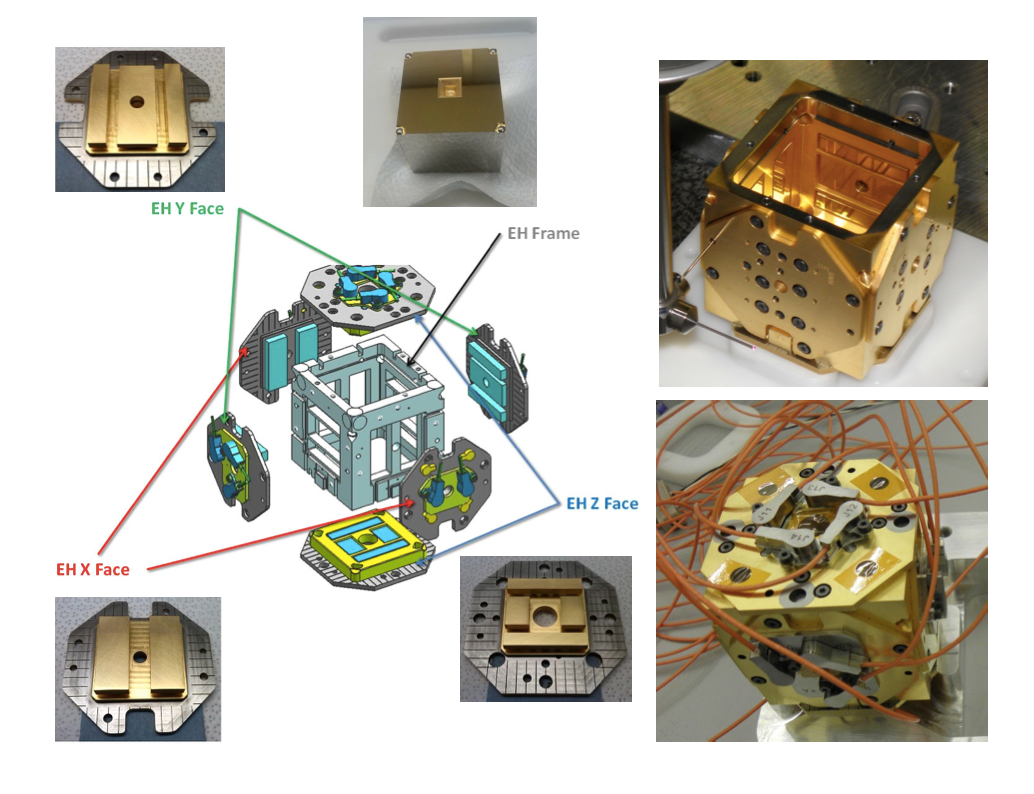}
\par\end{centering}
\caption{\label{fig:LISA-Gravitational-Reference}LISA test mass and test mass housing. 
Each LISA test mass will be a gold-coated gold-platinum 46\,mm large cube similar to the LPF test mass shown center-top in this figure.
During science mode, the gold-coated cube will float
inside the housing. Each surface of the housing holds gold-coated
sapphire electrodes which are mostly used to (a) sense the position
of the test mass and (b) apply the necessary forces and torques on
the test mass. Some of the electrodes are used to inject a 100~kHz
signal that is picked up by the sensing electrodes. The laser beam
for the local interferometer is injected through the hole in one of
the x-faces while the launch lock and release mechanism uses the z-surfaces. 
This figure uses parts of Figure 4 and 5 in \cite{LPF_Description}.} 
\end{figure}

Each test mass floats inside a hollow cube known as the GRS housing which protects the test mass
from external forces and torques and allows to sense and control its
position and orientation within the housing. Such a housing is shown
in figure \ref{fig:LISA-Gravitational-Reference}. All six surfaces
contain gold-coated sapphire electrodes; four on the z-surfaces, three
on the y-surfaces, and two on the x-surfaces which are normal to the
optical axis. The hole in the shown x-surface allows the laser beam from
the optical bench to pass through while the hole in each z-surface
is used to grab the test mass during launch and release it with a
very small residual velocity below a few tens of $\mu{\rm m}/\text{s}$. 

The two outer electrodes on the other plates are the sensing and actuation
electrodes while the middle electrodes inject signals. The various
electrodes and the test mass surfaces they face form a set of parallel
and serial capacitors where each capacitance can be approximated by
a parallel plate capacitor of some area $A$ and distance $d=d_{0}+\delta$:
\begin{equation}
C=\epsilon_{0}\frac{A}{d}=\epsilon_{0}\frac{A}{d_{0}+\delta}\approx C_{0}\left(1-\frac{\delta}{d_{0}}\right)
\label{eq:accel_noise}
\end{equation}
where $d_{0}$ is the nominal distance; typically a few mm. $\delta$
is the displacement from this nominal position and has to be kept
below a few tens of $\mu\text{m}$ to minimize for example tilt to length
coupling. 

Two identical AC signals are applied to opposing injection electrodes.
These will polarize the test mass, pull or push charges from the surfaces
normal to the injection axis to all other surfaces and change the electrical
potential of the sensing electrodes on the other surfaces. The transmission of the injected
signals to the sensing electrodes depend on the capacitances and therefor
the location of the test mass within the housing.
The differences in amplitudes on opposing electrodes is measured 
to determine the position and orientation of the test mass in all degrees of freedom. 
These signals are then used to either actuate
the test mass or command the thrusters to steer the spacecraft around
the two test masses. 

The actuation is based on electro-static forces where an external
electric field first polarizes the test mass and then attracts it.
This can be described as the attractive force between two capacitor plates:
\begin{equation}
F=-\nabla U=\frac{1}{2}V^{2}\frac{\epsilon_{0}A}{d^{2}}
\end{equation}
where $U$ is the energy of the field and $V$ is the potential difference between the plates. The force
does not depend on the polarity of the potential difference but scales
with its square and falls off with the distance squared. 
Applied for LISA, voltages need to be applied to specific electrodes to actuate the test mass in specific degrees of freedom. For example actuating the test mass in the positive x-direction requires
the application of a voltage to the two electrodes on the positive x-face of
the housing; actuating in the negative x-direction requires a voltage
on the negative x-face. Voltages applied to the upper electrode on
one of the z-faces and to the lower electrode on the other z-face
torques the test mass around the y-axis. 
The equation also shows that
the injection voltages have to be injected with ideally identical
amplitudes on the electrodes to not generate a
torque in addition to the commanded force or a force instead of the commanded torque on the test mass. 
This is one of the reasons why there are
no injection electrodes on the x-faces.

As in LPF, the capacitance in LISA will be on the order of a few pF
per surface resulting in a force of a few hundred $\text{pN}/\text{V}^{2}$
for the few mm gap sizes. Smaller gaps increase the sensing sensitivity
and also the applicable forces and torques but many of the noise sources
also scale with the inverse gap size. The gap size is one of the most
important optimization parameters of the GRS. Using this design, LPF
has demonstrated a residual relative acceleration noise between two
test masses of \cite{LPF_Results18}
\begin{equation}
\delta\tilde{a}(f)<\frac{2.4\,\text{fm}}{\text{s}^{2}\sqrt{\text{Hz}}}\sqrt{1+\left(\frac{0.4\,\text{mHz}}{f}\right)^{2}}\sqrt{1+\left(\frac{f}{8\,\text{mHz}}\right)^{4}}
\end{equation}
which exceeded even the most optimistic expectations and paved the
way for LISA; note that the increase in noise above 8 mHz is believed to be a limitation of the interferometric measurement system and is not a real acceleration noise. 

\begin{figure}
\begin{centering}
\includegraphics[width=8cm]{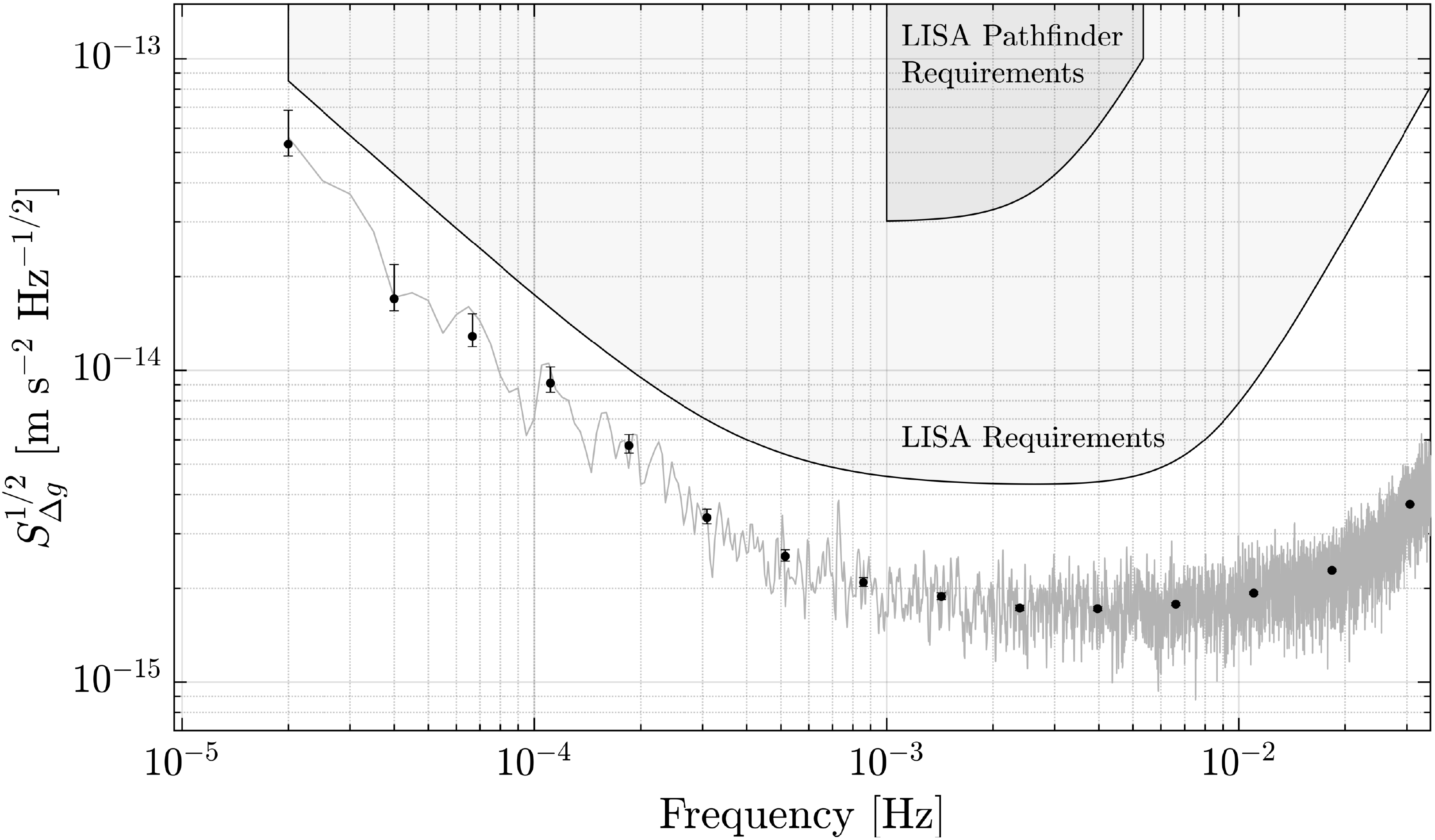}
\par\end{centering}
\caption{\label{fig:LISA-Pathfinder-final} As published in \cite{LPF_Results18} and reproduced in here, the original requirement for the LPF mission was roughly
an order of magnitude worse than the LISA requirement and limited to
frequencies above 1~mHz. After a year of operation, the relative
acceleration noise of LISA Pathfinder surpassed the LISA requirements by about half a
decade at all frequencies.}
\end{figure}

This performance is not limited by any uncertainty principle but by a long list of environmental and technical
noise sources. The most dominant ones are listed here in no particular order: 
\begin{itemize}
\item The mass distribution within the moving MOSAs and within the surrounding spacecraft have to be optimized to minimize the gravitational force on and the gravity gradient at the test mass. LISA typically requires the DC acceleration to stay below a few tens $\rm{nm}/\rm{s}^2$ while the effect of the gradient, or the change in acceleration per spacecraft motion also called the stiffness, has to be below $10^{-6}/\rm{s}^2$ for a spacecraft motion of $1\,\rm{nm}/\sqrt{\rm{Hz}}$ along the sensitive axis. 
\item Residual gas molecules will hit the test mass surface from both sides
and cause a Brownian motion of the test mass: 
\begin{equation}
\delta\tilde{a}(f)=\left[\left(1+\frac{\pi}{8}\right)\frac{A_{TM}}{M_{TM}^{2}}p\left(\frac{512m_{0}k_{B}T}{\pi}\right)^{1/2}\alpha\right]^{1/2}
\end{equation}
where $m_{0}$ is the mass of the dominant molecule and $p$ its partial
pressure. This well understood noise is amplified by a factor $\alpha\approx13$
due to correlations caused by gas molecules which bounce back and
force between the test mass and the housing. This factor depends on
the gap size and larger gaps would reduce the noise \cite{Cavalleri_09}.
\item The actuation in the other five degrees of freedom will always also
have a force component in the sensitive direction. This cross-talk 
can be minimized by optimizing the position and orientation of the
test mass within the housing; however, the test mass also has to be
aligned with respect to the optical axis defined by the telescope which places tight requirements
on the initial overall alignment of all components. 
\item Charges on the test mass will couple to electric and magnetic fields
via the Lorenz force. Early gravitational reference sensors used a
thin wire to discharge the test mass. This wire couples the motion
of the spacecraft to the test mass and would be orders of magnitude
too noisy. LPF and LISA use the photo-electric effect to discharge
the test mass. UV-light will be directed to the test mass to move
electrons from it to the housing and directed to the housing to move
electrons to the test mass. 
\item Magnetic fields and their gradients also couple to permanent magnetic
moments $\vec{m}_{i}$ of ferromagnetic inclusions, to the magnetic
susceptibility $\chi$, and any macroscopic currents $\vec{J}$ such
as eddy currents induced by time dependent magnetic fields and create
a force on the test mass:
\begin{equation}
F_{B,x}=\sum_{i}\vec{m}_{i}\cdot\left(\frac{\partial\vec{B}}{\partial x}\right)_{i}+\frac{\chi}{\mu_{0}}\int_{V}\vec{B}\cdot\frac{\partial\vec{B}}{\partial x}dV+\int_{V}\vec{J}\times\vec{B}dV
\end{equation}
\end{itemize}
These are just few of many noise sources which are part of the acceleration noise budget. The work on LPF allowed to gain a very good understanding of these noise sources and the knowledge how to minimize them. However, it should be noted that there is still some excess noise in LPF that is not yet understood but as it is below the LISA requirements, it is not preventing LISA from going forward.

\section{Data Analysis of Space-based Observatories}
\label{sec:space_gw_dataanalysis}

\subsection{Time-Delay Interferometry}

\label{sec:TDI}

\subsubsection{Introduction }

\begin{figure}

\begin{centering}
\includegraphics[width=10cm]{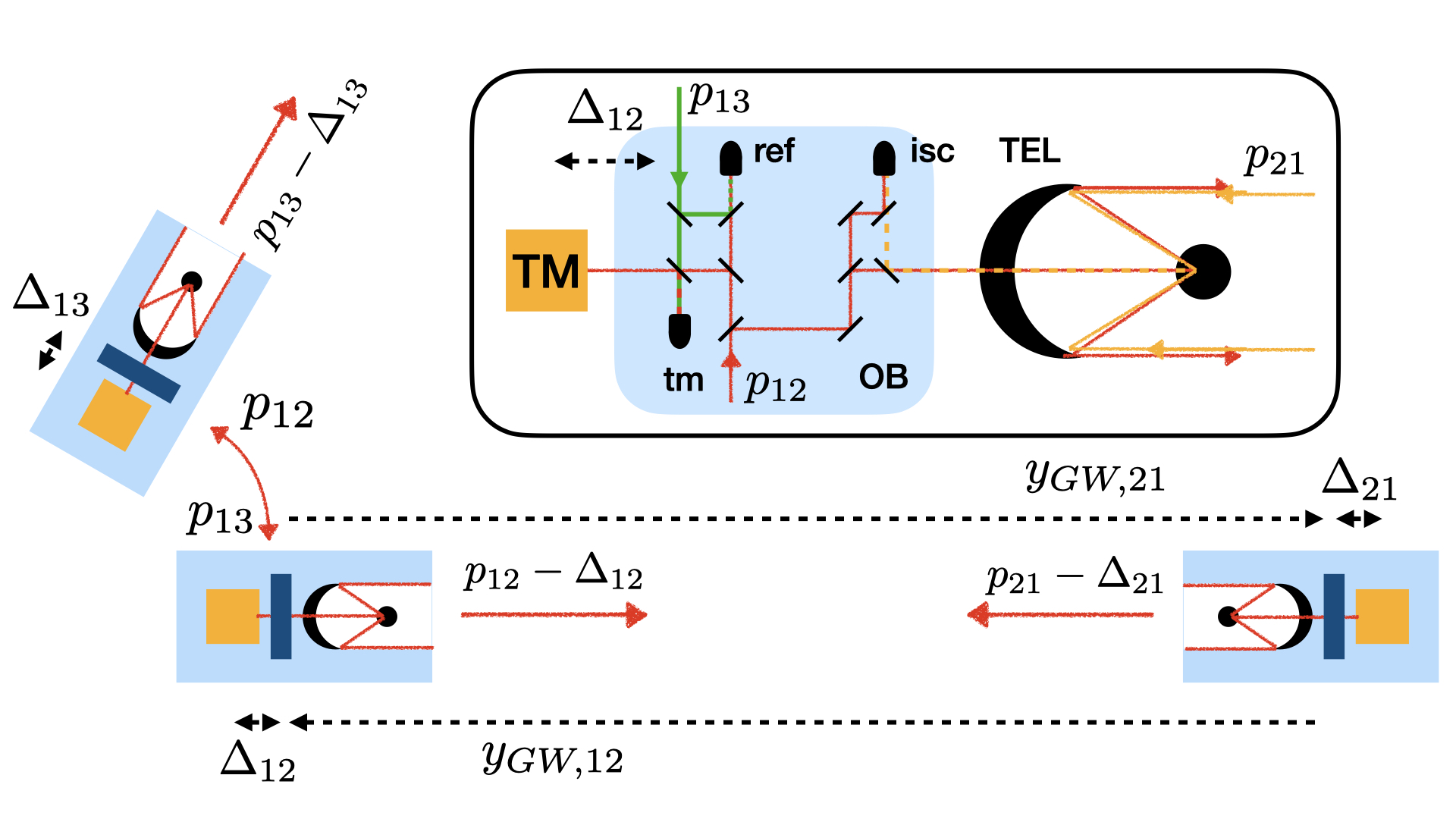}
\par\end{centering}
\caption{\label{fig:The-LISA-arm}TDI is an algorithm that uses three interferometer
signals in each MOSA to cancel laser frequency noise $p_{ij}$, spacecraft
motion $\Delta_{ij}$, and clock noise (not shown here) to extract
the gravitational wave information $y_{GW,ij}$. The inlet in the
upper right corner shows the relevant components of MOSA~{12}. The associated
laser (red) with phase noise $p_{12}$ is injected from the bottom,
split up into one beam that is sent into the telescope, one beam acts
as the local oscillator in the inter-spacecraft interferometer (isc) where
it beats against the received field with phase noise $p_{21}$ from
the far S/C. The local laser is also used to measure the test-mass
(TM) position using the beat signal with the third field (phase noise
$p_{13}$) which is injected from MOSA~{13}. This test-mass interferometer
(tm) is compared with the reference interferometer beat signal (ref).
Each of the six MOSAs produces these three signals.}
\end{figure}

\label{sec:TDI_IntroModel}

As explained in the previous section, the core measurements of LISA
are phase evolutions measured from multiple heterodyne interferometers.
This is shown in more detail in figure \ref{fig:The-LISA-arm}. Each
MOSA produces three laser beat signals: the inter-spacecraft (or science)
interferometer (isc) measures changes $y_{GW}$ in the distance between
the two optical benches (OB) on widely separated spacecrafts, the
test-mass (tm) interferometer measures the OB motion -- a.k.a. spacecraft
jitter -- $\Delta$ around the test-mass (TM). The reference
interferometer is needed to measure the differential phase noise between
the two lasers on each spacecraft. These three signals on MOSA~{12}
are:

\begin{eqnarray}
\text{isc}_{12} & = & \propdelay{12}\left[p_{21}-\Delta_{21}\right]+y_{GW,12}-\left[p_{12}+\Delta_{12}\right]+N_{\text{isc},12}\\
\text{tm}_{12} & = & \left(p_{13}+fn_{13}\right)-\left[p_{12}+2\left(\Delta_{12}-\delta_{TM,12}\right)\right]\\
\text{ref}_{12} & = & \left(p_{13}+fn_{13}\right)-p_{12}
\end{eqnarray}
where 12 refers to\emph{ located on spacecraft 1 and facing spacecraft
2.} $\propdelay{rs}x(t)=x(t-L_{rs}/c)$ is the propagation delay operator
which takes into account the propagation time between the spacecrafts;
in this case, the laser phase noise of the laser on the sending spacecraft
$s$ needs time before it is received on the receiving spacecraft
$r$. $N_{isc,12}$ is the interferometer noise which includes shot
noise, noise in the phasemeter measurement system and length changes
on the OB or within the telescopes which are ideally all below shot
noise. $\delta_{TM,12}$ is the residual acceleration noise of the test-mass 
expressed as displacement noise $\tilde{\delta}=\delta\tilde{a}/4\pi^{2}f^{2}$.
These two terms limit the sensitivity of LISA. The local exchange
of the laser beams between the two local MOSAs through the back link
fiber adds additional phase noise $fn_{13}$ to the field from the
other bench.

The formulation for the MOSA $23$ and $31$ are obtained by circular
permutation of the indices ($1\rightarrow2\rightarrow3\rightarrow1$).
For the adjacent MOSA $13$, the model is: 
\begin{eqnarray}
\text{isc}_{13} & = & \propdelay{13}\left[p_{31}-\Delta_{31}\right]+y_{GW,13}-\left[p_{13}+\Delta_{13}\right]+N_{\text{isc},13} \label{eq:isc}\\
\text{tm}_{13} & = & \left(p_{12}+fn_{12}\right)-\left[p_{13}+2\left(\Delta_{13}-\delta_{TM,13}\right)\right] \label{eq:tm}\\
\text{ref}_{13} & = & \left(p_{12}+fn_{12}\right)-p_{13}\label{eq:ref}
\end{eqnarray}
The formulation for the MOSA $21$ and $32$ are obtained again by
circular permutation of indices. 

\subsubsection{Spacecraft jitter measurement and subtraction}

\label{sec:TDI_SCjitter}

As discussed before, in a drag free system like LISA, the spacecraft
follows the motion of the test-mass along the sensitive axis using
the measured position of the test-mass with respect to the optical
bench as the sensor or error signal. The $\mu$N-thrusters which actuate
the heavy spacecraft have a limited actuation bandwidth and are expected
to reduce the spacecraft motion to a few $\text{nm}/\sqrt{\text{Hz}}$
in the LISA band, two to three orders of magnitude larger than the
sensitivity goal of LISA expressed in displacement noise. The test
mass interferometer measures the separation between test-mass and
optical bench but its signal also depends on the differential phase
noise of the two laser fields which is also measured by the reference
interferometer. The difference between the two reveals the spacecraft
jitter:
\[
\text{tm}_{12}-\text{ref}_{12}=2\left(\Delta_{12}-\delta_{TM,12}\right)
\]
A similar measurement on spacecraft 2, properly delayed by the light
travel time between the spacecrafts, together with the inter-spacecraft
interferometer measurement can be combined to form two new quantities
which are free of any spacecraft jitter along the sensitive axis:
\begin{eqnarray}
\xi_{12} & = & \text{isc}_{12}-\frac{\text{tm}_{12}-\text{ref}_{12}}{2}-\frac{\tdidelay{12}\ \left[\text{tm}_{21}-\text{ref}_{21}\right]}{2}\\
\xi_{13} & = & \text{isc}_{13}-\frac{\text{tm}_{13}-\text{ref}_{13}}{2}-\frac{\tdidelay{13}\ \left[\text{tm}_{31}-\text{ref}_{31}\right]}{2}
\end{eqnarray}
with $\tdidelay{rs}x=x\left(t-\frac{L_{rs}(t)}{c}\right)$ being the
TDI delay operator along arm $rs$. Note that these linear combination
will be formed in post-processing on the ground from the six times
three different MOSA measurements. In contrast to the propagation
delay, $ \propdelay{rs}$ which depends on the physical distance between the spacecraft,
the TDI delay, $\tdidelay{rs}$, is applied on-ground and depends on the knowledge of
the light travel time between the spacecraft.

Spacecraft jitter also has rotational components and lateral components
in the LISA band along the other five degrees of freedom. As discussed
before, these couple to other forms of misalignments and wavefront
errors and are responsible for what is known as tilt-to-length (TTL)
noise. This spacecraft jitter will also be measured -- either with
wavefront sensing or with capacitive sensors -- and actively suppressed
as much as possible. The residual spacecraft jitter and the TTL coupling
coefficient for each degree of freedom will be measured to later subtract
also TTL noise in post-processing on the ground.

\subsubsection{Suppressing laser frequency noise}

\label{sec:TDI_laser}

The suppression of the laser frequency noise requires that the back
link fiber noise is reciprocal $fn_{12}=fn_{13}$, at least to better
than $1\,\text{pm}/\sqrt{\text{Hz}}$. This allows to eliminate laser
frequency noise in post-processing by subtracting first the two reference
measurements taken on the same spacecraft from each other to eliminate
the fiber noise. This difference is then subtracted from the $\xi_{rs}$
to cancel the three laser frequency noises $p_{13}$, $p_{21}$ and
$p_{32}$: 
\begin{eqnarray}
\eta_{12} & = & \xi_{12}+\frac{\tdidelay{12}\left[\text{ref}_{21}-\text{ref}_{23}\right]}{2}\nonumber\\
\eta_{13} & = & \xi_{13}-\frac{\text{ref}_{13}-\text{ref}_{12}}{2} \label{eq:eta}
\end{eqnarray}
As a net result all $\eta_{rs}$ combinations only contain one laser
frequency noise ($p_{12}$, $p_{23}$ and $p_{31}$) per spacecraft;
this is the step where the artificial beam splitter is created.

\begin{figure}
\begin{centering}
\includegraphics[width=8cm]{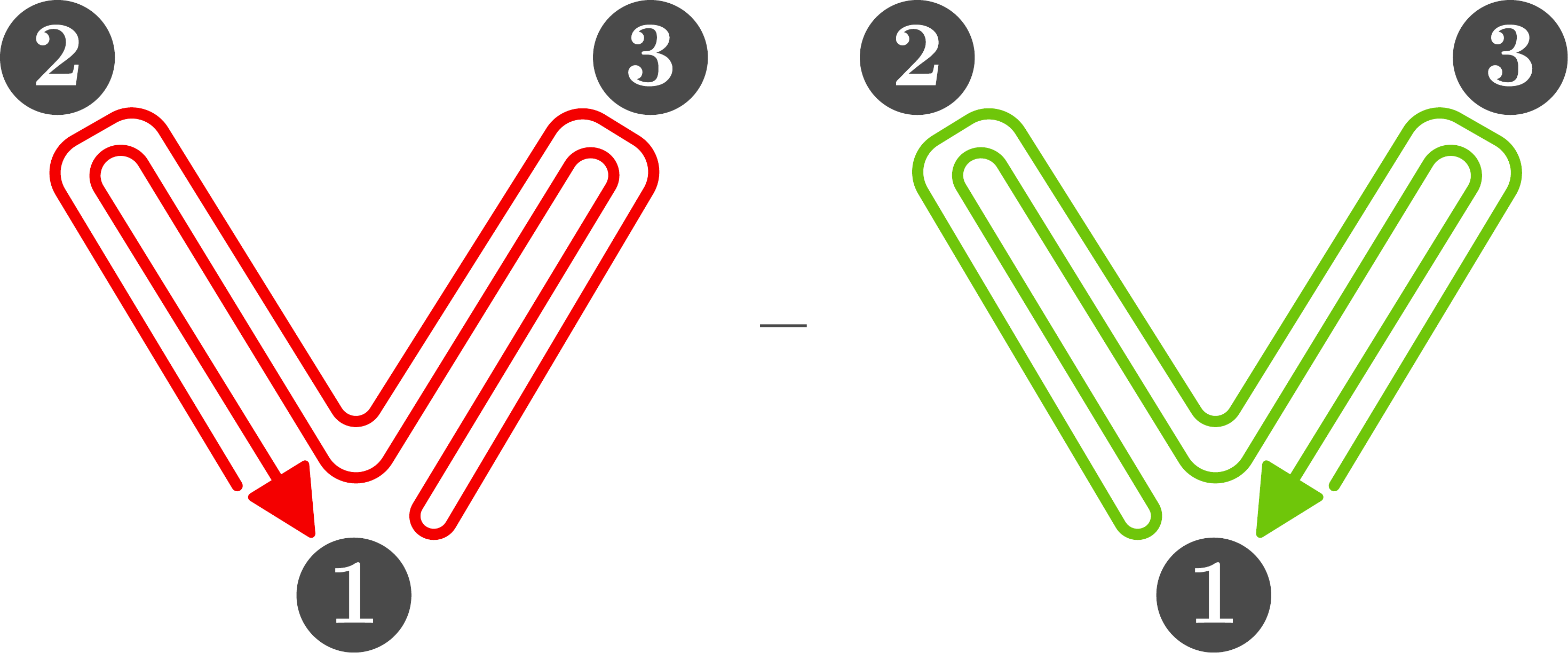} 
\par\end{centering}
\caption{\label{fig:TDI2}Geometric representation of the TDI combination $X_{2.0}$,
combination~\eqref{eq:X20-1}. Figure reproduced from~\cite{PhD_Bayle}.}
\end{figure}

Then, in order to suppress the three remaining laser frequency noises,
complex combinations are formed which are known as TDI generators.
They have different sensitivities to different gravitational waves,
some are even mostly insensitive and allow to measure instrumental
noise~\cite{TDI_LRR,TDI_ClockJitterReduction}. We will restrict
our discussion on the Michelson TDI generator X; the Y and Z Michelson
generators can be generated from it by cycling the indices or by renaming
the central spacecraft in figure~\ref{fig:TDI2}.

A TDI generator can be seen as the interference between two virtual
beams~\cite{TDI_Geometric}. For the classical static equal arm Michelson
interferometer in which the far MOSA act as mirrors or noise-free
transponders, $\propdelay{12}=\propdelay{21}=\propdelay{31}=\propdelay{13}$, 
$p_{21}=\propdelay{21}p_{12}$ and $p_{31}=\propdelay{31}p_{13}$,
laser frequency noise cancels at recombination. However, in LISA the transponders are not noise-free $p_{rs}\neq\propdelay{rs}p_{sr}$ and, in a triangular configuration, it is not possible that all MOSAs act as mirrors. For the still
static equal arm Michelson interferometer, the interference between
the beam doing the loop between spacecraft 1 and 2, i.e. $\eta_{12}+\tdidelay{12}\eta_{21}$
and the beam doing the loop between spacecraft 1 and 3, i.e. $\eta_{13}+\tdidelay{13}\eta_{31}$,
the laser frequency noise would still cancel if the arms are equal
in length $\propdelay{13}=\propdelay{31}=\propdelay{12}=\propdelay{21}$
and if the propagation delay is known such that we can set the applied
time delay $\tdidelay{rs}=\propdelay{rs}$. 

Unfortunately, the arms in LISA are not equal, here for example $\propdelay{31}\neq\propdelay{12}$,
and we have to create an artificial equal arm Michelson interferometer.
The TDI generator $X_{1.5}$ lets each of the two beams do an additional
virtual loop in the other arm such that the combination of real and
virtual paths for each beam is the same: 

\begin{eqnarray}
X_{1.5} & = & \eta_{13}+\tdidelay{13}\eta_{31}+\tdidelay{13}\tdidelay{31}\eta_{12}+\tdidelay{13}\tdidelay{31}\tdidelay{12}\eta_{21}\nonumber \\
 & - & \eta_{12}-\tdidelay{12}\eta_{21}-\tdidelay{12}\tdidelay{21}\eta_{13}-\tdidelay{12}\tdidelay{21}\tdidelay{13}\eta_{31}\label{eq:X15}
\end{eqnarray}
Similar TDI generators exist for Y and Z where the other spacecraft
are the central spacecraft. However, this is still not sufficient
for LISA as the arms change their length with $\text{m}/\text{s}$
velocities and the constellation is rotating around a common center
such that even $\propdelay{13}\neq\propdelay{31}$. A constant length
change can be compensated by doing two additional loops as shown in
figure~\ref{fig:TDI2}. This $X_{2.0}$ TDI generator:

\begin{eqnarray}
X_{2.0} & = & \eta_{13}+\tdidelay{13}\eta_{31}+\tdidelay{13}\tdidelay{31}\eta_{12}+\tdidelay{13}\tdidelay{31}\tdidelay{12}\eta_{21}+\tdidelay{13}\tdidelay{31}\tdidelay{12}\tdidelay{21}\eta_{12}\nonumber \\
 & + & \tdidelay{13}\tdidelay{31}\tdidelay{12}\tdidelay{21}\tdidelay{12}\eta_{21}+\tdidelay{13}\tdidelay{31}\tdidelay{12}\tdidelay{21}\tdidelay{12}\tdidelay{21}\eta_{13}+\tdidelay{13}\tdidelay{31}\tdidelay{12}\tdidelay{21}\tdidelay{12}\tdidelay{21}\tdidelay{13}\eta_{31}\nonumber \\
 & - & \eta_{12}-\tdidelay{12}\eta_{21}-\tdidelay{12}\tdidelay{21}\eta_{13}-\tdidelay{12}\tdidelay{21}\tdidelay{13}\eta_{31}-\tdidelay{12}\tdidelay{21}\tdidelay{13}\tdidelay{31}\eta_{13}\nonumber \\
 & - & \tdidelay{12}\tdidelay{21}\tdidelay{13}\tdidelay{31}\tdidelay{13}\eta_{31}-\tdidelay{12}\tdidelay{21}\tdidelay{13}\tdidelay{31}\tdidelay{13}\tdidelay{31}\eta_{12}-\tdidelay{12}\tdidelay{21}\tdidelay{13}\tdidelay{31}\tdidelay{13}\tdidelay{31}\tdidelay{12}\eta_{21}\nonumber\\
 \label{eq:X20-1}
\end{eqnarray}
is finally sufficient for LISA. Note that $X_{2.0}$ is only one of
several second generation TDI generators which cancel laser frequency
noise and have different sensitivities to different gravitational
wave polarizations and propagation directions. Furthermore, Sagnac-like
combinations exist which are mostly insensitive to gravitational waves at low-frequency
and laser frequency noise but allow to estimate uncorrelated instrumental
noise.

As discussed in equation \ref{eq:TDI_Ranging}, the application of TDI 2.0 still requires to know the propagation delays and apply them as matching TDI-delays in post-processing. The spacecraft ranging system uses pseudo-random noise (PRN) codes
which will be modulated onto the laser beams. Alternatively, a signal
pure region of the spectrum could be used to minimize the noise \cite{TDI_Mitryk}. Another
set of problems is associated with the need to downsample the data
on each spacecraft as the possible data rate from space to ground
is severely limited. The LISA data rate is expected to be a few samples
per second which will be created by downsampling the much faster phasemeter
data. This data is then upsampled again on ground to at least 0.3\,sns
sampling rates using precise interpolation filters to time shift the
data sets as accurate as needed \cite{TDI_FlexingFiltering}. Several
experiments have shown that 8 to 10 orders of magnitude common mode
suppression of laser frequency noise using TDI is possible.

\subsubsection{Suppressing clock noise and TTL using TDI}

\label{sec:TDI_clock}

Also already mentioned in section \ref{sec:space_gw_design}, the phasemeter is measuring the phase of the signal with respect to the onboard USO. Since USOs are not perfect and not in the same gravitational potential, they fluctuate adding errors of two types to the measurements: i) an additional phase noise corresponding to the clock jitter noise and ii) a time stamping error. Both noise sources are mostly relevant as a differential noise source. To correct for the additional clock jitter noise, LISA upconverts each 10 MHz clock signal to order 2 GHz and modulates the phase of each laser via an Electro-Optic Modulator. These sidebands generate beat signals similar to the interferometric beat signals. The phase noise of these beat signals is identical to the phase noise in the main beat signals with the addition of the upscaled differential USO noise. These measured phase noise terms also enter the TDI generators to eliminate their impact on the sensitivity. Time stamps are transmitted with the downsampled phasemeter data to ground. Time stamping errors are similar to ranging errors and would influence our ability to cancel laser frequency noise in post-processing. More details can be found in~\cite{TDI_ClockJitterReduction}.

Tilt-to-length (TTL) coupling includes for example piston effects in the local test-mass interferometer where an angular tilt of the test-mass couples to lateral spacecraft motion or where a lateral shift between the test-mass center of mass and the optical axis couples to angular spacecraft jitter. As discussed before, the relative alignment between the spacecraft and the test-mass is measured either interferometrically or capacitively and its TTL coupling will be calibrated and subtracted from the local interferometer data before it enters in equations \eqref{eq:eta}. However, the TTL noise in the inter-spacecraft interferometer includes contributions caused by angular jitter of the far spacecraft which turns into TTL due to the wavefront error. This jitter will be measured using wavefront sensing and properly calibrated and delayed before it also enters the TDI-generator.

\subsubsection{Impact on unsuppressed noises}

Only the spacecraft jitter noises including TTL, the laser frequency noises and
the clock jitter noises are suppressed by TDI (see sections~\ref{sec:TDI_SCjitter},
\ref{sec:TDI_laser}, \ref{sec:TDI_clock}) and, if the suppression works as expected, should not limit LISA. Instead, LISA will again be limited by noise sources which are typically uncorrelated such as shot noise, acceleration noise not associated with spacecraft motion, phasemeter noise, scattered light and many others. The TDI generators also affect their shape in the TDI data, how they will limit the sensitivity and our ability to extract gravitational waves from it. In order to compute the noise budget, it is necessary to compute the transfer function for each noise source. 

We will use the acceleration noises $\delta_{rs}$ and the interferometer noises, $N_{isc,rs}$,
in the simplified model introduced in section~\ref{sec:TDI_IntroModel} and the $X_{2.0}$ TDI generator to discuss this. We propagate the noises through the set of equations presented above ($\rm{isc}_{rs},\rm{tm}_{rs},\rm{ref}_{rs}\rightarrow\xi_{rs}\rightarrow\eta_{rs}\rightarrow X_{2.0}$) and arrive at:
\begin{eqnarray}
X_{2.0} & = & (1-D_{12}D_{21}D_{13}D_{31})((N_{isc,13}+D_{13}N_{isc,31}+(1+D_{13}D_{31})\delta_{TM,13}-2D_{13}\delta_{TM,31})\nonumber \\
 &  & +D_{13}D_{31}(N_{isc,12}+D_{12}N_{isc,21}-(1+D_{12}D_{21})\delta_{TM,12}+2D_{12}\delta_{TM,21}))\nonumber \\
 &  & -(1-D_{13}D_{31}D_{12}D_{21})((N_{isc,12}+D_{12}N_{isc,21}-(1+D_{12}D_{21})\delta_{TM,12}+2D_{12}\delta_{TM,21})\nonumber \\
 &  & +D_{12}D_{21}(N_{isc,13}+D_{13}N_{isc,31}+(1+D_{13}D_{31})\delta_{TM,13}-2D_{13}\delta_{TM,31}))
\end{eqnarray}
Then, we compute the Power Spectral Density (PSD), i.e. the Fourier
transform of the autocorrelation function of $X_{2.0}$,
$PSD_{X_{2.0}}=\left\langle \tilde{X}_{2.0}\ \tilde{X}_{2.0}^{*}\right\rangle $,
where the tilde denotes the Fourier transform or the earlier introduced linear spectral density. 

\begin{figure}[htbp]
\centering \includegraphics[width=0.8\textwidth]{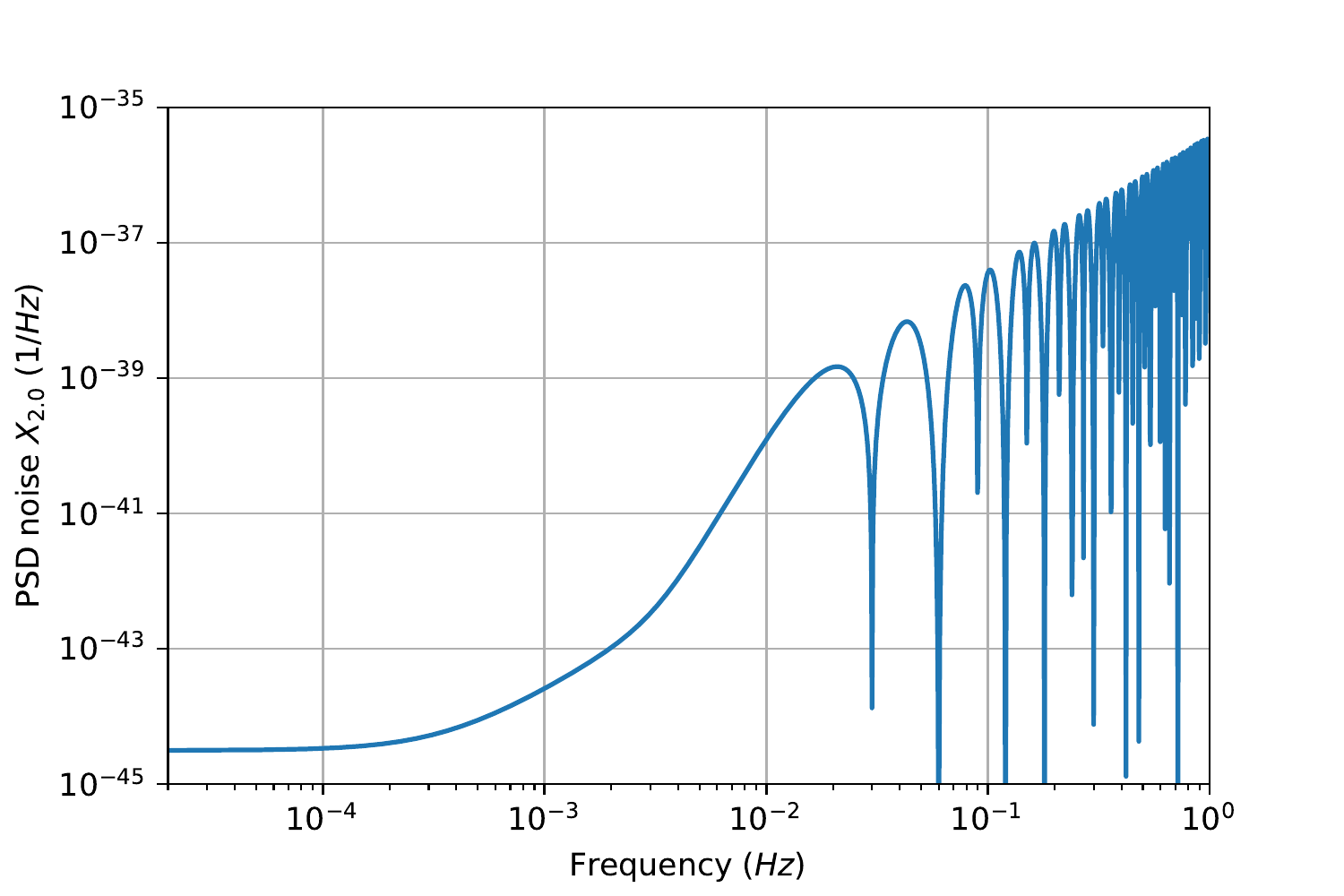}
\caption{PSD of noises for TDI $X_{2.0}$ in relative frequency unit for the LISA configuration as defined in the ESA LISA Science Requirement Document~\cite{SciRD}. 
Details are given in \cite{LISASensitivitySNR}.}
\label{fig:PSDX20_Noises_An} 
\end{figure}

Here we have to make certain assumptions about the noise. We will assume no correlation between noise sources such that all cross-terms will vanish, i.e. $\tilde{N}_{isc,rs} \tilde{\delta}_{TM,rs}=0$
and $\tilde{N}_{isc,rs}\tilde{N}_{isc,kl}=\tilde{\delta}_{TM,rs}\tilde{\delta}_{TM,kl}=0$
when $rs\neq kl$. While this is certainly true for shot noise in each isc interferometer or Brownian motion of the test-masses discussed earlier and measured in each tm interferometer, it is not obvious for example for noise sources which are driven by temperature changes or charging events caused by solar activities. However, here we limit the discussion to uncorrelated noise and the PSD will be:
\begin{eqnarray}
PSD_{X_{2.0}} & = & 16\sin^{2}\left(\omega\frac{L_{12}+L_{21}+L_{13}+L_{31}}{2c}\right)\nonumber \\
 &  & \times\left(\sin^{2}\left(\omega\frac{L_{12}+L_{21}}{2c}\right)\left(S_{N_{13}}+S_{N_{31}}+4\left(S_{\delta_{31}}+\cos^{2}\left(\omega\frac{L_{13}+L_{31}}{2c}\right)S_{\delta_{13}}\right)\right)\right.\nonumber \\
 &  & \left.+\sin^{2}\left(\omega\frac{L_{13}+L_{31}}{2c}\right)\left(S_{N_{12}}+S_{N_{21}}+4\left(S_{\delta_{21}}+\cos^{2}\left(\omega\frac{L_{12}+L_{21}}{2c}\right)S_{\delta_{12}}\right)\right)\right)\nonumber \\
\end{eqnarray}
where $\omega=2\pi f$ and $f$ is the Fourier frequency. $S_{\delta}=\left\langle \tilde{\delta}_{TM,rs}\tilde{\delta}_{TM,rs}^{*}\right\rangle $
and $S_{N}=\left\langle \tilde{N}_{isc,rs}\tilde{N}_{isc,rs}^{*}\right\rangle $ are the PSDs of the isc and tm noise terms introduced in equations \eqref{eq:isc} and \eqref{eq:tm}. Furthermore, using also the approximations that all armlengths are equal (or equal enough) and all noises of the same type have the same PSD in all six MOSAs, the PSD simplifies to: 
\begin{eqnarray}
PSD_{X_{2.0}} & = & 64\sin^{2}\left(\frac{2\omega L}{c}\right)\sin^{2}\left(\frac{\omega L}{c}\right)\left(S_{N}+2\left(1+\cos^{2}\left(\frac{\omega L}{c}\right)\right)S_{\delta}\right)\nonumber \\
\label{eq:PSDNoiseX20}
\end{eqnarray}
This PSD corresponds to the noise in the $X_{2.0}$ output of the instrument caused by $N_{isc}$ and $\delta_{TM}$. This exercise will have to be repeated for all other noise sources which can not be lumped into $N_{isc}$ and $\delta_{TM}$ and add to the overall noise of LISA. The resulting PSD  from ESAs LISA Science Requirement document~\cite{SciRD} for one of these TDI generators is shown in figure~\ref{fig:PSDX20_Noises_An}. The PSDs for the $Y_{2.0}$ and $Z_{2.0}$ are similar while the other generators will have different shapes.

\subsubsection{Noise quasi-uncorrelated TDI generators}

The noises in the Michelson TDI generators X,Y, Z are correlated.
In GW data analysis, in particular for computing the likelihood (see
section~\ref{sec:DA} and equation~\eqref{eq:likelihood}), we have
to do computation with the noise matrix: 
\[
\left(\begin{array}{ccc}
PSD_{X}(f) & CSD_{XY}(f) & CSD_{XZ}(f)\\
CSD_{YX}(f) & PSD_{Y}(f) & CSD_{YZ}(f)\\
CSD_{ZX}(f) & CSD_{ZY}(f) & PSD_{Z}(f)
\end{array}\right)
\]
In order to simplify the computation, this matrix can be diagonalised.

In the approximation that all the noise contributions of the same kind are equal in magnitude,
all PSD and all CSD are equal and the noise matrix can be
diagonalised~\cite{TDI_Prince}. We get the quasi-uncorrelated TDI
generators: 
\begin{eqnarray}
A=\frac{Z-X}{\sqrt{2}},\quad E=\frac{X-2Y+Z}{\sqrt{6}},\quad T=\frac{X+Y+Z}{\sqrt{3}}
\end{eqnarray}
A, E and T are widely used for GW data analysis. T is usually called
the ``null'' channel since the GW signal at low frequencies mostly vanishes.

\subsection{Instrument response to GW and sensitivity}

\label{sec:Sensitivity}

In the previous section, we expressed the noise level in the output
of the instrument. In this section we will expressed the GW signal
in this output and then the sensitivity.

The GW acts on the laser beams travelling between spacecrafts (link).
Its effect is obtained by integrating the metric perturbation $h_{ij}$
along the link. For a beam emitted by the sender $s$ and received
by $r$, it corresponds to 
\begin{eqnarray}
y_{GW,rs} & \approx & \frac{\Phi_{rs}(t-\hat{k}.\vec{R}_{s}/c-L_{rs}/c)-\Phi_{rs}(t-\hat{k}.\vec{R}_{r}/c)}{2(1-\hat{k}.\hat{n}_{rs})}\label{eq:GWRespLink}
\end{eqnarray}
where $\vec{R}_{s/r}$ is the vector position of a sender/receiver,
$\hat{n}_{rs}$ the unit vector connecting sender and receiver and
$\hat{k}$ the direction of GW propagation expressed as 
\begin{eqnarray}
\hat{k}=-\{\cos{\beta}\cos{\lambda},\cos{\beta}\sin{\lambda},\sin{\beta}\}.
\end{eqnarray}
where $\beta$ is the ecliptic latitude and $\lambda$ the ecliptic
longitude of the source. The projection of the GW strain, $h_{ij}$,
on the link is: 
\begin{eqnarray}
\Phi_{rs} & = & \hat{n}_{rs}\ h_{ij}^{SSB}\ \hat{n}_{rs}
\end{eqnarray}
The polarization basis of the GW in Solar System Barycenter reference
frame (SSB) is chosen as 
\begin{eqnarray}
\hat{u}=\frac{\partial\hat{k}}{\partial\lambda}\quad\textrm{and}\quad\hat{v}=\frac{\partial\hat{k}}{\partial\beta}
\end{eqnarray}
In this basis, 
\begin{eqnarray}
h_{ij}^{SSB} & = & (h_{+}\cos{2\psi}-h_{\times}\sin{2\psi})\epsilon_{ij}^{+}+(h_{+}\sin{2\psi}+h_{\times}\cos{2\psi})\epsilon_{ij}^{+}\\
\epsilon_{ij}^{+} & = & (\hat{u}\otimes\hat{u}-\hat{v}\otimes\hat{v})_{ij}\\
\epsilon_{ij}^{\times} & = & (\hat{u}\otimes\hat{v}+\hat{v}\otimes\hat{u})_{ij}
\end{eqnarray}
where $\psi$ is the polarization angle and $h_{+}$ and $h_{\times}$,
the polarization in the source frame. The GW response in TDI is obtained
by propagating the expression~\eqref{eq:GWRespLink} in TDI: $y_{GW,rs}\rightarrow isc_{rs},tm_{rs},ref_{rs}\rightarrow\xi_{rs}\rightarrow\eta_{rs}\rightarrow X,Y,Z$.

\begin{figure}[htbp]
\centering \includegraphics[width=0.8\textwidth]{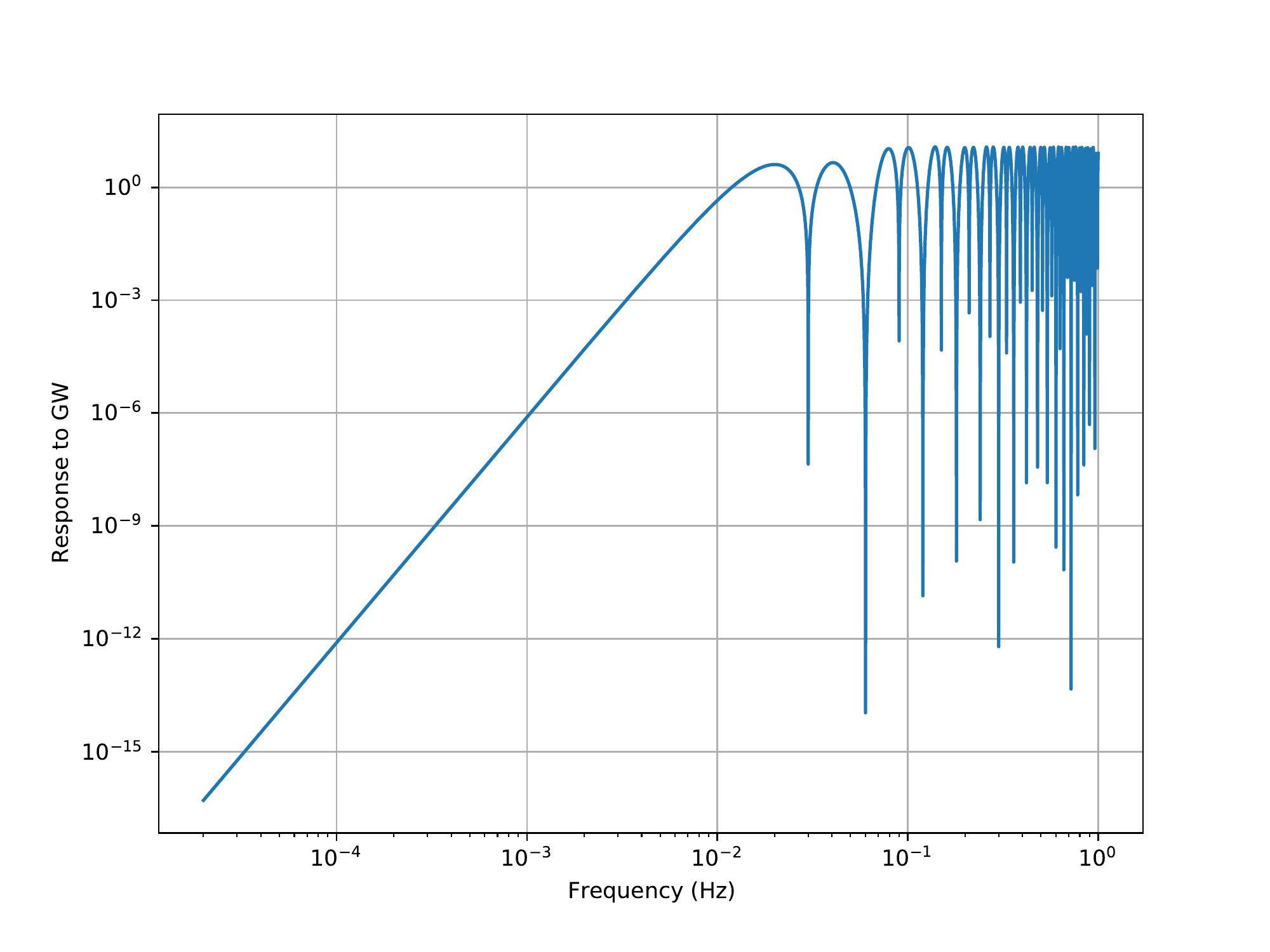}
\caption{Response of $X_{2.0}$ of LISA to GW sources.
Details are given in \cite{LISASensitivitySNR}.}
\label{fig:PSDX20_RespGW_An} 
\end{figure}

The response of the instrument depends of the source sky localisation
and polarization. To estimate the global response of the instrument
it is usually convenient to average it over sky and polarization.
Since there is no full analytical version of this average response,
it is computed numerically or analytically within some approximations
as for example the long wavelength limit that gives for TDI $X_{2.0}$:
\begin{eqnarray}
R_{LW,X_{2.0}} & = & \frac{48}{5}\left(\frac{\omega L}{c}\right)^{2}\sin^{2}\left(\frac{\omega L}{c}\right)\sin^{2}\left(\frac{2\omega L}{c}\right)
\end{eqnarray}
The response is shown on figure~\ref{fig:PSDX20_RespGW_An} .

\begin{figure}[htbp]
\centering \includegraphics[width=0.8\textwidth]{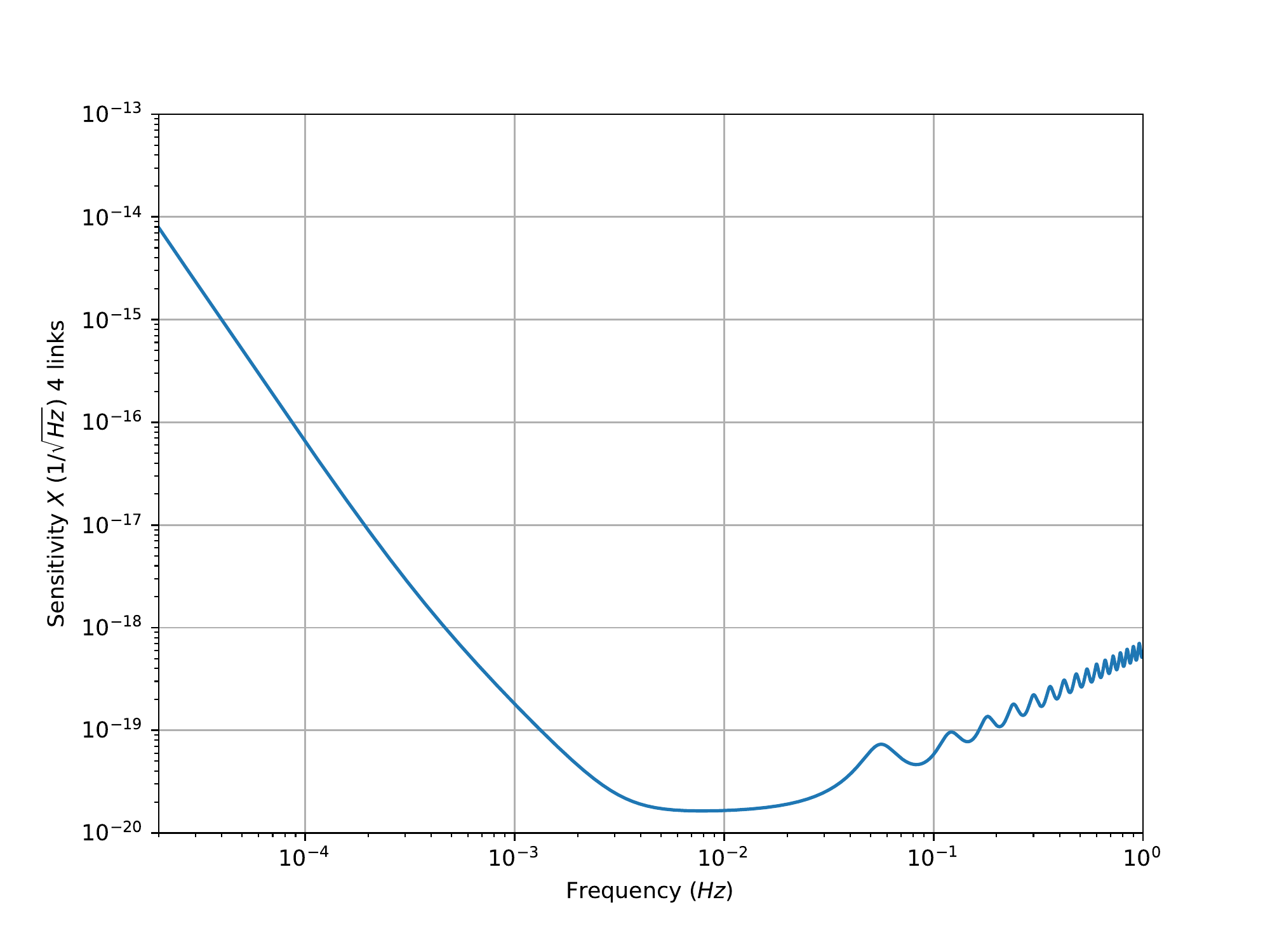}
\caption{Sensitivty of $X_{2.0}$ of the LISA configuration as defined
in the ESA LISA Science Requirement Document~\cite{SciRD}. 
Details are given in \cite{LISASensitivitySNR}.}
\label{fig:PSDX20_Sens_An} 
\end{figure}

The sensitivity of LISA is obtained
by dividing the PSD of the noise (example~\eqref{eq:PSDNoiseX20})
by the response of the instrument. An illustration is shown on figure~\ref{fig:PSDX20_Sens_An}

\subsection{Data analysis strategies}

\label{sec:DA}

We begin this section by summarising a few assumptions that are commonly
made when thinking about data analysis strategies for gravitational
wave detectors in space, or on the ground. The fundamental assumption
is that each data channel, $\mathbf{d}_{i}$, is a linear combinations
of signal, $\mathbf{h}_{i}$, and noise, $\mathbf{n}_{i}$. The transfer
function relating the incident gravitational wave strain to the content
of each of the TDI data channels was described above. The noise properties
were also characterised by the PSD. While a PSD can be defined for
any random process, for gravitational wave detectors the additional
assumption is typically made that the noise is \textit{stationary}
and \textit{Gaussian}. For a stationary process, noise fluctuations
at different frequencies obey the relation 
\begin{equation}
\langle\tilde{n}^{*}(f)\tilde{n}(f')\rangle=\delta(f-f')S_{n}(f)\label{eq:statnoise}
\end{equation}
where $S_{n}(f)$ is the PSD of the random process. In other words,
the fluctuations at different frequencies are independent and have
variances given by the PSD. For a Gaussian random process, all noise
components follow Normal distributions. A Normal distribution is completely
characterised by its covariance matrix and so the PSD completely specifies
the statistical properties of the noise. We introduce the noise-weighted
inner product 
\begin{align}
\left(\mathbf{a}|\mathbf{b}\right) & =\int_{-\infty}^{\infty}\frac{\tilde{a}^{*}(f)\tilde{b}(f)+\tilde{a}(f)\tilde{b}^{*}(f)}{S_{n}(f)}\;{\rm d}f\nonumber \\
 & =4\mbox{Re }\int_{0}^{\infty}\frac{\tilde{a}^{*}(f)\tilde{b}(f)}{S_{n}(f)}\;{\rm d}f\label{eq:innprod}
\end{align}
where the latter equality holds if $\mathbf{a}(t)$ and $\mathbf{b}(t)$
are both real time series. Under the assumptions of stationarity and
Gaussianity the probability distribution of the noise can be written
in terms of this inner product as 
\begin{equation}
p(\mathbf{n})\propto\exp\left[-\frac{1}{2}(\mathbf{n}|\mathbf{n})\right].
\end{equation}
The likelihood of the data, $p(\mathbf{d}_{i}|\vec{\theta})$, as
a function of the parameters, $\vec{\theta}$, of the gravitational
wave sources present is thus 
\begin{equation}
p(\mathbf{d}_{i}|\vec{\theta})=p(\mathbf{n}=\mathbf{d}_{i}-\mathbf{h}(\vec{\theta}))\propto\exp\left[-\frac{1}{2}(\mathbf{d}_{i}-\mathbf{h}(\vec{\theta})|\mathbf{d}_{i}-\mathbf{h}(\vec{\theta}))\right].\label{eq:likelihood}
\end{equation}

\subsubsection{Matched filtering and Bayesian analysis}

\label{sec:matchedfilt} Writing the signal component of the data
stream as $\mathbf{h}(\vec{\theta})=A\hat{\mathbf{h}}(\vec{\theta})$,
where $A$ is the ``amplitude'' of the waveform and $\hat{\mathbf{h}}(\vec{\theta})$
is a ``normalised'' template satisfying $(\hat{\mathbf{h}}|\hat{\mathbf{h}})=1$,
the log-likelihood can be expressed as 
\begin{equation}
\ln(p(\mathbf{d}_{i}|\vec{\theta}))\propto-\frac{1}{2}\left\{ \left[A-(\mathbf{d}_{i}|\hat{\mathbf{h}})\right]^{2}+(\mathbf{d}_{i}|\mathbf{d}_{i})-(\mathbf{d}_{i}|\hat{\mathbf{h}})^{2}\right\} .
\end{equation}
For given parameters $\vec{\theta}$, this is maximized for $A=(\mathbf{d}_{i}|\hat{\mathbf{h}})$,
and the maximum log-likelihood is $(\mathbf{d}_{i}|\hat{\mathbf{h}})^{2}/2$.
Therefore, the maximum likelihood parameters are those which maximize
$(\mathbf{d}_{i}|\hat{\mathbf{h}})^{2}$, the inner product of a normalised
template waveform with the data. This argument assumes that the amplitude
can be varied independently of the other parameters. This is typically
true for gravitational wave sources, since the amplitude is directly
proportional to the luminosity distance to the source, and the latter
does not otherwise alter the gravitational waveform.

The notion of identifying the parameters of the source by maximizing
the overlap $(\mathbf{d}_{i}|\hat{\mathbf{h}})$ also follows from
the theory of \textit{filtering}. We consider replacing the time series
$d_{i}(t)$ by a filtered series 
\begin{equation}
w(t)=\int_{-\infty}^{\infty}\,d_{i}(t')K(t-t')\;{\rm d}t',
\end{equation}
where $K(t)$ is some Kernel function. From the convolution theorem
for Fourier transforms $\tilde{w}(f)=\tilde{K}(f)\tilde{d}_{i}(f)$,
where $\tilde{K}(f)$ and $\tilde{d}_{i}(f)$ are the Fourier transforms
of $K(t)$ and $d_{i}(t)$ respectively. The expectation value of
the zero-lag output of the filter, $w(0)$, is 
\[
S=\langle w(0)\rangle=\int_{-\infty}^{\infty}\tilde{K}(f)\tilde{h}(f)\;{\rm d}f,
\]
where we use the fact that the noise has zero mean. The variance of
the zero-lag filter response in the absence of a signal is 
\[
N^{2}=\left\langle \bigg|\int_{-\infty}^{\infty}K(-t')n(t'){\rm d}t'\bigg|\right\rangle =\int_{-\infty}^{\infty}|\tilde{K}(f)|^{2}S_{n}(f)\,{\rm d}f,
\]
where we have made use of Eq.~(\eqref{eq:statnoise}). The ratio of
the expected zero-lag filter response in the presence of a signal
to the root-mean-square expected response in the absence of a signal,
$S/N$, is called the \textit{signal to noise ratio}. It is straightforward
to see that the signal-to-noise ratio is maximized by the choice of
a filter function 
\begin{equation}
\tilde{K}(f)\propto\frac{\tilde{h}(f)}{S_{n}(f)}.
\end{equation}
This is often referred to as a \textit{matched filter} because the
filter function takes the same form as the embedded signal, weighted
by the noise variance. The signal-to-noise ratio of this \textit{optimal}
filter is $(\mathbf{h}|\mathbf{h})$. The matched filter is only optimal
if the parameters of the signal are known. When the signal parameters
are unknown then typically multiple possible matched filters are considered,
forming a \textit{template bank} of possible waveforms. Every waveform
in the template bank is used to filter the data and the template with
the largest filter output provides a best-guess to the signal parameters.
As described above, this is equivalent to maximizing the likelihood.

Maximum likelihood evaluation and matched filtering provide point
estimates of the signal parameters, but a point estimate is usually
insufficient for useful scientific inference, unless it is accompanied
by some kind of estimate of the uncertainty in the estimate. While
it is possible to estimate uncertainties for point estimators, the
preferred approach in the gravitational wave community is to use Bayesian
inference for parameter estimation. In Bayesian inference the parameters
of the system are not regarded to be fixed and unknown, but to be
random variables described by a probability distribution. Before observing
any data, the state of knowledge of the system parameters is described
by a \textit{prior distribution}, $\pi(\vec{\theta})$. This can also
be thought of as a statement about the distribution of parameters
over the population of sources of that type. After observing the data,
the state of knowledge is udpated to a \textit{posterior distribution},
$p(\vec{\theta}|\mathbf{d}_{i})$, using the likelihood of the data
generating process in an application of \textit{Bayes' Theorem} 
\begin{equation}
p(\vec{\theta}|\mathbf{d}_{i})=\frac{p(\mathbf{d}_{i}|\vec{\theta})\pi(\vec{\theta})}{p(\mathbf{d}_{i})},\label{eq:BayesTh}
\end{equation}
where $p(\mathbf{d}_{i})=\int p(\mathbf{d}_{i}|\vec{\theta})\pi(\vec{\theta})\,{\rm d}\vec{\theta}$
is the \textit{Bayesian evidence}, and all probability distributions
are as previously defined. The Bayesian evidence plays a crucial role
in model selection but for parameter estimation it is just a normalising
constant and is typically ignored. The posterior distribution is usually
not available in a closed analytic form, and it is common to represent
the posterior by a (sufficiently large) set of samples drawn randomly
from it. Fast and efficient computational methods for drawing samples
from posterior distributions are an active area of research. Common
approaches employed in space-based gravitational wave data analysis
include Markov Chain Monte Carlo (MCMC) and nested sampling~\cite{MultiNest_LISA}.

The posterior distribution encodes all information that the data provides
about the signal, but it is computationally expensive to characterise.
For carrying out studies of expected parameter precision over the
full parameter space of signals it is advantageous to have faster
methods. In this context it is most common to compute the \textit{Fisher
matrix}, which is given in terms of the noise-weighted inner product
as 
\begin{eqnarray}
\Gamma_{ij}=\left(\frac{\partial\mathbf{h}}{\partial\theta^{i}}\bigg|\frac{\partial\mathbf{h}}{\partial\theta^{j}}\right).
\end{eqnarray}
The square root of the diagonal elements of the inverse Fisher matrix
provide a guide to the precision with which each parameter would be
expected to be measured, $\Delta\theta^{i}\sim\sqrt{\Gamma_{ii}^{-1}}$.
This can be justified in a number of ways, for example the Fisher
matrix appears as the Cramer-Rao lower-bound of the covariance of
an unbiased estimator of the parameters $\vec{\theta}$ and also as
the asymptotic covariance matrix characterising the limiting normal
distribution when performing multiple repeated measurements of the
parameters. However, the easiest way to see its relevance is to carry
out an expansion of the likelihood in the vicinity of the true parameters
of the signal, $\vec{\theta}_{0}$. We write 
\begin{equation}
\mathbf{h}(\vec{\theta})=\mathbf{h}(\vec{\theta}_{0})+\frac{\partial\mathbf{h}}{\partial\theta_{i}}\Delta\theta^{i}+\cdots,
\end{equation}
which is known as the \textit{linear signal approximation}. Substituting
into the likelihood for the data stream $\mathbf{d}_{i}=\mathbf{n}+\mathbf{h}(\vec{\theta}_{0})$
we obtain 
\begin{align}
p(\mathbf{d}_{i}|\vec{\theta}) & \propto\exp\left[-\frac{1}{2}(\mathbf{n}-\partial_{i}\mathbf{h}\Delta\theta^{i}|\mathbf{n}-\partial_{j}\mathbf{h}\Delta\theta^{j})\right]\nonumber \\
 & =\exp\left\{ -\frac{1}{2}\left[(\mathbf{n}|\mathbf{n})-2(\mathbf{n}|\partial_{i}\mathbf{h})\Delta\theta^{i}+(\partial_{i}\mathbf{h}|\partial_{j}\mathbf{h})\Delta\theta^{i}\Delta\theta^{j}\right]\right\} \nonumber \\
 & =\exp\left[-\frac{1}{2}(\mathbf{n}|\mathbf{n})\right]\exp\left[-\frac{1}{2}\left(\Delta\theta^{i}-(\Gamma^{-1})_{ik}(\mathbf{n}|\partial_{k}\mathbf{h})\right)\Gamma_{ij}\left(\Delta\theta^{j}-(\Gamma^{-1})_{jl}(\mathbf{n}|\partial_{l}\mathbf{h})\right)\right]\nonumber \\
 & \hspace{2cm}\times\exp\left[-\frac{1}{2}(\mathbf{n}|\partial_{i}\mathbf{h})(\Gamma^{-1})_{ij}(\mathbf{n}|\partial_{j}\mathbf{h})\right],
\end{align}
in which we are using the shorthand notation $\partial_{i}\mathbf{h}\equiv\partial\mathbf{h}/\partial\theta^{i}$
and all derivatives are evaluated at $\vec{\theta}_{0}$. Ignoring
the last term as it is lower order we see that this is a Gaussian
distribution with covariance given by the Fisher matrix. Thus the
Fisher matrix can also be seen as a leading-order approximation to
the log-likelihood. In this context it is easy to see that the Fisher
matrix is a better approximation at high signal-to-noise ratio, when
the expected uncertainties in the parameters are smaller and hence
the linear signal approximation is more likely to be valid. The Bayesian
posterior is not equal to the likelihood, but instead includes the
prior term. The Fisher matrix is still a good approximation to the
width of the posterior distribution if the prior is slowly varying
over the typical width of the likelihood. If that is not the case,
then the analogous result for the posterior distribution can be found
by replacing $\Gamma_{ij}$ with $\Gamma_{ij}+P_{ij}$, where $P_{ij}=(\partial\pi/\partial\theta^{i})(\partial\pi/\partial\theta^{j})$.

The Fisher matrix is proportional to the square of the waveform and
hence to the signal-to-noise ratio squared. Thus the expected precision
of parameter estimation, in the range of validity of the Fisher matrix,
scales like one over the signal-to-noise ratio. In section~\ref{sec:space_gw_sources}
we will estimate the precision with which LISA can measure source parameters using the Fisher matrix approach. While these results depend on the shape of the sensitivity curve and will in principle be different for different observatories, the results obtained for LISA can be used to roughly estimate the precision for other missions by computing the ratio of the signal-to-noise ratios in two different
gravitational wave detectors.


\subsubsection{Global fit strategies}

For a given GW source observed with LISA, the likelihood surface (likelihood
as a function of parameters) can be quite complex with multiple modes
(degeneracies) and very narrow peaks. Therefore extracting one source
is already not easy and requires the evaluation of a large number
of templates and likelihoods in order to sufficiently explore the parameters
space.

Compared to LIGO/Virgo/KAGRA, the additional complexity of LISA is
the large number of sources of various types that will be observed
with signals partially overlapping in time. In a matched filtering
approach, the optimal template would include all of the sources in the data, but this is not practical as the number of parameters to search for would be too large.
To solve this challenge, refereed to as the  \textit{global fit}, a number of different approaches
are currently being investigated. One approach is to first identify the loudest
sources in the data, subtract them, and then search for the loudest sources remaining in
the residual, continuing iteratively in this way. An alternative approach is to divide the data, either in the frequency or time domains, or perhaps the wavelet domain, into smaller sub-domains, and independently explore each part to find the sources~\cite{2010PhRvD..81f3008B}.

A common feature of all data analysis methods is their reliance on the existence of models for the gravitational waves as a function of the system parameters. Any analysis will require the evaluation of the likelihood at a large number of points in parameter space, which will require evaluation of a waveform template. Therefore, very fast computation of templates is necessary. A template depends on both the GW strain and the response of the instrument to a GW, which includes evaluation of TDI.
Many proposed approaches to LISA data analysis use Bayesian techniques~\cite{MultiNest_LISA,2007CQGra..24.5729C,2009CQGra..26m5004B}, but other approaches that have been suggested include genetic algorithms~\cite{GeneticAlgo_SMBHB,GeneticAlgo_GBs} and machine learning.

Since 2005 there has been intense research into LISA data analysis. One of the
driving elements was the Mock LISA Data Challenge (MLDC), which operated between 2005
and 2011~\cite{MLDCFirst,MLDC1b,MLDC34,2017JPhCS.840a2026B} and
now the LISA Data Challenge (LDC), which has operated since 2017. These comprise regular
releases of simulated data of increasing complexity both for the GW
sources and for the realism of the instrument. The results from many
participants are collected and analysed to compare efficiency of diifferent methods.

Almost all methods are using the TDI data as the input. However some
recent studies have investigated the possibility of using the interferometer
measurements directly , marginalising over the laser frequency noises~\cite{TDIInfinity,TDIPCABaghi}.

\subsubsection{Robust analysis and other analysis}

While the matched filtering methods are efficient, they strongly rely
on the knowledge of the signal we are looking for. Alternative methods
with minimum assumptions on the signal are also developed to complement
the matched filtering approaches. Since they are not optimal, they
are less efficient to extract small signals in the noise but they are more robust. For example some of them are looking for excess power~\cite{2020CQGra..37u5017K} and others at sparsity in the data~\cite{2020PhRvD.102j4053B}.

\subsection{Instrumental artefacts and noise characterisation}

In a space-based mission, it is impossible to modify the hardware
after the launch. We are also limited in our possibilities to investigate
problems and noise sources. It is therefore crucial to build an instrument
as robust as possible including a large variety of measurements and
to prepare calibrations procedures which will be used during the commissioning
phase but also to characterise the noises during the science operations.
In order to characterize the instrument and to check its performance,
we will also make use of the ``verification binaries'' which are
guaranteed sources with a well known GW emission. The LISA data will
probably be dominated by powerful GW sources hiding the other sources
and the noises. Since it is not possible to switch-off the GW sources,
we will have to rely on auxiliary measurements and on a precise noise
budget. The noise characterization will also be very important for
searching for stochastic sources as the potential GW emission from
the very Early Universe which appears as correlated noises between
various TDI channels.

In addition to ``standard noises'' several artefacts are expected
and the data analysis strategies have to be adapted to them. For example: 
\begin{itemize}
\item {\bf Glitches}: in LISA Pathfinder glitches (transient instrumental events)
have been observed with a duration between a few seconds to a few hours.
The characteristic time between two glitches is a day. These instrumental events are not
fully understood yet but we are expecting similar events in LISA. 
\item {\bf Gaps}: there will be gaps in the data. Some of them are required for maintenance
(ex: antenna repointing, change
of laser frequencies, etc). 
\item {\bf Non-stationarity}: fluctuations in the instrument and the environment such as changes in temperature distributions, fluctuations of the pressure around the test masses, aging and even failure of components such as the lasers will change the noise spectrum.
\end{itemize}

\subsection{Ground segment design}

The Ground Segment (GS) is a key element in most space missions
and this is particularly true for GW space based observatories. In LISA it is considered to be a part of the instrument since it cleans the data and produces the final interference measurements, the TDI
data. The GS will also perform the GW searches which will require large computing resources.

The design of the LISA GS is an ongoing effort. It will have to perform the
following processes: 
\begin{itemize}
\item \textbf{Calibration}: As for any instrument, measurements will have
to be calibrated. Dedicated operations on ground before the launch
or in space could be necessary to measure the calibration coefficients
as for example the actuation gain.
\item \textbf{Forces subtraction} : As for LISA Pathfinder, we are expecting
to have various forces acting on the test-mass and spacecraft. These
forces will have to be subtracted from the data.
\item \textbf{Clock synchronisation}: As described before (see section~\ref{sec:TDI_clock}),
each spacecraft has an Ultra Stable Oscillator used to time tag all
measurements. Since USO are not perfect and are in different gravitational
potential, the data from the three spacecraft are not synchronised
with each other and with the Coordinated Universal Time. It is therefore
necessary to synchronise the different data on the same time reference.
The Kalman filtering also used for ranging (see next item) will contribute
to this synchronisation.
\item \textbf{Ranging}: In order to perform the suppression of dominating
noises with TDI, it is crucial to have a good knowledge of the armlength.
While the analysis of the PRN with the phasemeter will provide a first
estimate of the armlength, it is necessary to refine it to reach the
precision required for TDI. In LISA, this refinement will be done
via Kalman filtering.
\item \textbf{TDI}: Several dominant noises (the spacecraft jitter noises,
the laser frequency noises and the clock jitter noises) have to be
strongly suppressed by TDI (see section~\ref{sec:TDI}).
\item \textbf{Low latency analysis}: The first step of the GW analysis
strategy is the low latency pipeline for performing a fast analysis
of the data and provide as soon as possible alerts to
the broad community. These alerts are the detection of new strong
sources with their localisation parameters or the refinement of the
localisation parameters of sources already detected.
\item \textbf{Full analysis}: The full analysis will generate the primary scientific results for LISA. Using multiple pipelines (chains of analysis processes) it will investigate deeply the data to extract
as much information as possible, about all the GW sources in the data that pass some detectability threshold, and also the characteristics of the instrument.
\end{itemize}
LISA is the first mission of this kind and there are a number of uncertainties
both for instrument and GW sources (populations, new unexpected sources,
etc). Therefore it is crucial to allow some flexibility in the data
analysis pipeline to quickly adapt them. It is also important to use
in parallel different methods to cross-check and consolidate the results.

\section{GW Sources for Space-based Observatories}
\label{sec:space_gw_sources}
In this section we will give an overview of the likely sources of gravitational waves for space-based gravitational wave detectors. There are other chapters in this book that provide more detailed expositions of the astrophysics of these various types of sources, so we will only briefly discuss this, but will also discuss the precision with which space-based gravitational wave detectors will be able to characterise the properties of the systems they observe. We start with some general observations. The following results follow from straightforward physical arguments which we will provide, but related results and arguments can be found in~\cite{PhysRev.136.B1224,Flanagan_1998}.

The primary source of gravitational waves for all types of gravitational wave detectors are binary star systems containing two compact objects. The dominant gravitational emission from a binary is at twice the orbital frequency, $f_{\rm gw} = 2 f_{orb}$. We will now derive some results for a Newtonian binary, but the scalings also apply in the relativistic limit. The orbital frequency of a Newtonian binary with component masses $M_1$ and $M_2$ and a semi-major axis $a$ is 
\begin{equation*}
    2 \pi f_{\rm orb} = \sqrt{\frac{G (M_1+M_2)}{a^3}}.
\end{equation*}
The semi-major axis of a compact binary just before merger is
\begin{equation}
    a_{\rm merg} = k \frac{G (M_1 + M_2)}{c^2}
\end{equation}
where $k$ is a constant that depends on the mass ratio and eccentricity of the binary and the spin of the individual binary components. 
The gravitational wave frequency at merger is thus
\begin{equation}
    f_{\rm gw,merg} = 4.4 {\rm mHz} \left(\frac{6}{k} \right)^{\frac{3}{2}} \left(\frac{10^6 M_\odot}{(M_1+M_2)} \right).
\end{equation}
In the point-particle limit, $M_2 \ll M_1$, and for a non-spinning primary black hole and a circular orbit, $k = 6$. 
Ground-based gravitational wave detectors operating in the frequency band from a few to a few thousand Hertz can thus observe the mergers of compact binaries with masses between $\sim 1 M_{\odot}$ and $\sim 1000 M_\odot$. Space-based gravitational wave detectors, on the other hand, operate in the millihertz frequency range and so cannot observe such mergers, but are instead sensitive to gravitational waves emitted during the merger of much more massive systems, typically with total mass in the range $\sim 10^4 M_\odot$--$10^7 M_\odot$. 

Merging binaries containing MBHs 
are important sources for space-based detectors and will be discussed in sections~\ref{sec:mbh} and~\ref{sec:emri}. This is not the whole story, however. Binaries emit not only at the merger frequency but at all frequencies below the merger frequency. To understand if the early phases of the binary emission are detectable we need to understand how the gravitational wave amplitude changes as the binary evolves. The detectability of a gravitational wave source may be characterised by its matched filtering signal-to-noise ratio, as defined and described in section~\ref{sec:matchedfilt}. Explicitly this is given by
\begin{equation}
    \rho^2 = 4 \int_0^\infty \frac{|\tilde{h}(f)|^2}{S_n(f)} \, {\rm d}f.
    \label{eq:snrsq}
\end{equation}
For an evolving binary the Fourier transform is related to the amplitude of the radiation, $h$, and the rate of change of frequency via the stationary-phase approximation, $\tilde{h} \sim h/\sqrt{\dot{f}}$. At leading order, the gravitational wave strain from a source at distance $D$ is given by the second time derivative of the quadrupole moment of the source, $h \sim \ddot{I}/D$. For a Newtonian binary with orbital frequency $\omega/(2\pi)$, the quadrupole moment can be estimated as
$$
I\sim \mu r^2 \cos2\omega t \sim \frac{M_1 M_2}{(M_1+M_2)^{\frac{1}{3}}} \omega^{-\frac{4}{3}},
$$
where $\mu = M_1 M_2/(M_1 + M_2)$ is the reduced mass. The gravitational wave strain can therefore be estimated as
$$
h \sim \frac{\ddot{I}}{D} \sim \frac{1}{D} \frac{M_1 M_2}{(M_1+M_2)^{\frac{1}{3}}} \omega^{\frac{2}{3}}.
$$
The rate of energy loss scales like the third time derivative of $I$ squared and so this has the scaling
$$
\dot{E} \sim -\dddot{I}^2 \sim -\mu^2 M^{\frac{4}{3}} \omega^{\frac{10}{3}}.
\label{eq:Edot}
$$
Finally, we need to know how the energy relates to the orbital separation or equivalently the orbital frequency. In the Newtonian limit this follows from
$$
E = -\frac{M\mu}{2r} = -\frac{\mu (M\omega)^{\frac{2}{3}}}{2}
$$
from which we deduce
\begin{equation}
\dot{E} \sim -\mu M^{\frac{2}{3}} \omega^{-\frac{1}{3}} \dot\omega.
\label{eq:EdotCBC}
\end{equation}
Combining this with expression~(\ref{eq:Edot}) we obtain
$$
\dot{\omega} \sim \mu M^{\frac{2}{3}} \omega^{\frac{11}{3}} = \frac{M_1 M_2}{(M_1+M_2)^{\frac{1}{3}} } \omega^{\frac{11}{3}}  = M_c^{\frac{5}{3}} \omega^{\frac{11}{3}}
$$
where we have introduced the chirp mass
$$
M_c = \frac{M_1^{\frac{3}{5}} M_2^{\frac{3}{5}}}{(M_1+M_2)^{\frac{1}{5}} }.
$$
We deduce that the Fourier domain strain scales as
$$
\tilde{h}(f) \sim \frac{h}{\sqrt{\dot f}} \sim \frac{1}{D} \frac{M_c^{\frac{5}{3}} f^{\frac{2}{3}}}{M_c^{\frac{5}{6}} f^{\frac{11}{6}}} = \frac{1}{D} M_c^{\frac{5}{6}} f^{-\frac{7}{6}}.
$$
A more careful calculation, no longer making a Newtonian assumption, but working only to leading order, (see, for example,~\cite{Berti:2004bd}) gives the prefactor in this expression
\begin{equation}
\tilde{h}(f) = \frac{c}{\sqrt{30} \pi^{\frac{2}{3}}} \frac{1}{D} \left(\frac{G M_c}{c^3}\right)^{\frac{5}{6}} f^{-\frac{7}{6}}.
\label{eq:FDstrain}
\end{equation}
We note that for sources at a cosmological redshift $z$, the correct mass to use in these expressions is the \textit{redshifted mass}, i.e., $M_{c,z} = (1+z) M_c$, and the correct distance is the luminosity distance. This distinction is not important for the following few sections, but it will be relevant when we discuss the science applications of the various sources. The Fourier domain amplitude in Eq.~(\ref{eq:FDstrain}) increases at lower frequencies, but the available bandwidth for observation, i.e., the range of frequencies over which a source is observed, typically decreases. 
For sources for which the full inspiral to merger is observed, the bandwidth can be approximated by the merger frequency, and $|\tilde{h}(f_{\rm merg}|^2 f_{\rm merg} \sim M_c^{5/3} (M_1+M_2)^{4/3}/D^2 = \eta (M_1+M_2)^3/D^2$, where $\eta=\mu/(M_1+M_2)$ is the symmetric reduced mass ratio. As argued above, merging sources for space-based detectors are typically $10^4$ times heavier than those for ground based detectors, providing a factor of $10^{12}$ in this expression for sources at the same distance. The signal-to-noise ratio depends also on the PSD of the detectors, and this is typically a few orders of magnitude  larger for space-based detectors. Nonetheless, these simple arguments suggest that merging sources will have signal to noise ratios several orders of magnitude larger at the same distance. More careful calculations show that this is indeed the case and space-based detectors are expected to see mergers at very high redshifts, if such sources exist.

For sources that do not inspiral completely over the observation, we can approximate the bandwidth by $\Delta f \sim \dot{f} T$, where $T$ is the length of observation. In this case, the square root of the numerator of the integrand of Eq.~(\ref{eq:snrsq}) can be seen to scale as $\tilde{h}(f) \sqrt{\Delta f} \sim M_c^{5/3} f^{2/3} \sqrt{T}$. A binary with mass of $1$ solar mass would be observed by ground-based detectors merging at about $1$kHz, and would be in the band of ground-based detectors for $\sim 100$s. A space-based detector could observe the same source at $\sim 1$mHz for several years, $\sim 10^8$s, and so the signal-to-noise ratio would only be a factor of $\sim 10$ lower, if the strain sensitivities expressed as PSDs were comparable in the respective frequency bands. The difference in PSD changes this to a factor of one thousand or more, for sources at the same distance. The detection horizon for such systems in ground-based detectors is tens of megaparsecs, so these arguments suggest that space-based detectors could detect the same systems in the early inspiral phase at distances of tens of kiloparsecs. This encompasses our galaxy, in which such early-inspiral compact binaries are expected to be abundant. Such compact binaries in the Milky Way are another important source for space-based detectors, which are discussed in section~\ref{sec:WDbinaries}.

We can repeat the above arguments for systems with mass around $100M_\odot$. These will be observed for $\sim 0.1$s up to merger at $\sim 50$Hz by ground-based detectors, while a space-based detector could observe these at frequencies of $\sim 10$mHz for several years. The same argument as before suggests the square root of the numerator of Eq.~(\ref{eq:snrsq}) is a few tens higher for the space-based detector. Accounting for PSD differences we might therefore expect such sources to be visible with similar signal-to-noise ratios in ground and space-based instruments. More careful calculations confirm this is indeed the case, and it is now expected that the more massive stellar origin compact binaries in the population being observed by ground-based detectors could also be observed several years earlier by space-based detectors. This will be discussed in section~\ref{sec:sobhb}.

So far we have concentrated on the observability of individual sources, but there is also the prospect of detecting a stochastic background of gravitational wave radiation, which could be generated by a superposition of a larger number of individual sources, or via processes occurring in the early Universe. The latter source of gravitational waves will be discussed in section~\ref{sec:cosback}.

\begin{figure}
\begin{centering}
\includegraphics[width=10cm]{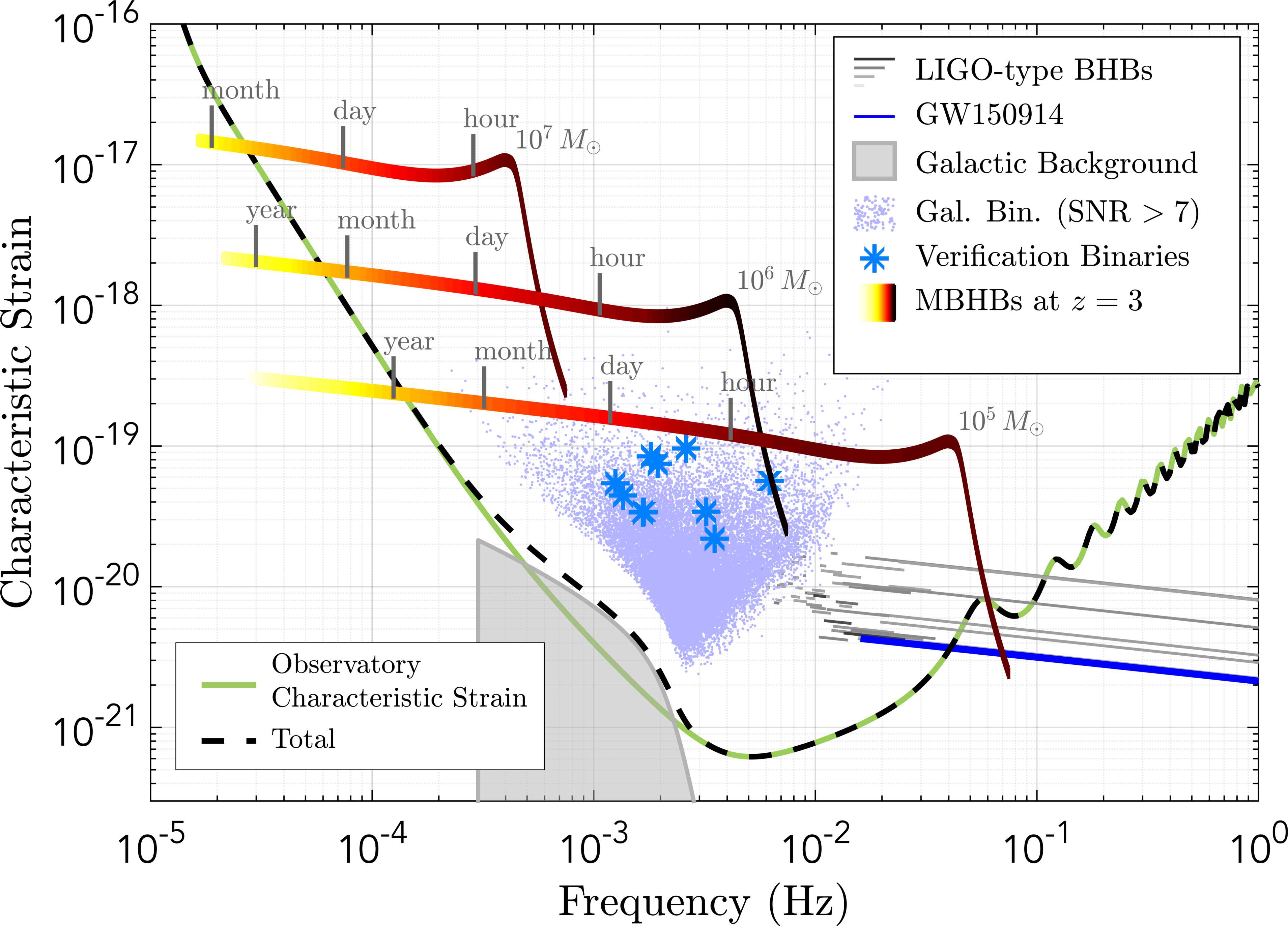}
\par\end{centering}
\caption{\label{fig:sources}LISA sensitivity curve (in green), and a selection of astrophysical sources for space-based gravitational wave detectors. Sources illustrated include resolved and unresolved galactic binaries (see section~\ref{sec:WDbinaries}), massive black hole mergers (see section~\ref{sec:mbh}) and stellar-origin black hole binaries (see section~\ref{sec:sobhb}). The tracks show the evolution of the system in the observation band, while the height above the instrumental sensitivity is a measure of the signal-to-ratio of the system.}
\end{figure}

Figure~\ref{fig:sources} shows a representation of a number of sources for space-based gravitational wave detectors, overlaid over the sensitivity curve of the LISA instrument~\cite{LISA_Proposal2017}.

\subsection{Compact binaries in the Milky Way}
\label{sec:WDbinaries}
The majority of stars in the Universe are found in binaries, and the end point of stellar evolution is the formation of a compact object, either a white dwarf, a neutron star or a black hole. If the binary survives the formation of the compact objects, then once the binary has decayed to the point that the orbital frequency is of the order of an hour, the binary will be generating gravitational waves at millihertz frequencies~\cite{1990ApJ...360...75H,2001A&A...368..939N}. As argued above these should be detectable by space-based gravitational wave detectors for binaries within the Milky Way. There are tens of millions of such ``ultra-compact'' hour-period binaries in the Milky Way, the vast majority of which are double white dwarf systems. There are so many binaries that they will not all be individually resolvable, but the majority will form a stochastic confusion foreground~\cite{1987ApJ...323..129E}. A significant number of the ultra-compact binaries will be evolving. Some binaries will be chirping to higher frequency due to the decay of the orbit through the emission of gravitational waves, but other binaries will be moving to lower frequency as a result of mass transfer between the binary components driving an increase in the orbital separation. Several ultra-compact binaries are already known through electromagnetic observations which are sufficiently close and at high enough frequency that the gravitational waves they are emitting will be quickly detected by space-based observatories~\cite{2006CQGra..23S.809S,2018MNRAS.480..302K}. These ``verification binaries'' could play an important role in assessing the performance of the LISA mission.

 A first-generation space-based detector such as LISA would be expected to resolve between five and ten thousand ultra-compact binaries, and also detect the astrophysical foreground from the unresolved population~\cite{1990ApJ...360...75H,Nissanke_2012,2019MNRAS.490.5888L}. The exact number depends on the sensitivity of the instrument and on currently unknown details of the astrophysical population. About one quarter if the resolved systems will show sufficient frequency evolution for the frequency derivative to be measured and for a handful of binaries the second time derivative of frequency could be measured, allowing a determination of the mass ratio of the system.

The ability of space-based detectors, in particular LISA, to constrain the parameters of ultra-compact binaries has been assessed using Fisher matrix methods, and also verified within the framework of the LISA (Mock) Data Challenges~\cite{MLDCFirst}. The latter have confirmed using realistic datasets containing an entire simulated galaxy of signals, that thousands of ultra-compact binaries can be individually resolved~\cite{2010PhRvD..81f3008B,MLDC1b,MLDC34}. As the sources are essentially monochromatic, the precision of frequency estimation is determined by the resolution of Fourier frequency bins, and so is of the order of $1/(1 {\rm year}) \sim 10^{-8}$s$^{-1}$ (for a $\sim 3$ year observation). The precision of measurements of the frequency derivative can be similarly estimated to be $\Delta f/T \sim 10^{-16}$Hz/s, set by the requirement that the frequency changes by one frequency bin over the observation time. Estimates of sky location and distance precision from the Fisher matrix suggest typical sky location accuracies of a tens of square degrees, and distances to ten percent. However, about $20\%$ of detected systems will be well localised on the sky (less than $10$ square degrees) and about $5\%$ of systems will have both good sky localisation and accurate distance determination (less than $1\%$)~\cite{2013MNRAS.429.2361L}. 

\subsection{Stellar-origin black hole binaries}
\label{sec:sobhb}
The gravitational wave source GW150914, was transformational not only because it was the first gravitational wave source detected by man-made observatories~\cite{PhysRevLett.116.061102,Abbott_2016}, but because it provided the first ever direct constraint on the binary black hole population. The system was surprisingly massive, with components of mass $\sim 29M_\odot$ and $\sim 36M_\odot$ respectively. As argued earlier in this section, such systems can be observed with comparable signal-to-noise ratio by space-based detectors, a few years before the source is observed to merge by a ground-based detector. The space-based observation not only probes an earlier phase in the inspiral, but also provides pre-warning of the time and sky location of merger events to facilitate joint observations in multiple wave and frequency bands. Combined observations offer a unique opportunity to probe the astrophysics of binary black hole systems as they approach merger.

After the announcement of GW150914 it was shown that a space-based detector like LISA could detect a few tens of similar stellar-origin black hole binary (SOBHB) systems in the early stages of inspiral~\cite{PhysRevLett.116.231102}. This assumed a threshold signal-to-noise ratio of $\sim 8$ would be needed for detection, and that the source could be observed by the space-based detector for $\sim 5$years. The exact number of observed events depends crucially on the details of the astrophysical population, and on the assumed high-frequency sensitivity of the space-based interferometer~\cite{2019MNRAS.488L..94M,2019PhRvD..99j3004G}. GW150914 was a threshold event for LISA, meaning that any source with lower mass or at greater distance would not be detectable. Therefore the number of observed events depends on the details of the black hole population at masses greater than GW150914. Subsequent LIGO/Virgo observations have included a number of additional high mass systems, including some which would also have been observable by a LISA-like detector~\cite{Abbott_2019_GWTC1,Abbott_2020_GWTC2}. These observations also showed evidence for the hypothesized ``mass-gap'', the absence of black holes with masses between $\sim 50M_\odot$ and $\sim 150M_\odot$ due to the onset of pair-instability during the collapse of the parent star at the end. The mass gap reduces the number of potential high mass LISA sources. However, LIGO/Virgo have also seen one other event, GW190521~\cite{Abbott_2020_GW190521}, that is consistent with at least one component lying within the mass gap. While the exact nature of this system is under debate, it could hint at the existence of another population of higher mass systems that are potential sources for space-based detectors~\cite{2020arXiv201006056T}.

Space-based detectors observe the early inspiral of SOBHB systems, where the system is evolving slowly, but they can observe them for many years, observing many cycles of the phase evolution. This facilitates extremely precise measurements of the system parameters. The space-based observation alone can determine the individual masses in the binary to better than $1\%$, the sky location to within 10 square degrees, and the time of coalescence to within $10$s, months to years before the coalescence is due to take place~\cite{PhysRevLett.116.231102,2020arXiv201006056T}. If the SOBHB orbit is eccentric at a frequency of $0.01$Hz, the eccentricity can be measured. Eccentricities as small as $10^{-3}$ can be distinguished in the observation~\cite{PhysRevD.94.064020}, while larger eccentricities, $e \sim 0.1$, can be measured with uncertainties of $\sim 10^{-7}$. These eccentricity measurements are important for distinguishing SOBHB formation channels, which will be discussed in section~\ref{sci:astro:sobh}.


\subsection{Massive black hole binaries}
\label{sec:mbh}
It is now well established that galaxies and massive black holes formed very early in the evolution of the Universe. Galaxies have been found at redshifts greater than $10$~\cite{2013ApJ...762...32C} and accreting supermassive black holes have been observed at redshifts greater than $7.5$~\cite{2021ApJ...907L...1W}. There is also growing evidence of the presence of lower mass accreting black holes at high redshift~\cite{Matsuoka:2017frx}. Most galaxies appear to host black holes at their centres~\cite{Kormendy:2013dxa} and these are very massive, $\sim10^6M_\odot$--$10^{10}M_\odot$, but more and more low mass black holes are now being discovered~\cite{Greene_2020}. Over cosmic history, galaxies merge and it is expected that, following such mergers, the massive black holes at their centres will also merge via gravitational wave emission. Lower mass galaxies tend to have lower mass black holes in their centres, and as we go back in time galaxies were less massive than today, so many of these gravitational waves will come from systems with mass in the range $10^4M_\odot$--$10^7M_\odot$ and will thus be in the millihertz range observable to space-based detectors. These lighter black holes are hard to observe electromagnetically and so there are a number of viable models for the formation of massive black hole seeds that are consistent with current EM observations. In ``light seed'' models, black holes with mass $\sim100M_\odot$ form at high redshift through the collapse of the first generation of low-metallicity high-mass stars~\cite{1999ApJ...527L...5B}. In ``heavy seed'' models, black holes of higher mass, $\sim10^5 M_\odot$ form later through the direct collapse of most of the gas in the galaxy into a single supermassive star~\cite{2003ApJ...596...34B}. Other formation channels include the formation of seed black holes of $\sim10^3M_\odot$ 
via run-away stellar or black hole collisions in young, dense star clusters~\cite{10.1111/j.1365-2966.2012.20406.x,2012ApJ...755...81M}. Gravitational wave observations with space-based detectors will directly probe the first epoch of massive black hole (MBH) mergers and hence help to distinguish between these different models and shed light on the early growth of structure in the Universe.

\begin{figure}
\begin{centering}
\includegraphics[width=10cm]{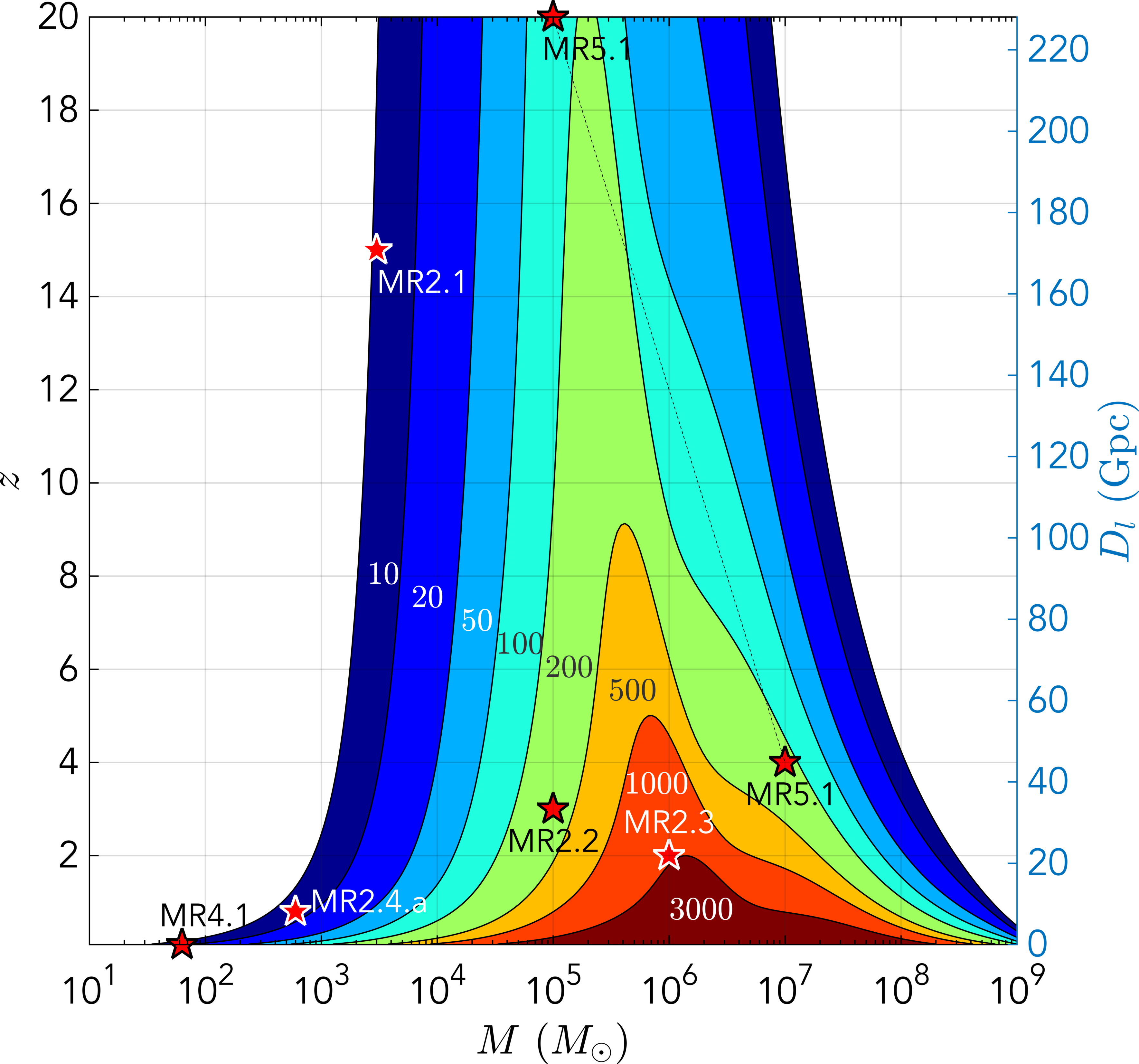}
\par\end{centering}
\caption{\label{fig:mbhsnr}Contours of constant signal-to-noise ratio for observations of MBH mergers with mass ration $1:5$ with the space-based detector LISA, as a function of the redshifted total mass of the binary (horizontal axis) and distance/redshift (vertical axis). Space-based gravitational wave detectors will be able to observed MBH mergers of the right mass to very high redshift. Stars indicate points for which LISA mission requirements were set in order to ensure detection of these sources.}
\end{figure}

Space-based GW detectors will observe MBH mergers with very high signal-to-noise ratio out to very high redshift. Figure~\ref{fig:mbhsnr} shows the signal-to-noise ratio with which the LISA detector would be able to resolve MBH mergers with mass ratio of $1:5$, as a function of the mass of the object and the distance/redshift of the source. Despite these high signal-to-noise ratios, the number of events that will be observed is somewhat uncertain, driven by the uncertainties in the astrophysical population that were described above. 
In addition to the uncertainties in the mass of the black hole seeds, a significant source of uncertainty is in the unknown delay time between the merger of the host galaxies and the subsequent merger of the MBH binary~\cite{Antonini_2015}. Nonetheless, most models predict that a LISA-like space-based detector would observe between about $10$ and about $100$ MBH mergers per year~\cite{Klein_2016}, assuming that a signal-to-noise ratio of $8$ is required for a confident detection. 
Of these observed mergers approximately half will be seen at high redshift $z > 7$, except when delay times are long and mergers occur at later times. 
The number of events observed in the heavy-seed models is largely independent of the exact configuration of the space-based detector, since these events are so loud, but some light-seed models predict many light mergers at high redshift, so a more sensitive detector can detect significantly more of these, with important astrophysical implications~\cite{Klein_2016}.

Space-based gravitational wave observations will constrain the intrinsic parameters of observed MBH binaries to very high precision, driven by the fact that the signals can be observed with very high signal-to-noise ratio and for many thousands of cycles of phase evolution. The (redshifted) masses of the binary components can be determined to a precision better than $1\%$ for about half of the observed events, and the spin magnitude of the primary (secondary) black hole measured to $1\%$ ($10\%$) for about ten percent of the observed events
~\cite{Klein_2016}. 
A significant fraction of systems will have the primary spin magnitude and the misalignment of the spin of the binary components with respect to the orbital plane of the binary constrained to the percent level. The fraction of observed systems for which this is possible could be as high as $\sim 25\%$, but this is dependent on the details of the spin distribution of astrophysical MBHs, which is extremely uncertain. 
A few events per year should also have well measured spins for the remnant black hole created during the merger~\cite{Klein_2016}. The typical expected precision of sky localisation is tens of square degrees and that of luminosity distance is tens of percent. However, as many as a few tens of events per year at $z < 5$ could be localised to better than $10$ square degrees and $10\%$ in distance~\cite{Klein_2016}. These well localised events at lower redshift are good targets for follow-up electromagnetic observations, to identify any counterparts. It is also expected that there will be at least one event, and perhaps a few tens, that will be at high redshift, $z > 7$, and have distance measured to $\sim 30\%$~\cite{Klein_2016}. These events are important as the distance precision is enough that we will be sure they are at high redshift, and hence provide important constraints on models of MBH 
formation and evolution.

\subsection{Extreme-mass-ratio inspirals}
\label{sec:emri}
The MBHs 
in the centres of galaxies that were described in the previous section are typically surrounded by clusters of stars. Stars in these clusters follow the usual evolutionary path, leading to the eventual formation of a compact remnant, which will be a black hole, neutron star or white dwarf, depending on the mass and the metallicity of the original star. These galacto-centric stellar clusters are dense, and the stars within them undergo frequent encounters which can leave these compact objects on orbits that pass very close to the central MBH. 
Such objects can get captured onto orbits bound to the central MBH and then gradually inspiral into the MBH via emission of gravitational waves~\cite{2007CQGra..24R.113A}. Typically the ratio of the mass of the stellar-origin compact object that is falling into the MBH to the mass of the MBH is $\sim10^{-5}$, so these events are called \textit{extreme-mass-ratio inspirals} or EMRIs.

Over the past two decades, observations of the stellar cluster around the black hole in the centre of the Milky Way have revealed a number of unexpected features~\cite{2008ApJ...689.1044G,2009ApJ...692.1075G}, indicating that the physics of stellar clusters around MBHs is poorly understood. EMRI observations will explore a much larger sample of these stellar environments in the Universe. In addition, EMRIs offer an exciting new way to probe fundamental physics. Due to the extreme mass-ratio, each EMRI emits detectable gravitational waves for hundreds of thousands of waveform cycles while the small object is in the strong gravitational field close to the central MBH. The emitted gravitational waves encode a map of the spacetime structure that can be used to test general relativity~\cite{Gair_2013}. This will be discussed further in section~\ref{sec:fp:mapping}. 

The capture scenario for the formation of an EMRI described above is the ``standard'' formation channel~\cite{1995ApJ...445L...7H,1997MNRAS.284..318S} and leads to EMRIs that have moderate eccentricity ($\sim 0.2$) at plunge, and are on orbits that are inclined with respect to the orbital plane of the central MBH. However, a number of alternative scenarios have also been suggested. Binary stars in the vicinity of the MBH can come close enough to the MBH to undergo a three-body interaction that splits the binary and leaves one component bound to the MBH~\cite{Miller_2005}. 
Massive stars that similarly come close to the MBH can have their outer envelope stripped, leaving the white dwarf core bound to the central objects~\cite{Davies_2005}. In both these scenarios, the compact object is captured with random inclination, but sufficiently far from the central MBH that the orbit will have circularised before the object enters the band of space-based detectors as an EMRI. A final alternative scenario is the formation of compact objects in an accretion disc around a MBH. In this scenario, parts of the disc collapse to form massive stars which then evolve as normal and leave compact remnants in orbits around the MBH, that eventually inspiral as EMRIs~\cite{2007MNRAS.374..515L}. As in the previous two scenarios, EMRIs formed in this way are predicted to be on circular orbits, but now also in the equatorial plane of the MBH. The relative importance of these various scenarios in the Universe is currently unknown, but gravitational wave observations could elucidate the different channels through measurements of the orbital properties of the objects. We refer the reader to~\cite{2007CQGra..24R.113A} and references therein for more details.

The complexity of the physics of stellar clusters means that the rate at which EMRIs occur in the Universe is very uncertain. Of particular importance is the fraction of compact object captures that lead to gradual inspiral into the MBH, which would be observable as EMRIs, versus those that plunge directly into the MBH, the poorly known scaling of the EMRI rate with MBH mass, and the uncertain number of MBHs in the range relevant to space-based gravitational wave detectors. The impact of these various uncertainties was extensively explored in~\cite{Babak:2017tow}, where it was shown that the number of EMRIs observed by LISA could be anywhere between $1$ and several thousand per year. The most pessimistic and the most optimistic models were deliberately chosen to be extreme, but the more reasonable models spanned a range from a few tens to almost a thousand events per year. These rates assume that a signal-to-noise ratio of $\sim 20$ is required for the detection of an EMRI, which is somewhat larger than the $\sim 8$ that is typically assumed for MBH binaries and other sources. This is driven by the expected complexity of the very long EMRI waveforms, which means that the number of independent waveform templates across the parameter space is very large~\cite{Gair_2004}. Preliminary results from the Mock LISA Data Challenges suggest that this threshold might be pessimistic~\cite{2009CQGra..26m5004B,MLDC34}, but those uncertainties are negligible compared to the much greater astrophysical uncertainties. If the rate of EMRI events is at the high end of predictions then in additional to these individually resolvable EMRI events there could be a stochastic foreground generated by the population of unresolved EMRIs, similar to the expected foreground from ultra-compact binaries in the Milky Way~\cite{PhysRevD.70.122002,Bonetti_2020}.

\begin{figure}
\begin{centering}
\includegraphics[width=10cm]{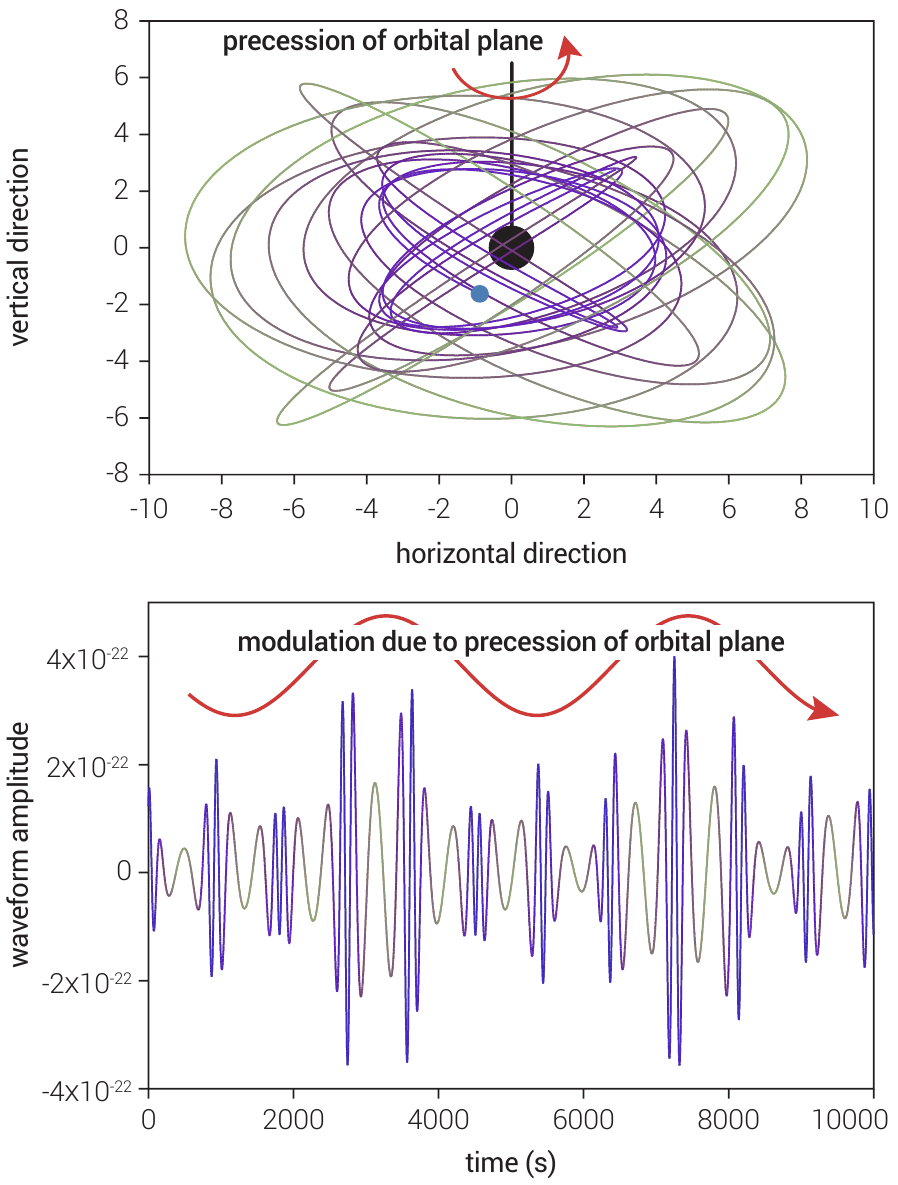}
\par\end{centering}
\caption{\label{fig:emriwaveform}Snapshot of the orbit of an EMRI viewed from the side (top panel) and the corresponding gravitational waveform (bottom panel). The colour coding identifies points of the orbit with the corresponding part of the GW. When the object is close to the black hole the emission is higher amplitude and frequency than when the object is far away. The overall waveform is modulated by precession of the orbital plane with respect to the line of sight to the observer. Figure taken from~\cite{LISA_Proposal2017}, adapted from~\cite{2008RSPTA.366.4365G}.}
\end{figure}

Due to the eccentricity and inclination of the orbits, EMRI waveforms show a very rich structure that is a superposition of the orbital frequency and precession frequencies of the periapse and orbital plane. This is illustrated in Figure~\ref{fig:emriwaveform}. This complexity, combined with the long duration and hence large number of cycles observed for a typical EMRI facilitate extremely accurate measurements of the parameters of the source. Using a Fisher matrix approach it was shown that a single EMRI observation will typically provide estimates of the masses of both components, the spin of the MBH and the eccentricity of the orbit to fractional accuracies of $\sim10^{-6}$--$10^{-5}$~\cite{Babak:2017tow}. The location of the EMRI on the sky can be determined to better than $10$ square degrees in most cases, and less than a square degree in a good fraction of cases. The luminosity distance of the EMRI will typically be measured to a precision of $\sim1\%$--$10\%$. If the MBH has near-extremal spin, i.e., the rotation rate is close to the maximum allowed value of $1$, the spin measurement improves by another order of magnitude~\cite{Burke_2020}, allowing the confident identification of near-extremal systems if they exist. These precise measurements have been verified through Bayesian posterior estimation within the context of the Mock LISA Data Challenges~\cite{2009CQGra..26m5004B,MLDC34} and have important implications for science with EMRIs, which will be discussed in section~\ref{sec:space_gw_science}.

We conclude this section we mentioning a couple of related sources which could also be observed with space-based gravitational wave detectors. In the standard EMRI formation picture, the inspiralling object begins on a highly eccentric orbit, with periapse quite close to the MBH. Until the source has inspiralled sufficiently to be radiating continually in the millihertz GW band, the periapse remains approximately fixed, while the apoapse decays. The emission in this period is characterised by periodic bursts of gravitational waves emitted each time the compact object passes the MBH. These GW bursts from systems in this early stage of inspiral in the centre of the Milky Way~\cite{Berry_2012}, or nearby galaxies~\cite{Berry_2013}, could potentially be seen by a space-based detector, but this is dependent on the properties of the MBHs in the local Universe and on the astrophysical EMRI event rate. Another related source has been termed an \textit{extremely-large mass-ratio inspiral} or ``XMRI''~\cite{Amaro_Seoane_2019}. These are the inspirals of brown dwarfs, which have mass a few hundredths of a solar mass. Brown dwarfs are more abundant than compact objects in galactic centres, and inspiral more slowly, so there could be  many  of these in the process of inspiral at any given time. These are only detectable in the local Universe and the exact number that will be observed is highly uncertain, but a space-based detector like LISA could observe $O(10)$ signals from such systems.

\subsection{Cosmological sources}
\label{sec:cosback}
Processes occurring at high energies in the early Universe can generate stochastic backgrounds of gravitational waves. A detection of this radiation would be of great importance in understanding early Universe cosmology, since it can freely propagate from earlier times than the cosmic microwave background, which is the earliest that can be probed with EM observations. The frequency of GWs generated by cosmological processes is determined by the horizon scale, and hence temperature, of the Universe at the time of production, and by the amount of expansion of the Universe between the time of production and today. A number of physical processes have been proposed that could generate GWs in the $0.1$mHz -- $100$mHz
band. These include cosmological first-order phase transitions at energy scales of $0.1$ to $100$TeV. Such phase transitions happen when the plasma in the universe changes phase via bubble nucleation. These bubbles expand, perturb the plasma, and collide, creating in this way gravitational waves at close to the horizon scale~\cite{PhysRevD.30.272,1986MNRAS.218..629H,PhysRevD.66.103505,Kosowsky_2002}. A detection of these gravitational waves with a space-based detector could provide constraints on new physics, such as self-coupling of the Higgs field, supersymmetry or conformal dynamics~\cite{Caprini:2015zlo,Caprini:2019egz}. Another scenario is the existence of extra space-time dimensions. At the TeV scale the Hubble length is about 1mm, so GWs on this scale could probe the dynamics of warped extra-dimensions, as predicted in some string theory scenarios~\cite{Hogan:2000is,Randall_2007}.

The Planck scale is not far above the TeV scales in some braneworld scenarios, in which case space-based GW detectors could probe inflationary reheating~\cite{Khlebnikov_1997,Easther_2006,Felder_2007}. GWs at millihertz frequencies could also be produced through the amplification of quantum vacuum fluctuations in some unconventional inflationary models, such as the pre-big bang and bouncing brane scenarios~\cite{BRUSTEIN199545,Buonanno_1997,Bartolo:2016ami}. A final mechanism for producing stochastic gravitational waves in the millihertz backgrounds is through the interactions of cosmic string networks. Cosmic strings are topological defects created by phase transitions, initially on microscopic scales, which are then stretched to astronomical scales by cosmological expansion~\cite{Copeland_2004,Binetruy_2012}. These strings can interact, forming cusps and loops that decay through emission of GWs. The emitted gravitational waves will form a background that is distinct from backgrounds generated via any other source, with nearly constant energy per logarithmic interval in frequency over many decades in frequency~\cite{Binetruy_2012}. Space-based detectors are the most sensitive probes for these objects~\cite{Auclair:2019wcv}. If strings are not too light, GW bursts from individual cosmic string cusps could also be detected, providing firm evidence of the cosmic string origin of the cusp.

All of the scenarios outlined above are somewhat speculative, so there is no guarantee that a space-based GW detector will see a stochastic cosmological background. Nonetheless, if it did the implications for the physics of the early Universe would be profound. Detection of a stochastic background in a gravitational wave detector is somewhat different to detection of individual sources. It will rely on cross-correlation of two data channels with independent noise. For space-based interferometers like LISA, the TDI channels are noise-independent and so can be used for this purpose. The idea is that in the cross-correlation the cosmological noise component combines constructively, while the instrumental noise does not. Detectability also depends on the shape of the stochastic background, since the stochastic signal is broadband and can be integrated over the range of sensitivity of the detector. The detectability of backgrounds of various shapes and for various specific scenarios was explored in detail in~\cite{Caprini:2015zlo, Bartolo:2016ami,  Caprini_2016, Caprini:2019egz,  Auclair:2019wcv, Caprini:2019pxz,  Flauger:2020qyi}. Broadly speaking, a space-based detector like LISA would be able to detect any background which contains more than $\sim10^{-5}$ of the closure energy density of the Universe~\cite{LISA_Proposal2017}. Broad-band backgrounds with logarithmic energy density at $1$mHz in excess of a few$\times 10^{-14}$ should be detectable. For more detailed results under specific assumptions, we refer the reader to~\cite{Caprini:2015zlo, Bartolo:2016ami,  Caprini_2016, Caprini:2019egz,  Auclair:2019wcv, Caprini:2019pxz,  Flauger:2020qyi}. 

\section{Science with Space-based Observatories}
\label{sec:space_gw_science}
In this section, we will highlight some of the science applications of observations with space-based gravitational wave detectors, which range across astrophysics, cosmology and fundamental physics.


\subsection{Astrophysics}
\subsubsection{Compact binaries in the Milky Way}
The formation of ultra-compact binaries depends on various astrophysical processes, such as stellar formation and binary stellar evolution, including the poorly-understood common envelope phase~\cite{2013A&ARv..21...59I}. Characterising the ultra-compact binary population will thus shed light on open astrophysical questions about the Milky Way stellar population. Currently only a few tens of ultra-compact binaries are known, and only a couple of these have periods shorter than ten minutes~\cite{Marsh_2011}. A space-based gravitational wave detector such as LISA will discover several thousand additional systems, expanding the known population by two orders of magnitude, and making a complete survey of such systems at the shortest periods. These observations will provide key insights into the total number of such systems and hence their merger rates. The observed distribution of individually resolved systems and the distribution of the unresolved population inferred from the modulation of the stochastic foreground will resolve the structure of the Milky Way, including the thin and thick disc, the halo and globular clusters~\cite{PhysRevD.86.124032,2019MNRAS.483.5518K}. Gravitational waves provide a unique probe for this as they do not suffer from dust obscuration and can thus ``peer through''; the galactic centre. Finally, joint observations of ultra-compact binaries with gravitational waves and electromagnetic observations at high signal-to-noise ratio will provide key insights into the complex physics of interacting binaries, including tidal interactions and mass transfer~\cite{2014ApJ...791...76S}.

\subsubsection{Stellar-origin black hole binaries}
\label{sci:astro:sobh}
By identifying the time of merger and approximate sky location of SOBHB systems well in advance of the event, space-based observatories can trigger follow-up observations with electromagnetic telescopes and ground-based GW detectors~\cite{PhysRevLett.116.231102}. GW detectors are not pointable, so the pre-localisation on the sky is not important, but pre-determination of time of coalescence would allow the ground-based detectors to avoid scheduling maintenance at the time of the event. Triggers to EM facilities would allow deep searches for associated EM emission both pre- and post-merger. Detection of any EM emission would reveal properties of the material in the vicinity of the SOBHB, and hence shed light on the astrophysical environment of such systems.

Detection of residual eccentricity in the binary would provide crucial clues as to the origin of such systems. SOBHBs could form in the field as the end-point of isolated binary evolution, but could also form in the dense environments of globular clusters through dynamical capture, or in the vicinity of a MBH 
through Kozai-Lidov  hardening of binaries created via mass segregation~\cite{Antonini_2012}. The residual eccentricity would be larger in the latter two cases, and could be detectable in an observation by a space-based detector, but would be too small by the time the source reached the band of ground-based detectors to be measurable. It was shown in~\cite{PhysRevD.94.064020,10.1093/mnras/stw2993} that just a handful of SOBHB observations with a space-based detector could identify binaries formed in the vicinity of a MBH, so this should certainly be possible. Several tens of systems would be needed to distinguish the isolated binary and dynamical capture scenarios, so this will only be possible if the number of SOBHB systems observed is at the high end of current ranges. The formation channels of SOBHBs is currently hotly debated, and the information obtained from space-based observatories could be crucial to resolving it~\cite{2016ApJ...830L..18B}.

\subsubsection{Massive black hole binaries}
As described in section~\ref{sec:mbh}, massive black holes are observed to exist very early in cosmic history, and it is generally assumed that these black holes start as seed black holes and then grow through mergers and accretion. However, there are a number of plausible models for the formation of those seeds that are consistent with current data. 
A space-based GW detector operating in the millihertz band will measure the masses and spins of MBHs in merging binaries out to very high redshift, directly probing the epoch of formation and early evolution of black holes. This epoch cannot be easily probed with EM observations, so GWs may provide unique insight into the nature of black hole seeds, and their early growth through accretion. It was shown in~\cite{Sesana_2011} that LISA will be able to distinguish between a wide variety of seed black hole models, and identify mixed populations, determining the mixture fraction up to a precision of $\sim \pm 0.2$~\cite{Sesana_2011,amaroseoane2012elisa}.

In addition to the properties of the individual black holes, the number and redshift at which the MBH binaries are observed to merge encodes important astrophysical information. MBH mergers follow mergers between their host galaxies, so the merger distribution tracks mergers between galaxies and the early growth of structure~\cite{2014ARA&A..52..589H,Kormendy:2013dxa}. LISA observations of MBH mergers out to high redshift will provide indirect constraints on the rate of galaxy mergers in the early Universe and the relative fraction of ``major'' or ``minor'' mergers, i.e., the fraction of mergers in which the MBH have similar masses or not. Measurements of the spin of these black holes will provide clues to the nature of accretion in galaxy halos at early and later times~\cite{Berti_2008}. The distribution of events in redshift will encode clues to the delay time between the galaxy merger and the MBH merger, and hence the efficiency with which the MBH binary is brought to the centre of the merged galaxy. Finally, if these binaries are observed to have significant residual eccentricity, it could suggest the presence of a third MBH in the vicinity of the binary, which can excite eccentricity through the Kozai-Lidov resonance ~\cite{2019MNRAS.486.4044B}. The fraction of LISA mergers observed to occur in triple systems is another important clue to build up a picture of the early evolution of cosmic structure.

The astrophysical impact of GW observations of MBH binaries will be significantly enhanced if multi-messenger observations are made~\cite{2020NatAs...4...26M}. Observing an EM counterpart to the GW event will provide complementary information about the material in the vicinity of the black hole. The environment of a black hole plays a key role in its evolution, driving spin and mass growth through accretion, and can play a role in driving the coalescence of MBH binaries~\cite{2019ApJ...871...84M}. While it is not certain that EM counterparts will be detected for any MBH binary, the discovery potential of multi-messenger observations is huge, as already demonstrated by joint observations of the binary neutron star merger GW170817 with ground-based detectors~\cite{Abbott_2017_GW170817MM}.

\subsubsection{Extreme-mass-ratio-inspirals}
EMRI observations will probe MBHs of similar mass to MBH mergers, but the primary black holes in EMRI events will be ``quiescent'' MBHs at lower redshifts, rather than MBHs undergoing highly dynamical interactions during mergers. Comparing the properties of the quiescent MBH population to the dynamical MBH population will provide further clues to the evolution of the MBH population. The observation of EMRI events would provide constraints on the MBH population in the interval where EM observations are poor or missing~\cite{PhysRevD.81.104014}. EMRI observations could also constrain the occupation fraction of MBHs in low mass galaxies without relying on accretion signatures~\cite{Volonteri_2009}. EMRI measurements of black hole spins will constrain the spin distribution of low-mass black holes up to moderate redshift~\cite{Babak:2017tow}, providing a more complete census than can be obtained through, for example, accretion disc measurements, which are restricted to actively accreting MBHs, which are a minority.

EMRI observations will also provide precise measurements of the masses and orbital properties of compact objects in galactic nuclei~\cite{Babak:2017tow}. These observations will reveal the mass spectrum of stellar-origin black holes in galactic nuclei, which can be compared to the corresponding mass spectrum observed in SOBHBs, which is now being constrained by ground-based GW detectors~\cite{Abbott:2020gyp}. Differences or similarities between the observed populations will shed further light on stellar evolution in different astrophysical environments and on the origin of the SOBHB population. The number of EMRI events observed as a function of black hole mass will encode information about mass segregation in galactic nuclei~\cite{2011CQGra..28i4017A}, while the observed eccentricity and inclination distributions provide direct constraints on the EMRI formation channel~\cite{2007CQGra..24R.113A}. Taking together, EMRI observations will build up a comprehensive picture of the complex physical processes that govern the dynamics of stars in galactic nuclei~\cite{Amaro_Seoane_2018}.

EMRI-like systems in which the smaller object is an intermediate mass black hole (IMBH) of $\sim 10^2$ -- $10^4M_\odot$ could also be observed by space-based detectors~\cite{2018PhRvD..98f3018A}. These are often called \textit{intermediate-mass-ratio inspirals} or IMRIs. Binaries of two IMBHs could also potentially be observed. The existence of IMBHs is not yet conclusively established observationally~\cite{Greene_2020}, but space-based GW detectors would provide mass measurements that are precise enough to robustly identify black holes that lie in the IMBH range. GW observations of such systems from space and from the ground thus offer a unique way to understand the astrophysics of these objects if they exist~\cite{Gair_2010}.

\subsection{Cosmology}
\subsubsection{Probes of the early Universe}
The detection of a stochastic background of gravitational waves 
and the measurement of its amplitude and slope would have profound implications for our understanding of the early Universe. As described in section~\ref{sec:cosback} there are a number of non-standard scenarios that could produce such a background, including first-order phase transitions~\cite{PhysRevD.30.272,1986MNRAS.218..629H,PhysRevD.66.103505,Kosowsky_2002,Caprini:2015zlo,Caprini:2019egz}, warped extra dimensions~\cite{Hogan:2000is,Randall_2007}, inflationary reheating in braneworld scenarios~\cite{Khlebnikov_1997,Easther_2006,Felder_2007}, non-standard inflation including pre-big bang and bouncing brane scenarios~\cite{BRUSTEIN199545,Buonanno_1997,Bartolo:2016ami} or cosmic string networks~\cite{Copeland_2004,Binetruy_2012}. The spectra generated under these various scenarios are distinct and can be constrained by space-based GW observations~\cite{Caprini:2015zlo, Bartolo:2016ami,  Caprini_2016, Caprini:2019egz,  Auclair:2019wcv, Caprini:2019pxz,  Flauger:2020qyi}, so if a background were to be detected it would provide insight into this new physics. In addition, because the GW background is generated before Big Bang Nucleosynthesis (BBN), it can probe earlier epochs than any that have been constrained so far, even indirectly. In cosmological models that differ from the standard model prior to BBN, the GW background spectrum can change dramatically, providing a smoking gun for new physics during that epoch~\cite{Auclair:2019wcv}.

\subsubsection{Cosmography with standard sirens}
An important application of gravitational wave observations of SOBHBs, MBH binaries and EMRIs is for cosmography, i.e., to probe the expansion of the Universe over cosmological history. To probe the expansion of the Universe we need to measure the rate of expansion of the Universe, characterised by redshift, as a function of distance, characterised by luminosity distance. The luminosity distance-redshift relation depends on the cosmological model and the matter and energy content of the Universe, and hence can be used to constrain these properties. Various EM sources, including type IA supernovae~\cite{Riess:2016jrr}, have been used for this purpose. These sources are referred to as \textit{standard sirens}, since the basis of the approach is to assume that the intrinsic luminosity of the source is known and hence the observed luminosity provides a measure of distance. Redshift can be measured directly from the shift in frequency of spectral lines. The notion of using GW sources as \textit{standard sirens} for the same purpose was first suggested in~\cite{Schutz:1986gp}. As described earlier, the strain of a GW source scales with the ratio of its (redshifted) mass to its luminosity distance. The redshifted mass also impacts the GW phasing and so can typically be measured very accurately from the GW data, so the observed amplitude gives a direct measurement of the luminosity distance. This is appealing since, in contrast to EM probes of cosmology, these measurements do not need to be calibrated to the local distance ladder. However, GW observations do not provide direct measurements of redshift.

If the GW event has an EM counterpart, the redshift can be obtained from the EM observation. This was exploited for GW170817, the first binary neutron star merger observed by ground-based interferometers~\cite{TheLIGOScientific:2017qsa}, for which a kilonova counterpart was observed~\cite{Abbott_2017_GW170817MM}, enabling the first gravitational wave constraint on the Hubble constant~\cite{Abbott:2017xzu}. For space-based detectors, the only source for which counterparts are thought to be possible are MBH mergers~\cite{2020NatAs...4...26M}. MBH mergers at low redshift, $z \sim 1$--$2$, can be localised to a few square degrees, permitting searches for EM counterparts~\cite{Klein_2016}. These sources will have luminosity distance measurements of $\sim 1\%$, so any event with an associated counterpart will provide a percent-level constraint on cosmological parameters. An EMRI in which the smaller object is a white dwarf and the larger object is a low mass, rapidly spinning black hole, could generate an observable counterpart when the white dwarf is tidally disrupted toward the end of the inspiral~\cite{Sesana_2008,Maguire_2020}. However, the event rate of such systems is likely to be very low~\cite{2007CQGra..24R.113A,Babak:2017tow}. 

In the absence of a counterpart, cosmological constraints can be obtained statistically by comparing the locations of observed GW events with catalogues of galaxy redshifts. This has also been done using observations with ground-based detectors~\cite{Fishbach:2018gjp,Soares-Santos:2019irc,Abbott:2019yzh}. It has been shown that this statistical approach could yield constraints on the Hubble constant at the level of $1\%$, if $20$ EMRIs are observed at redshift $z < 0.5$ with a LISA-like space-based GW detector~\cite{MacLeod:2007jd}. That analysis assumed somewhat optimistic EMRI localisation volumes, but this will be partially compensated by the larger number of EMRI events predicted in current models~\cite{Babak:2017tow}. Statistical cosmological constraints using observations of SOBHBs will achieve comparable precision on $H_0$, if the number of observed events is at the higher end of predictions~\cite{Kyutoku:2016zxn,DelPozzo:2017kme}, while observations of MBH mergers will achieve slightly worse precision on the Hubble parameter, but will permit estimates of the matter content of the Universe and the equation of state of dark energy~\cite{Petiteau:2011we}.

There is a third approach to cosmology with gravitational wave sources, which is to use the GW measurement of the redshifted mass to estimate the redshift of the source. This can be done if assumptions are made about the distribution of masses of the observed signals. This was initially proposed in the context of observations of binary neutron star mergers with ground-based detectors, where it is justified by the narrow observed mass distribution of neutron stars in compact binaries~\cite{Taylor_2012,Taylor:2012db}. The mass distributions of EMRIs and MBHs are not expected to be sufficiently compact to permit interesting constraints in this way. However, the same procedure can be applied when the distribution has a sharp feature, such as the presence of the mass gap in SOBHBs. Exploiting this feature with observations of SOBHBs in future ground-based detectors could yield interesting cosmological constraints that are independent of all EM information~\cite{Farr:2019twy}, and so it is possible that something could also be done with SOBHB observations by space-based detectors. However, the lower expected number of events and evidence that the mass distribution is more complicated than a truncated power law~\cite{Abbott_2020_GW190521,Abbott:2020gyp} suggest that, for space-based detectors,  this approach will not be competitive with the counterpart or statistical approaches. 

To finish this section, we note that the cosmological constraints described here, although competitive with current EM constraints, will probably be surpassed by EM data obtained between now and the launch of LISA. These measurements are nonetheless interesting as they provide a completely independent verification of the EM results, and are subject to a completely different set of systematic errors. In addition, space-based GW observations are one of the few approaches that can obtain constraints over a wide range of redshifts, probing redshift values that cannot be measured by electromagnetic probes. This could be crucial for resolving current tensions between low redshift~\cite{Riess:2016jrr,Riess_2019} and high redshift~\cite{Planck2020} cosmological measurements.

\subsection{Fundamental Physics}
Gravitational wave sources observed by both ground-based and space-based detectors can provide powerful tests of fundamental physics, i.e., whether the evolution of the binary and the observed gravitational wave emission are consistent with the predictions of general relativity. These tests are possible because GW observations provide very precise measurements of the waveform phase, and hence can identify very small changes to the phasing arising from new physics. Observations with space-based detectors are particularly powerful, because the sources typically have very high signal-to-noise ratio and are also long-lived. There are three distinct types of tests of gravitational physics that have been proposed, which we now briefly summarise.

\subsubsection{Elucidating dark-matter}
Only $\sim15\%$ of the Universe is composed of ``normal'' baryonic matter~\cite{Tanabashi:2018oca}. For decades astronomers and particle physicists have been struggling to understand the nature of the other $85\%$ of ``dark matter'' (DM). Observations with GW detectors will be able to shed light onto the nature of dark matter in a number of ways. Measurements of the distribution of masses and spins of MBHs 
can reveal the existence of DM due to the effect of DM interactions on these distributions (see~\cite{Brito:2015oca} for a review and further references). If EMRIs or MBH mergers are taking place in an environment containing significant amounts of DM, this will impact the observed phasing of the emitted GWs in a measurable way~\cite{Barausse_2014}. The emitted waveforms can also be used to identify if the central MBH is in fact a self-gravitating DM structure ~\cite{Cardoso:2019rvt,Liebling:2012fv}. Clouds of ultra-light DM particles around spinning black holes can also generate GWs, either continuously or as bursts, that could be directly detected by GW detectors~\cite{Brito_2017a,Brito_2017b}. Finally, DM interacting directly with the space-based interferometer could lead to measurable signatures~\cite{Nagano:2019rbw,Grote:2019uvn}.

\subsubsection{Testing the foundations of the gravitational interaction}
Departures in the physics of gravity away from the predictions of general relativity can lead to differences in how gravitational waves are generated and how they propagate through the Universe. These differences change the phasing of the GWs, which can be detected in observations with gravitational wave detectors. A large number of alternatives to general relativity have been proposed, each varying one or more of the physical assumptions that underlie GR. Examples of alternatives that lead to measurable deviations in gravitational waveforms include massive gravity theories~\cite{Will:1997bb,Mirshekari:2011yq}, the existence of large or compact extra spacetime dimensions~\cite{Yagi:2011yu,Cardoso:2019vof,Du:2020rlx}, variation of Newton's coupling constant over cosmic time~\cite{Yunes:2009bv,Tahura:2018zuq}, violations of parity or chirality~\cite{Alexander:2007kv,Yagi:2012vf}, violations of Lorentz invariance~\cite{Hansen:2014ewa,Zhang:2019iim}, violations of the Equivalence Principle~\cite{Berti:2004bd,Yagi:2009zm,Yagi:2011xp} or the existence of additional scalar or vector fields that generate alternative polarisation states for gravitational waves~\cite{Tinto:2010hz}. We refer the reader to~\cite{Gair_2013,Berti_2015,Barack:2018yly} for reviews and further references.

There are two basic approaches to constraining these physical effects with space-based observations of GWs. The first is to construct model waveforms in these alternative physical scenarios, and use them within the framework of Bayesian inference calculations to place constraints on the parameters that occur in these specific theories. The alternative approach is ``model-free'' in the sense that no reference is made to a specific alternative theory. Instead generic modifications are made to the waveform models, which are then constrained with observations. Two different formalisms have been proposed in the literature. The first is to directly measure, or constrain modifications to, the post-Newtonian phase coefficients, i.e., the numerical factors multiplying different powers of the GW frequency in an expansion of the GW phase~\cite{Arun_2006}. Space-based detectors should be able to identify departures from the GR values in observed MBH merger waveforms at the level of one tenth of a percent in the low order coefficients. Alternatively, generic additional terms can be included in the amplitude or phase of the gravitational waveform, allowing deviations in not only the size of the terms but also their frequency dependence. This is termed the \textit{parameterised post-Einsteinian (ppE) formalism}~\cite{Yunes_2009_ppE}. MBH merger observations with space-based detectors will be able to place constraints across a wide range of the ppE parameter space, with particularly strong constraints at higher orders in frequency which cannot be well constrained by EM observations of Newtonian binaries in the Newtonian regime~\cite{PhysRevD.84.062003,perkins2020probing}.

Similar tests are also possible with SOBHBs, by exploiting the possibility of observing the same system by both space-based and ground-based detectors. The two observations provide snapshots of the waveform at two different epochs. Small differences in the rate of inspiral evolution of the binary can lead to measurable changes in the time separation between the two observations. Multi-band SOBHB observations with LISA and ground-based detectors can provide constraints on various alternatives to general relativity that are much better than are currently available~\cite{perkins2020probing}. This includes a six orders of magnitude improvement in constraints on the emission of dipole radiation~\cite{Barausse_2016}; several orders of magnitude improvement in constraints on Brans-Dicke theory, Einstein-dilaton-Gauss-Bonnet gravity and dynamical Chern-Simons gravity~\cite{Gnocchi_2019}; and an order of magnitude improvement in merger-ringdown consistency checks (see next section)~\cite{Carson_2020}. 

To conclude this section we will mention the memory effect. A prediction of general relativity is that after a merger the spacetime will retain a permanent shift in its zero point and hence have a ``memory'' of the fact  that a merger took place. There is both linear and non-linear memory and while the final spacetime off-set is at zero frequency and hence unobservable, it is in principle possible to directly observe the build up of the memory through the GW observation. Space-based detectors will measure the nonlinear memory build-up in MBH coalescences with sufficient significance to also test this aspect of gravitational theory~\cite{Favata_2009,2019arXiv190611936I}.

\subsubsection{Testing the nature of black holes}
\label{sec:fp:mapping}
Black holes in general relativity are completely characterised by two parameters ---  a mass and a spin. All higher ``moments'' in an expansion of the gravitational field are determined by these two parameters and the resulting spacetime structure is described by the Kerr metric. This is often referred to as the ``no-hair property''. If general relativity does not describe the structure of black holes, or one of the auxiliary assumptions that lead to the uniqueness of the Kerr metric, such as the energy conditions or the formation of a horizon, are violated then the spacetime structure could have ``hair'' and deviate from the Kerr metric~\cite{2008RSPTA.366.4365G,Gair_2013,Cardoso:2019rvt}. GW observations can be used to construct a map of the spacetime in the vicinity of a black hole and hence test the no-hair property. For space-based detectors observations of EMRIs and of the ringdown signal following the merger of two MBHs provide the cleanest tests. For an EMRI observation the information comes from tracking the waveform phase over many hundreds of thousands of waveform cycles. Small changes in the multipole structure lead to changes in the rate of inspiral of the binary that accumulate over the observation. This has been studied extensively, starting from a direct extraction of successive multipole moments described in~\cite{Ryan:1997hg}. More recent work has focused on the ability of space-based detectors to quantify the size of ``bumps'' away from the Kerr metric. Relevant early works include~\cite{Collins_2004,Glampedakis_2006,Barack:2006pq,Gair_2008,PhysRevD.86.044010} but we refer the interested reader to~\cite{Gair_2013,2016CQGra..33e4001Y,Cardoso:2019rvt} for a comprehensive summary of the literature.

\begin{figure}
\begin{centering}
\includegraphics[width=8cm]{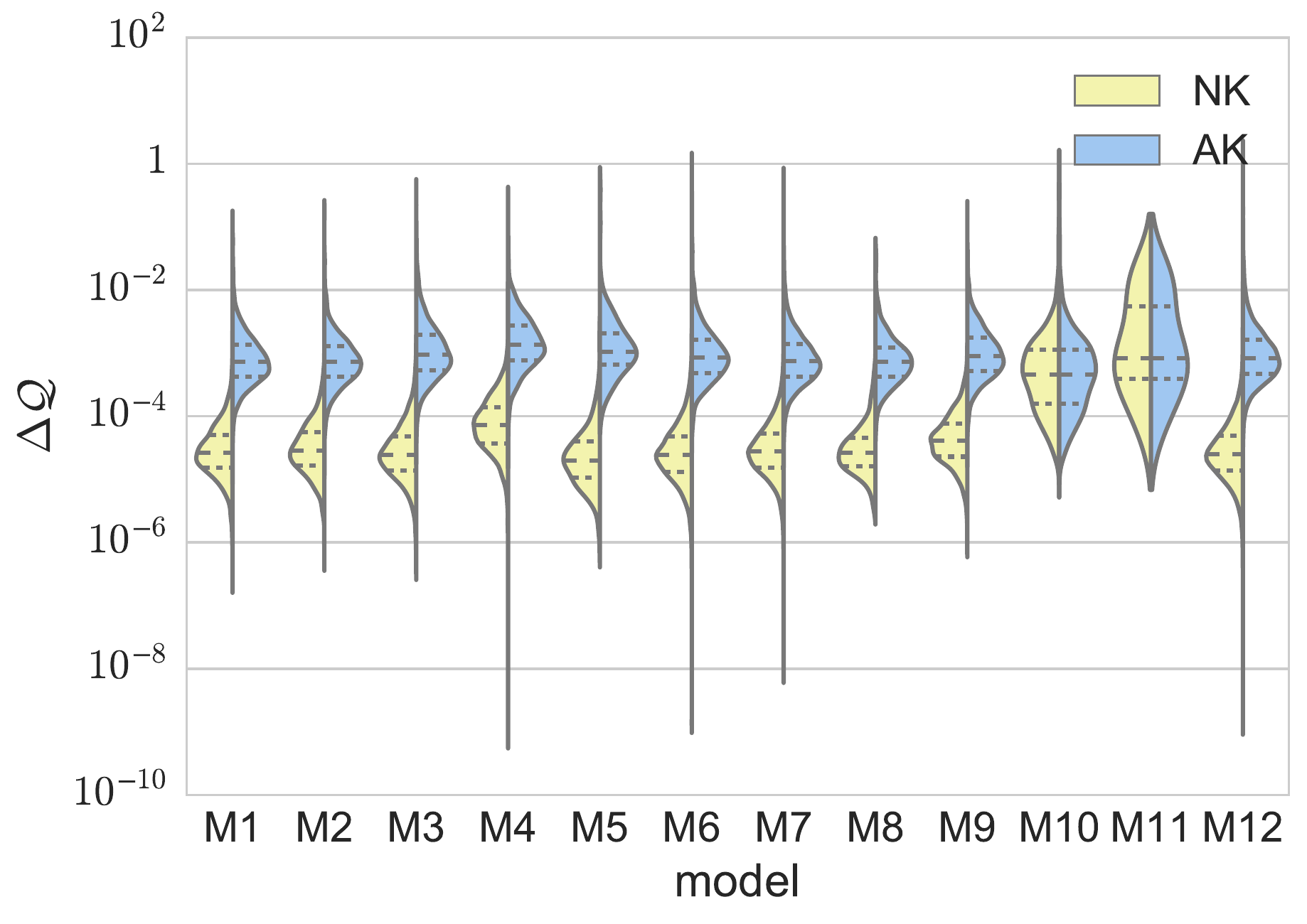}
\par\end{centering}
\caption{\label{fig:nohair}Distributions of accuracy with which EMRI observations can constrain deviations in the quadrupole moment of the central MBH 
from the value predicted by the Kerr metric. Results are shown for various models of the EMRI population, as described in~\cite{Babak:2017tow}. Figure also reproduced from~\cite{Babak:2017tow}.}
\end{figure}

Figure~\ref{fig:nohair} shows results, first reported in~\cite{Babak:2017tow}, on the precision with which EMRI observations can detect departures in the quadrupole moment, $\Delta Q$, of the central MBH from the value predicted by its mass and spin. These results are shown for twelve different models of the astrophysical population of EMRIs, labelled M1 -- M12, and for two different models of the EMRI waveform, labelled ``AK'' and ``NK''. The two EMRI waveforms both use the analytic kludge model described in~\cite{Barack_2004}, but differ in where the inspiral is terminated as the small object plunges into the MBH. 
Full details can be found in~\cite{Babak:2017tow}. Departures at the level of $10^{-4}$ can be detected if they are present. Other information about the nature of the object that can be extracted from EMRI observations includes the presence of a horizon~\cite{Kesden_2005}, the nature of the tidal-coupling interaction~\cite{PhysRevD.77.064022} and the influence of perturbing matter~\cite{Barausse_2014} or nearby stars~\cite{PhysRevD.83.044030}. These effects should be distinguishable from those arising from differences in the nature of the central object.

After a MBH merger, the remnant black hole that forms settles down from a highly perturbed state to a quiescent Kerr spacetime through a process called ringdown. The ringdown radiation is a superposition of damped sinusoids, the frequency and decay time of which are uniquely determined by the mass and spin of the remnant black hole. Observation of two or more ringdown modes allows a consistency check between the frequencies and damping times that directly tests the no-hair property of the remnant~\cite{Berti_2006,Berti_2016}. Recently a framework for model-independent ringdown constraints, similar to the ppE formalism, has been developed~\cite{Cardoso_2019,McManus_2019,Maselli_2020}. Ringdown constraints probe a different regime to inspiral constraints and are therefore somewhat complementary. Certain types of modification to the spacetime structure are better probed by one approach or the other~\cite{Berti_2018a,Berti:2018vdi}.

A final approach to testing the nature of the black holes in GW observations is to look for consistency in the properties of the black holes inferred from the inspiral and those inferred from the ringdown~\cite{Ghosh:2017gfp,Berti:2018cxi}. Any differences between the observed properties of the merger and ringdown and those predicted from the inspiral using general relativity would reveal new physics during the highly dynamical merger phase. This approach has been applied to observations with ground based detectors~\cite{TheLIGOScientific:2016src}, so far revealing no evidence for deviations from the predictions of general relativity. A closely related idea is to look for additional signals, or ``echoes'', in the data after an observed event. If they are seen, these echoes could indicate the existence of new physics near the horizon of black holes~\cite{Cardoso:2016oxy}. Claims have been made for evidence of these echoes in LIGO observations~\cite{Abedi:2016hgu}, but these are more likely to have been due to instrumental noise~\cite{Westerweck:2017hus}. Future space-based detectors will shed further light onto this ongoing debate.

\section{Prospects for Space-based Observatories}
\label{sec:space_gw_future}


LISA will certainly not be the last space-based gravitational wave
observatory. Missions beyond LISA have already been proposed
in white papers and peer-reviewed scientific journals since at least the first years of this century \cite{ALIA_2005,ALIA_2005_cqg, BBO}. Several new concepts were submitted to the Decadal 2020 review in the US \cite{Decadal_2020} and ESA's Voyage 2050 long term plan \cite{Voyage_2050,Sedda_2020,2019arXiv190811410B,baibhav2019probing,sesana2019unveiling}. They often target the frequency range between LISA and LIGO and
are optimized between 1\,mHz and 1\,Hz~\cite{Sedda_2020,2019arXiv190811410B}. 
This frequency range is very interesting for mergers between intermediate
mass black holes beyond LISA, typical LIGO mergers at higher redshifts
and for transient signals passing from the LISA band into this band and then finally merging in the ground-based band. This frequency range might also be the most promising range for detecting the gravitational wave background radiation formed just after the big bang. Most of these proposals are loosely based on LISA technologies and will likely be limited by the same noise sources than LISA; one notable exception is atom interferometer based observatories~\cite{El_Neaj_2020} which are outside of the scope of this chapter but are discussed in the chapter of this volume titled \textit{Atom Interferometer}. Here we want to look at ways to start the design of such missions, how basic mission parameters are used to derive target sensitivities,
and glance over technological improvements that will be required to
enable these missions. 

\begin{figure}
\begin{centering}
\includegraphics[width=8cm]{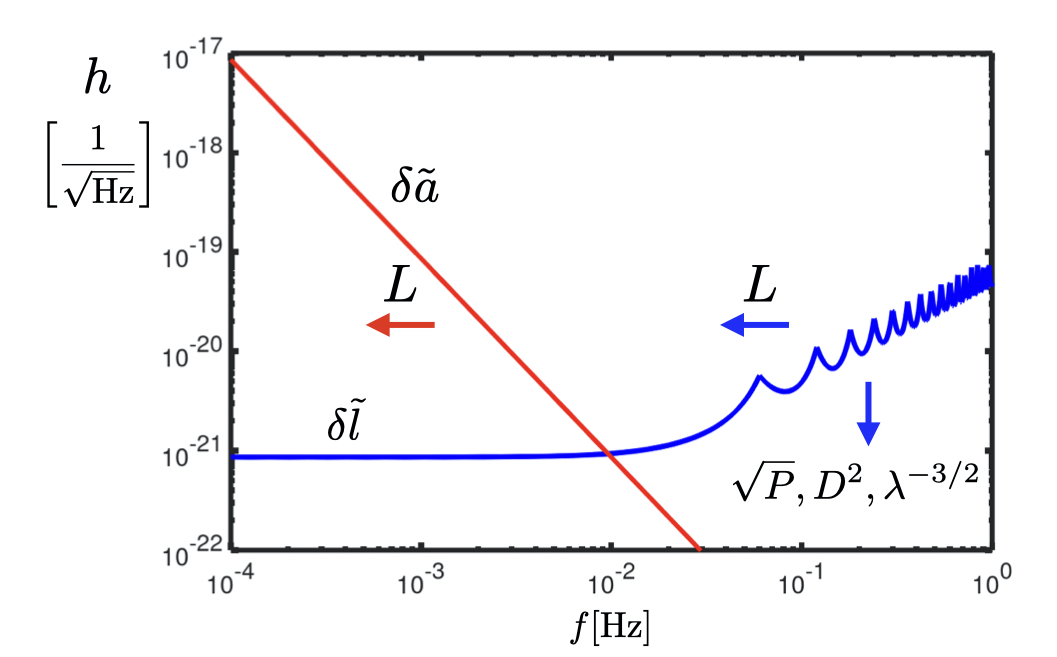}
\par\end{centering}
\caption{\label{fig:Generic-sensitivity}The generic sensitivity curve shows
the two standard constituents, acceleration noise $\delta\tilde{a}$
and interferometer noise $\delta\tilde{l}$, and how they limit the
strain sensitivity. Improvements in acceleration noise pushes the
red curve down and improve the sensitivity at low frequencies, improvements
in interferometer noise pushes the blue curve down and improve the
sensitivity at high frequencies. Increasing the arm lengths pushes
both curves to the left, reducing it to the right. Increasing the
laser power or the diameter of the telescopes and decreasing the wavelength
will reduce shot noise and $\delta\tilde{l}$ as long as the sensing system allows shot noise limited detection.}
\end{figure}

Figure \ref{fig:Generic-sensitivity} shows how acceleration noise
$\delta\tilde{a}$ and interferometer or sensing noise $\delta\tilde{l}$
define the sensitivity expressed as a linear spectral density of the
strain:
\begin{equation}
\tilde{h}(f)\sim\left<\frac{1}{L}\sqrt{\left(\frac{\delta\tilde{a}}{4\pi^{2}f^{2}}\right)^{2}+\left(\delta\tilde{l}\right)^{2}}\frac{\pi Lf/c}{\sin\left(\pi Lf/c\right)}\right>_{\rm sky\,\,locations}
\end{equation}
For simplicity and following most mission proposals, the acceleration
noise and the interferometer noise are assumed to be frequency independent.
As discussed before, frequency independent or white acceleration noise causes a displacement
of the test masses which scales with $1/f^{2}$ and limits the performance
at low frequencies. At higher frequencies, the periods of the gravitational
waves become comparable to the light travel time in the arms and the
stretching and squeezing of space time compensate each other. This
leads to a sinc function in the response of the instrument or an inverse
sinc function in the strain sensitivity for otherwise optimally aligned
gravitational waves when the propagation direction is normal to the
constellation. The zeros in the response at $f_{GW}=nc/L$ wash out
when the sensitivity is averaged over all possible sky locations which
turns the sharp peaks of the sinc into these wiggles. The sweet spot
is in the frequency range where acceleration noise and sensing noise
become comparable and before the sinc function starts to matter; between
approximately 3~mHz and 30~mHz in this generic LISA-like sensitivity curve.
Pushing this range to lower (higher) frequencies requires to lengthen
(shorten) the arms and, as long as all other basic parameters --
laser power, telescope diameter, and laser wavelength -- stay the
same and sensing noise is dominated by shot noise, both curves move
to the left (right) without changing their relative position. 

For shorter arms the received laser power increases and the shot noise
limit will decrease. As shown in equations \ref{eq:shotnoise} and \ref{eq:Prec},
a shorter wavelength improves the displacement sensitivity for the
same phase sensitivity. The light is also better collimated and the
amount of received light increases. However, as every photon is also
more energetic, the number of received photons only increases linearly
with the inverse wavelength and not quadratically. Therefore the sensitivity
scales with $\lambda^{-3/2}$ if the laser power stays the same. Increases
in the laser power without changing the wavelength improves the sensitivity
with $\sqrt{P}$ while the diameter $D$ of the telescope enters quadratically.
Future mission proposals which plan to take advantage of the lower shot noise have to 
assume that the technology progresses enough that the interferometer
measurement system continues to be limited by it and not by technical
noise. While this seems
to be overly optimistic, many technical noise sources are driven by
temperature changes which will be significantly smaller at higher
frequencies. The shorter arms will also reduce the dynamics within
the constellation which reduces the Doppler shifts and potentially
reduces the beat frequencies and the timing requirements within the
phasemeter. 

It is expected that significant improvements in acceleration noise
beyond what is shown in equation \ref{eq:accel_noise} are more likely
at higher frequencies than at lower frequencies. The reason is again
temperature which rises faster than $f^{-2}$ towards lower frequencies
and residual spacecraft motion which couples gravitationally to the
test mass and is also expected to be smaller at higher frequencies.
Also voltage noise in actuators and capacitive sensors improves at
higher frequencies while frequency independent noise sources typically
scale with somewhat controllable environmental parameters such as
pressure and absolute temperature. Many forces scale with the surface
and not the volume of the test mass and can be reduced by increasing
the mass of the test mass itself. Interferometric readouts can be significantly more sensitive than capacitive readouts. Employing those for all degrees of freedom 
would improve our ability to measure, calibrate and subtract spacecraft motion.

The ambitious goals spelled out in these proposals require a
broad range of technical improvements across several decades in frequency
space. They will not be easy to reach but that has always been the
case and compared to 30 years ago, the community has a much better
understanding of the challenges ahead of them.

\begin{figure}
\begin{centering}
\includegraphics[width=8cm]{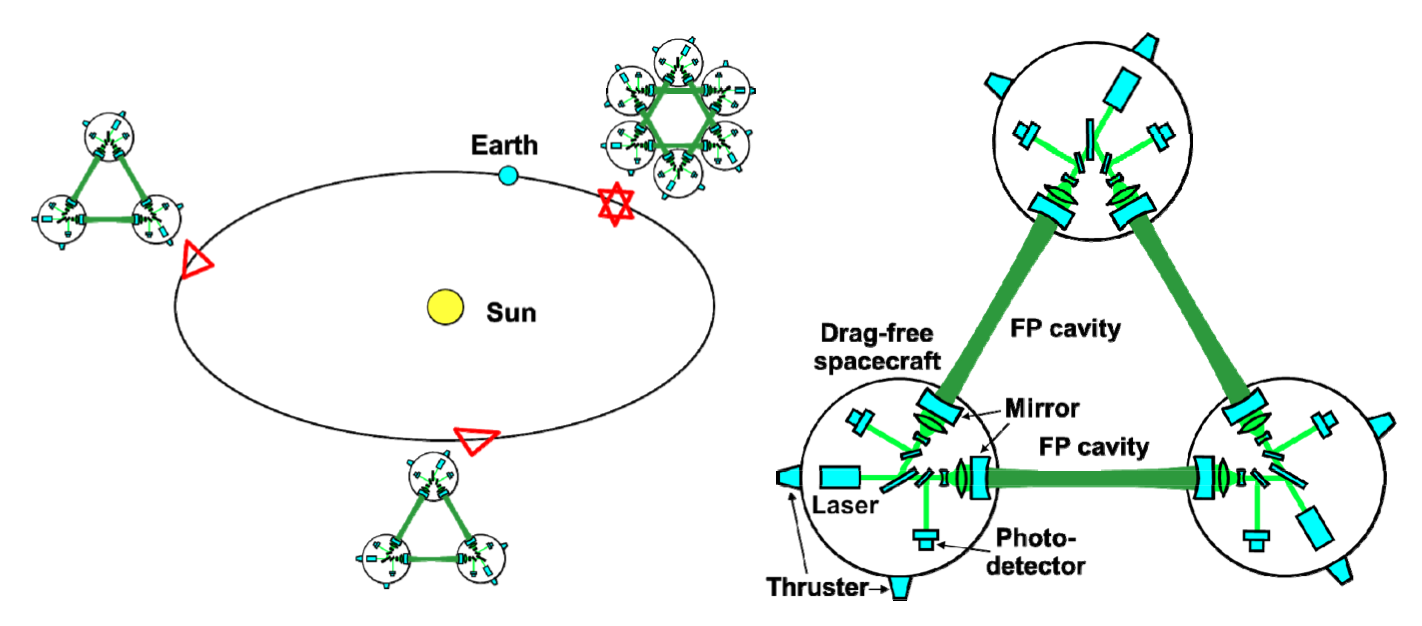}
\par\end{centering}
\caption{\label{fig:DECIGO-is-one}DECIGO is one of the most ambitious proposed
future missions. It uses twelve spacecrafts which form four equilateral
triangular clusters. Two clusters are collocated in their heliocentric
orbits while the two others are distributed around the sun as shown
on the left. The right graph shows a conceptual design of a single
constellation in which each arm is defined by two mirrors forming
an optical cavity~\cite{DECIGO}.}

\end{figure}

In the context of future mission proposals, DECIGO plays a somewhat
special role \cite{DECIGO}. It targets the 0.1 to 10~Hz frequency
range with a total of twelve spacecrafts to form four equilateral
triangular clusters. All clusters are placed in a heliocentric orbit,
the center of two of them are collocated but rotated by $60^{\circ}\text{deg}$
with respect to each other. The others are placed $120^{\circ}\text{deg}$
offset in heliocentric orbits as shown in figure \ref{fig:DECIGO-is-one}.
The Japanese project plans to form $10^{6}\,\text{m}$ long optical
cavities between free falling 1\,m diameter mirrors. The higher measurement
frequency allows the application of forces to the mirrors at lower frequencies that can be used for station keeping to keep the cavities on resonance with their interrogating
laser beams. For this reason, DECIGO is often seen as a mission which
brings ground-based technologies to space which might be needed for
orders of magnitude improvements in displacement sensitivity.

A last mission which we want to mention here is known as the Folkner
mission \cite{Folkner} which was proposed to NASA during the SGO
studies \cite{SGO_Report} six years prior to the LPF launch. In this
mission, the three spacecrafts are placed in heliocentric orbits separated
by $120^{\circ}\text{deg}$, similar to the locations of the DECIGO
clusters. The arm length of $250\,\text{Gm}$ was expected to compensate
for increased acceleration noise should LPF fail. Since LPF was successful
and a similar GRS could be used for the Folkner mission, the entire
sensitivity curve would (ideally) be shifted to the left and probe frequencies 100 times lower than current LISA. A similar proposal, $\mu$Ares, was submitted to the Voyage 2050 call~\cite{sesana2019unveiling}. These missions would bridge the gap between pulsar timing and LISA for super-massive black hole mergers. Galactic ultra-compact binaries are also expected to create a gravitational wave background in this frequency range which such missions could study, but this background would somewhat limit the distance to which massive black hole binaries could be resolved.

\section{Cross-References to other chapters in the Handbook of Gravitational Wave Astronomy}
This chapter describes space-based interferometers following mostly the LISA design. While the measurement principle, laser interferometry between free-falling macroscopic test masses, is similar to the principle used for ground-based interferometers which are described in the chapter \textit{Terrestrial Laser Interferometers}, the long arms together with celestial dynamics and the space environment require vastly different approaches and technologies. A somewhat different approach for ground and space-based observatories is described in the chapter \textit{Atom Interferometer}. This chapter also gives an overview of the signals that will be discovered and the science that will be enabled by LISA. More details on sources for space-based gravitational wave detectors can be found in the chapters on \textit{The gravitational capture of compact objects by massive black holes} (EMRIs), \textit{Supermassive black hole mergers}, \textit{LISA and the Galactic Population of Compact Binaries} and \textit{Stochastic gravitational wave backgrounds of cosmological origin}. More details on the tests of fundamental physics that will be facilitated by space-based gravitational wave detectors can be found in the chapter \textit{Testing the Nature of Dark Compact Objects with Gravitational Waves}. Finally, the analysis of data from space-based gravitational wave detectors will rely on the availability of waveform models, the construction of which is described in the book section \textit{Gravitational wave modeling} and will use some of the techniques described in the chapters in the book section \textit{Data analysis techniques}.

\section*{Acknowledgements}
We thank Emanuele Berti, Germano Nardini, Gijs Nelemans, Alberto Sesana, Robin T. Stebbins and Marta Volonteri for comments on the manuscript.


 \bibliographystyle{spmpsci}
 \bibliography{space_references}


\end{document}